\documentclass[11pt,a4paper]{article}%
\pdfoutput=1
\usepackage{jheppub}
\usepackage{amsfonts}
\usepackage{amsmath}
\usepackage{amssymb}
\usepackage{graphicx}%
\def\p{\partial}
\newcommand{\tr}{\mathop{\rm Tr}}
\def\expec#1{\langle #1 \rangle}

\newcommand{\cD}{{\mathcal{D}}}
\newcommand{\cE}{{\mathcal{E}}}

\newcommand{\cI}{{\mathcal{I}}}

\newcommand{\cL}{{\mathcal{L}}}
\newcommand{\cM}{{\mathcal{M}}}

\newcommand{\cO}{{\mathcal{O}}}

\newcommand{\cR}{{\mathcal{R}}}
\newcommand{\cS}{{\mathcal{S}}}

\newcommand{\cW}{{\mathcal{W}}}

\begin{document}

\title{Renormalized AdS gravity and holographic entanglement entropy of
even-dimensional CFTs}
\author[a]{Giorgos Anastasiou,}
\author[b]{Ignacio J. Araya,}
\author[c]{Alberto G\"uijosa}
\author[b]{and Rodrigo Olea}
\affiliation[a]{Instituto de F\'{i}sica, Pontificia Universidad Cat\'{o}lica de Valpara\'{i}so, 
Casilla 4059, Valpara\'{i}so, Chile}
\affiliation[b]{Departamento de Ciencias F\'{i}sicas, Universidad Andres Bello, Sazi\'{e}
2212, Piso 7, Santiago, Chile}
\affiliation[c]{Departamento de F\'{i}sica de Altas Energ\'{i}as, Instituto de Ciencias
Nucleares, Universidad Nacional Aut\'{o}noma de M\'{e}xico, Apartado Postal
70-543, CDMX 04510, M\'{e}xico}
\emailAdd{giorgos.g.anastasiou@gmail.com}
\emailAdd{araya.quezada.ignacio@gmail.com} 
\emailAdd{alberto@nucleares.unam.mx} 
\emailAdd{rodrigo.olea@unab.cl}

\abstract{
We derive a general formula for renormalized entanglement entropy in even-dimensional
CFTs holographically dual to Einstein gravity in one dimension higher.
In order to
renormalize, we adapt the Kounterterm method to asymptotically locally AdS manifolds with
conical singularities. On the gravity side, the computation considers
extrinsic counterterms and the use of the replica trick \textit{\`{a} la}
Lewkowycz-Maldacena. The boundary counterterm $B_{d}$ is shown to satisfy a key
property, in direct analogy to the Euler density: when evaluated on a conically singular manifold, it decomposes into a regular part plus a codimension-2 version of
itself located at the conical singularity. The renormalized entropy thus
obtained is shown to correspond to the universal part of the holographic
entanglement entropy, which for spherical entangling surfaces is proportional
to the central charge $a$ that is the subject of the $a$-theorem.
We also review and elucidate various aspects of the Kounterterm approach, including in particular its full compatibility with the Dirichlet condition for the metric at the conformal boundary, that is of standard use in holography.
}
\maketitle
\tableofcontents

\section{Introduction}

Entanglement, whose usefulness as an experimental resource has long been understood by the quantum information community, has deeply impacted theoretical high energy and gravity research in the past 15 years. On the one hand, it sheds valuable light on the dynamics of quantum field theories. Indeed, entanglement underlies correlation functions \cite{Wolf:2007tdq}, diagnoses phase transitions \cite{Osterloh:2002,Osborne:2002zz,Vidal:2002rm,Kitaev:2005dm,Levin:2006zz,Klebanov:2007ws,Huijse:2011ef}, and yields important insight on renormalization group flows \cite{Casini:2004bw,Myers:2010xs,Myers:2010tj,Jafferis:2011zi,Klebanov:2011gs,Casini:2012ei,Casini:2017vbe}. On the other hand, in the context of holographic duality \cite{Maldacena:1997re,Gubser:1998bc,Witten:1998qj}, a very profound link has been discovered between entanglement and gravity. This connection serves as a calculational tool for deducing the pattern of entanglement in certain strongly-coupled field theories \cite{Ryu:2006bv,Ryu:2006ef,Hubeny:2007xt},
and enables the emergence of a dynamical spacetime from degrees of freedom living on a lower-dimensional rigid geometry \cite{Swingle:2009bg,VanRaamsdonk:2010pw,Maldacena:2013xja}. For reviews on developments in these two directions, see e.g. \cite{Nishioka:2009un,Nishioka:2018khk} and \cite{VanRaamsdonk:2016exw,Rangamani:2016dms,Harlow:2018fse}, respectively.

In more detail, given a state $\varrho$ of a quantum system, and a subset $A$ of its degrees of freedom, entanglement between $A$ and its complement $A^\mathsf{c}$ is quantified by the entanglement entropy $S\equiv-\tr_A(\varrho_A\ln\varrho_A)$, with $\varrho_A\equiv\tr_{A^\mathsf{c}}\varrho$. In a quantum field theory, it is natural to choose $A$ to be a spatial region \cite{Bombelli:1986rw,Srednicki:1993im}.
$S$ is then extremely difficult to compute even at weak coupling, and can usually only be extracted, under favorable circumstances, using the replica trick \cite{Callan:1994py}. This is a procedure that employs the partition function $Z_n$ for the theory on a conically-singular manifold that stitches together $n$ replicas of the original spacetime, with $n\to 1$ at the end of the calculation. Remarkably, in strongly-coupled field theories with a holographic gravity dual, an effortless alternative is available. If one restricts attention to the regime of the `boundary' field theory where the `bulk' gravitational description involves classical Einstein gravity on an asymptotically locally anti-de Sitter spacetime (ALAdS) $\cM$, the entanglement entropy can be obtained through the simple relation
\begin{equation}\label{rt}
S=\frac{\mathrm{Area}[\Sigma]}{4G_{\mathrm{N}}}~.
\end{equation}
Here, $G_{\mathrm{N}}$ is Newton's constant, and the numerator refers to the area of the codimension-two surface $\Sigma\subset\cM$ that is minimal among all those that extend out to the conformal boundary $\partial \mathcal{M}$, are anchored there on (a locus conformal to) the entangling surface $\p A$, and can be continuously deformed to $A$. Relation (\ref{rt}) was proposed and motivated by Ryu and Takayanagi (RT) \cite{Ryu:2006bv,Ryu:2006ef}, partially justified later by Casini, Huerta and Myers in \cite{Casini:2011kv}, and more firmly established by Lewkowycz and Maldacena in \cite{Lewkowycz:2013nqa}, by applying the replica trick in the gravity side of the duality. Their procedure extracts $S$ from the replica partition function $Z_n$, computed via the standard GKPW recipe \cite{Gubser:1998bc,Witten:1998qj}. In the classical gravity limit, this boils down as usual to the exponential of the on-shell bulk action $I$. Relation (\ref{rt}) has been generalized to non-static configurations \cite{Hubeny:2007xt,Dong:2016hjy}, to higher-derivative theories of gravity \cite{Fursaev:2013fta,Dong:2013qoa,Camps:2013zua,Miao:2014nxa}, and to the domain where quantum corrections are incorporated in the gravitational description \cite{Faulkner:2013ana,Barrella:2013wja,Engelhardt:2014gca}.

Both sides of (\ref{rt}) are divergent, and must therefore be regularized. On the field theory side, this reflects the fact that the dominant contribution to the entanglement between $A$ and $A^\mathsf{c}$ arises from the UV degrees of freedom that are adjacent across $\p A$. On the gravity side, it is a consequence of the fact that the proper distance to the conformal boundary $\p\cM$ is infinite. A natural way to regularize is to place the boundary of $\cM$ at a large but finite distance, which is equivalent to introducing a UV cutoff in the field theory. Denoting the position-space version of this cutoff by $\varepsilon\to 0$, the entanglement entropy for a smooth region $A$ in a field theory on flat spacetime has the structure \cite{Liu:2012eea,Myers:2012ed}
\begin{align}
S =
\begin{cases}
s_{d-2} \left(\frac{L}{\varepsilon}\right)^{d-2} + s_{d-4} \left(\frac{L}{\varepsilon}\right)^{d-4} + \cdots
+s_1\, \frac{L}{\varepsilon}
+ {\sf s}_0 + {\cO}(\varepsilon^1)
\qquad  \ \text{for odd}\; d~,
\\
s_{d-2} \left(\frac{L}{\varepsilon}\right)^{d-2} + s_{d-4} \left(\frac{L}{\varepsilon}\right)^{d-4} + \cdots
+ {\sf s}_0 \,
\ln\left(\frac{L}{\varepsilon}\right) + {\cal O}(\varepsilon^0)
\qquad\;\text{for even} \; d~,
\end{cases}
\label{Sdiv}
\end{align}
where $d$ is the spacetime dimension of the field theory, and $L$ is a length scale that characterizes the size of $A$. In a conformal field theory (CFT), the coefficients $s_p$ are dimensionless numbers that depend only on the shape of the entangling surface $\p A$ \cite{Solodukhin:2008dh}. In a generic field theory, intrinsic mass scales $m_r$ are available to form dimensionless combinations $m_r\varepsilon$, and the $s_p$ are functions of these \cite{Liu:2012eea,Myers:2012ed}. All coefficients $s_p$ in (\ref{Sdiv}) depend on the regularization scheme and are therefore unphysical, except for ${\sf s}_0$. The latter coefficient is thus a \emph{universal} (i.e., scheme-independent) contribution to the entanglement entropy, which in general depends on the shape of $\p A$. In the remainder of the paper, the term in $S$ containing ${\sf s}_0$ will be denoted by $S_{\mathrm{univ}}$ (i.e., $S_{\mathrm{univ}}\equiv{\sf s}_0, {\sf s}_0\ln(L/\varepsilon)$ for odd and even $d$, respectively).

In the case of a CFT on flat spacetime where $A$ is chosen to be a ball of radius $L$, the universal coefficient ${\sf s}_0$ is believed to encode the irreversibility of renormalization group flow. Indeed, for even $d$ one finds that
${\sf s}_0=(-1)^{\frac{d}{2}+1}4 a$,
where $a$ is the central charge associated with the Euler density contribution to the conformal anomaly \cite{Ryu:2006ef,Solodukhin:2008dh,Myers:2010xs,Myers:2010tj,Casini:2011kv} (see Section \ref{anomalysec}).
A generalization of Zamolodchikov's $d=2$ $c$-theorem \cite{Zamolodchikov:1986gt} posits \cite{Cardy:1988cwa} that $a$ ought to decrease under renormalization flow between two fixed points, i.e.,
$a_{\mathrm{UV}}> a_{\mathrm{IR}}$.
For $d=4$, this $a$-theorem was proven in \cite{Komargodski:2011vj,Komargodski:2011xv}. For even $d\ge 6$, evidence for the $a$-theorem has been obtained with the aid of holography \cite{Myers:2010tj,Myers:2010xs,Liu:2012eea,Myers:2012ed} or supersymmetry \cite{Cordova:2015fha}.

For odd $d$, even though there is no conformal anomaly,  ${\sf s}_0$ still provides a direct analog of the central charge $a$. It can be shown \cite{Casini:2011kv} that
${\sf s}_0=(-1)^{\frac{d-1}{2}}2\pi F$,
where $F$ is the finite part of the free energy $-\ln Z$ associated with the Euclidean partition function $Z$ for the CFT on the $d$-sphere. For this quantity there also exists a statement of monotonicity, $F_{\mathrm{UV}}> F_{\mathrm{IR}}$. This $F$-theorem was proposed in \cite{Myers:2010tj,Jafferis:2011zi,Klebanov:2011gs}, and has received additional support from holography \cite{Myers:2010xs,Taylor:2016kic,Ghosh:2018qtg}, large-$N$ and $\epsilon$-expansion analysis \cite{Fei:2014yja}, as well as supersymmetric localization \cite{Jafferis:2012iv,Fluder:2018chf}. A quantity $\tilde{F}$ that usefully interpolates across dimensions between $a$ and $F$ was identified in \cite{Giombi:2014xxa}.

Entropic proofs of the $a$ theorem in $d=2,4$ and the $F$ theorem in $d=3$ were constructed respectively in \cite{Casini:2004bw},\cite{Casini:2017vbe} and \cite{Casini:2012ei}. For $d=2,3$ the proofs led to entropic $c$-functions, which interpolate monotonically between the corresponding values at the fixed points, but are in general not stationary there \cite{Casini:2005zv,Klebanov:2012va,Nishioka:2014kpa}, unlike Zamolodchikov's $c$-function. More recently, entropic reasoning based on tools that go beyond $S$ has led to additional monotonicity statements and other restrictions on renormalization flows \cite{Casini:2016fgb,Casini:2016udt,Casini:2018cxg}.

 In nearly all previous works, the universal term in the entanglement entropy, $S_{\mathrm{univ}}$, was extracted by hand from the divergent expression (\ref{Sdiv}). {}From the field-theoretic perspective, however, it is more natural to seek a renormalized version of entanglement entropy, just as one does for any other divergent quantity. A first attempt in this direction \cite{Liu:2012eea} involved a process of iterated differentiation of $S$ with respect to the size $L$. This recipe was shown to have some merit, but it is not connected with renormalization in the traditional sense. A different approach \cite{Casini:2015woa} is to sidestep the issue of renormalizing the entanglement entropy, considering instead the mutual information $\cI(A,B)\equiv S(A)+S(B)-S(AB)$, where UV divergences cancel out if the two regions $A$ and $B$ are disjoint.

 A systematic prescription for properly renormalizing $S$ was developed by Taylor and Woodhead in \cite{Taylor:2016aoi}, reasoning along two different lines. Most importantly, they noted that the connection established in \cite{Lewkowycz:2013nqa} between the entanglement entropy $S$ and the bulk gravitational action $I$ implies that the standard method of holographic renormalization \cite{Henningson:1998gx,Balasubramanian:1999re,Emparan:1999pm,deHaro:2000vlm,Skenderis:2002wp}, which eliminates divergences in $I$ by adding local counterterms defined on $\p \cM$, automatically induces the desired renormalization of $S$. Separately, Taylor and Woodhead worked out the counterterms on $\partial \Sigma$ needed to directly eliminate the divergences in the right-hand side of (\ref{rt}), and showed that this line of analysis leads to the same results as renormalization of $I$.

 The motivation for the present work stems from the existence of an interesting alternative method to renormalize the gravitational action $I$. In the usual approach, this action includes not only the bulk Einstein-Hilbert and cosmological constant terms, $I_{\mathrm{bulk}}$, but also the familiar York-Gibbons-Hawking boundary term, $I_{\mathrm{YGH}}$ \cite{York:1972sj,Gibbons:1976ue}. Beyond this, the local counterterms $I_{\mathrm{ct}}$ that are added to $I=I_{\mathrm{bulk}}+I_{\mathrm{YGH}}$ in the standard procedure of holographic renormalization depend only on quantities \emph{intrinsic} to the boundary $\p\cM$: the induced metric $h_{ij}$, and the associated Riemann tensor $\cR_{ijkl}$ and covariant derivative $\cD_i$. The required counterterms are obtained by detailed examination of the divergences that arise in $I$ when the bulk metric near $\p\cM$ is written out in the Fefferman-Graham (FG) expansion \cite{FG} (see Section \ref{fgsubsec}), and they need to be worked out separately for each bulk dimension $D=d+1$, or for diverse theories of gravity. An improved, Hamiltonian implementation of holographic renormalization for Einstein gravity was developed in \cite{Kraus:1999di,Papadimitriou:2004ap,Papadimitriou:2005ii}, where a recursive formula is available to generate the counterterms appropriate to any given dimension.

 A different strategy for renormalization was proposed in \cite{Aros:1999kt,Mora:2004rx,Olea:2005gb,Olea:2006vd}, which employs counterterms involving the \emph{extrinsic} curvature of $\p\cM$, $K_{ij}$, and is consequently known as the \emph{Kounterterm} method. Remarkably, this approach, without resorting to FG expansion, leads to a closed expression for the boundary terms $I_{\mathrm{Kt}}$ that need to be added to $I_{\mathrm{bulk}}$ in any dimension, and moreover, yields expressions that work equally well for Einstein gravity, for any one of its Lovelock generalizations in arbitrary dimension \cite{Kofinas:2007ns}, or for generic quadratic curvature corrections in $D=4$ \cite{Giribet:2018hck}. In each case, adjustment is needed only in a single overall coefficient. Another very attractive feature of the method is its deep connection with geometrical structures: topological invariants in the case of even $D$ \cite{Aros:1999kt,Olea:2005gb} (in which case the method is alternatively known as \emph{topological renormalization}), and transgression forms in the case of odd $D$ \cite{Olea:2006vd}. Also of interest is the fact that the Kounterterms lead to a renormalized action that makes direct contact with the definition of renormalized volume for AAdS previously studied in the mathematical literature \cite{Anastasiou:2018mfk}, and is intriguingly linked on-shell with critical gravity \cite{Miskovic:2014zja,Anastasiou:2017mag} and conformal gravity \cite{Maldacena:2011mk,Anastasiou:2016jix}.

At first sight, the use of extrinsic counterterms on $\p\cM$ would seem to run contrary to the boundary condition normally used for holography, and indeed this distinction was emphasized in the early literature. Recall that holography works with a Dirichlet condition on the non-normalizable mode of the bulk field in question, which serves as the external source for the dual field theory operator \cite{Gubser:1998bc,Witten:1998qj}. In the case of the $(d+1)$-dimensional metric $G_{\mu\nu}$, dual to the field theory stress-energy tensor $T_{ij}$, this means that we fix not the induced metric on $\p\cM$, $h_{ij}$, which is in fact divergent, but a conformally-rescaled version of it: the leading coefficient $g^{(0)}_{ij}$ in the FG near-boundary expansion (see Section \ref{fgsubsec}). This is interpreted as the rigid background metric on which the field theory lives, which indeed sources $T_{ij}$. The independent, normalizable mode of the bulk metric, $g^{(d)}_{ij}$, appears at higher order in the FG expansion, and determines the one-point function $\expec{T_{ij}}$ in the given state \cite{Balasubramanian:1999re,Emparan:1999pm,deHaro:2000vlm}. The counterterms of standard holographic renormalization are by construction compatible with a Dirichlet boundary condition for $g^{(0)}_{ij}$. In other words, $I+I_{\mathrm{ct}}$ constitutes a well-posed variational principle under the condition
$\delta g^{(0)}_{ij}=0$.

By contrast, the Kounterterms are naturally associated \cite{Miskovic:2006tm,Olea:2006vd} with a variational principle that exploits the ALAdS structure of $\cM$ to hold the extrinsic curvature fixed on $\p\cM$, in the sense that
$\ell K^i_j=\delta^i_j$ (where $\ell$ is the asymptotic radius of curvature), i.e., $\delta K^i_j=0$. If this requirement were imposed at a non-asymptotic radial depth, it would be a mixed (Robin-type) boundary condition involving both the non-normalizable and normalizable modes of the metric, and would thereby be incompatible with the standard Dirichlet condition. The crucial point, however, is that for $I_{\mathrm{bulk}}+I_{\mathrm{Kt}}$ to define a well-posed variational problem, the condition
$\ell K^i_j=\delta^i_j$ only needs to hold asymptotically, up to an order in the FG expansion that \emph{does not} involve the normalizable mode $g^{(d)}_{ij}$. This property will be demonstrated in Section \ref{Ksubsec}. There is thus no incompatibility between the Kounterterms and the usual Dirichlet boundary condition, because $\delta g^{(0)}_{ij}=0$ indeed implies $\delta K^i_j=0$ at the required level of accuracy. (Other works have explored the viability of different boundary conditions for ALAdS gravity, of Neumann \cite{Compere:2008us,Krishnan:2016dgy,Krishnan:2016mcj,Krishnan:2016tqj} or Robin \cite{Krishnan:2017bte} type.)

Given that both of these alternative prescriptions succeed in eliminating the divergences of the bulk gravitational action, and both are ultimately based on the same choice of boundary condition, one would expect them to secretly agree. In other words, upon FG expansion, $I_{\mathrm{Kt}}$ should coincide with $I_{\mathrm{YGH}}+I_{\mathrm{ct}}$ (up to scheme-dependent finite terms, if these are allowed).
And indeed, this agreement has been proven explicitly in previous works, first for Einstein gravity in $D=3$ \cite{Miskovic:2006tm}, then for a subclass of Lovelock theories (Chern-Simons-AdS for odd $D$ and Born-Infeld-AdS for even $D$) \cite{Miskovic:2007mg}, and finally for Einstein gravity in arbitrary dimension \cite{Anastasiou:2018the}. In the latter work, full agreement was shown for all ALAdS spacetimes with $D\le 5$. For
 the calculation to be manageable in higher dimensions, attention was restricted to the subclass of ALAdS spacetimes that are asymptotically conformally flat (as detailed below in Section \ref{Ksubsec} and Appendix \ref{C}), and agreement was demonstrated up to terms of eighth-order in derivatives, which first arise in $D=10$ or $11$.

 Even if the counterterm and Kounterterm methods were known to  agree exactly in all situations, their implementation is sufficiently different that it would no doubt still be interesting to compare their implications for the renormalization of diverse quantities. This brings us back to the entanglement entropy (\ref{rt}), whose standard holographic renormalization has been carried out in \cite{Taylor:2016aoi}. It is worth going over this same analysis using Kounterterms, to determine whether or not the most appealing properties of  $I_{\mathrm{Kt}}$, namely its compact and universal nature and its direct link with geometric structures, are directly inherited by the corresponding entropy Kounterterms, $S_{\mathrm{Kt}}$. Motivated by this question, a subset of us got started on this task in \cite{Anastasiou:2018rla}, examining the Kounterterm renormalization of entanglement entropy for Einstein gravity on spacetimes with even values of the bulk dimension $D$, or equivalently, odd values of the CFT dimension $d$. The results of that work answered the motivating question in the affirmative: it was found there that, for even $D$, $S_{\mathrm{Kt}}$ is indeed just as compact and universal as $I_{\mathrm{Kt}}$, and has precisely the same topological character. Moreover, just as expected, $S+S_{\mathrm{Kt}}$ yields precisely the universal term in (\ref{Sdiv}) relevant for the $F$-theorem, $S_{\mathrm{univ}}=(-1)^{\frac{d-1}{2}}2\pi F$, in agreement with \cite{Taylor:2016aoi}.


 The purpose of the present paper is to examine the  cases complementary to \cite{Anastasiou:2018rla}, by applying the Kounterterm method, still within Einstein gravity, to renormalize the entanglement entropy $S$ for odd $D$ (even $d$). {}From the preceding discussion, the main novelties that we expect to encounter in our analysis are the presence of a conformal anomaly in the CFT, and the fact that the renormalization method is no longer topological. And of course, one of our central goals will be to establish whether or not $S+S_{\mathrm{Kt}}$ again yields the result expected from (\ref{Sdiv}), $S_{\mathrm{univ}}=(-1)^{\frac{d-2}{2}}4a\ln(L/\varepsilon)$, with $a$ the central charge relevant for the $a$-theorem.\footnote{The use of boundary terms of topological origin to obtain the universal contribution to the entanglement entropy of spheres in even $d$ CFTs was also explored in \cite{Herzog:2015ioa,Herzog:2016kno}, albeit in a different context.}

The paper is organized as follows. We begin in Section \ref{antecedentssec} by providing the necessary antecedents: the Lewkowycz-Maldacena (LM) prescription in \ref{lmsubsec}, standard holographic renormalization in \ref{fgsubsec}, the Kounterterm method in \ref{Ksubsec}, and the result of \cite{Anastasiou:2018rla} for entanglement entropy in odd CFT dimensions in \ref{oddsubsec}.  We then proceed in Section \ref{evensec} to the main calculation of our paper, applying LM to the Kounterterm-renormalized bulk action
in even CFT dimensions. For this purpose, we give in \ref{bpqsubsec} a useful decomposition of the action Kounterterm $B_d$ into simpler building blocks whose behavior on the replica orbifold is determined in \ref{bpqsplitsubsec}. This allows us in \ref{bdsplitsubsec} to split  $B_d$ into its regular and conically-singular parts, leading us in \ref{srensubsec} to a concrete general expression for the renormalized entanglement entropy $S_{\mathrm{ren}}$, given in (\ref{sreneven}), which is our main result. In Section \ref{verifsec} we evaluate the Fefferman-Graham expansion of the two parts of $S_{\mathrm{ren}}$, namely the entropy Kounterterm $S_{\mathrm{Kt}}$ in \ref{sktsubsec} and the Ryu-Takayanagi term in \ref{rtsubsec}, which are then combined in \ref{cancellationsubsec} to explicitly verify the cancellation of the divergences. In Section \ref{anomalysec} we finally show that our result for $S_{\mathrm{ren}}$ indeed provides the desired universal part of the entanglement entropy,  $S_{\mathrm{univ}}$, allowing us to determine the $a$ central charge of the CFT. Conclusions and directions for future work are given in Section \ref{conclusionsec}. Additional material is included in four appendices. The last two of these deal with results that are central to our discussion of the validity of the Kounterterm method, but whose derivation is too lengthy to be included in the main text.

\section{Antecedents} \label{antecedentssec}

\subsection{Entanglement Entropy (EE) from the Gravitational Action}\label{lmsubsec}

As mentioned in the Introduction, Lewkowycz and Maldacena (LM) \cite{Lewkowycz:2013nqa} justified the Ryu-Takayanagi (RT) prescription for entanglement entropy (EE) by implementing the replica trick \cite{Callan:1994py} in the gravity side of the duality. Their approach starts from the usual presentation of the EE in the CFT as the $n\to 1$ analytic continuation of Renyi entropies $S_n$ obtained via path integration,
\begin{equation}
S=\lim_{n\to 1} S_n=-n\p_n\left[\ln Z(n)-n\ln Z(1)\right]|_{n=1}~,
\label{replica}
\end{equation}
where $Z(n)\equiv\tr(\varrho^n)$ is the partition function for the CFT on the $n$-replicated spacetime. LM then invoke the GKPW recipe \cite{Gubser:1998bc,Witten:1998qj} to evaluate (\ref{replica}) using the classical gravity approximation $Z(n)=\exp[-I(n)]$, with $I(n)$ the on-shell gravitational action for the replicated bulk geometry $\cM(n)$. If the $\bf{Z}_n$ replica symmetry is preserved in the bulk, we can write $I(n)=n I^{(n)}$, where $I^{(n)}$ is the action integrated over only one of the $n$ replicas, i.e., over the orbifold $\cM^{(n)}\equiv\cM(n)/{\bf Z}_n$. The EE can then be rewritten as
\begin{equation}
S=n^2\p_n I^{(n)}|_{n=1}~.
\label{lm}
\end{equation}

For Einstein gravity, $\cM^{(n)}$ is a squashed cone (i.e., a cone that in general lacks $U(1)$ isometry) with angular deficit
$2\pi\left(1-1/n\right)$,
sourced by a codimension-2 cosmic brane with tension $T=\frac{\left( 1-1/n\right)
}{4G_{\mathrm{N}}}$, anchored on the entangling surface $\p A\subset \p\cM$, and coupled to the bulk
metric through the Nambu-Goto action. In the tensionless limit
($n=1$), the location of the cosmic brane becomes the usual
 minimal surface $\Sigma$, and (\ref{lm}) yields \cite{Lewkowycz:2013nqa} the RT formula (\ref{rt}). For finite tension, there is backreaction of the brane on the ambient geometry, but the right-hand side of (\ref{lm}) still localizes on the brane, and yields a holographic prescription for computing Renyi entropies \cite{Lewkowycz:2013nqa,Dong:2016fnf}. In general, the
location of the cosmic brane, and therefore of the conical
singularity, corresponds to the fixed-point set of the replica symmetry.
(If the replica symmetry is broken, (\ref{lm}) does not hold, but the RT formula (\ref{rt}) can still be derived a la LM \cite{Camps:2014voa}.)

It is convenient to employ not the continued replica parameter $n$ but its inverse $\alpha\equiv 1/n$, such that the cone $\cM^{(\alpha)}$ has angular deficit $2\pi(1-\alpha)$. Equation (\ref{lm}) then becomes
\begin{equation}
S=-\partial_{\alpha}I^{\left( \alpha\right)}  \vert _{\alpha=1}.
\label{lm2}
\end{equation}
This connection between EE and the gravitational action makes it clear that renormalization of the latter will induce renormalization of the former \cite{Taylor:2016aoi}. The standard framework for analyzing this is the Fefferman-Graham expansion for the bulk metric, which we briefly review in the following subsection.

\subsection{The Fefferman-Graham Expansion and Holographic Renormalization} \label{fgsubsec}

We work with the Einstein-Hilbert action in $D=d+1$ dimensions,
\begin{equation}
I_{\mathrm{bulk}}=\frac{1}{16\pi G_{\mathrm{N}}}
{\displaystyle\int\limits_{\cM}}
d^{d+1}x\sqrt{-G}\left( R-2\Lambda\right)~,
\label{eh}
\end{equation}
with negative cosmological constant $\Lambda= -d(d-1)/2\ell^2$, which sets the asymptotic radius of curvature $\ell$. For the standard formulation of the Dirichlet boundary problem, (\ref{eh}) is complemented with the York-Gibbons-Hawking (YGH) boundary term
\begin{equation}
I_{\mathrm{YGH}}=-\frac{1}{8\pi G_{\mathrm{N}}}
{\displaystyle\int\limits_{\p\cM}}
d^{d}x\sqrt{-h}K~,
\label{ygh}
\end{equation}
where $h_{ij}$ is the induced metric on the boundary, and
$K$ is the trace of the extrinsic curvature.

For an ALAdS$_{d+1}$ spacetime, the metric near the conformal boundary can be expanded in the Fefferman-Graham (FG) form \cite{FG}
\begin{equation}
ds^2=G_{\mu\nu}(x,z)dx^{\mu}dx^{\nu}=
\frac{1}{z^2}
\left(\ell^2 dz^2+g_{ij}(x,z)dx^i dx^j\right)~,
\label{fgmetric}
\end{equation}
where
\begin{equation}
g_{ij}(x,z)=g^{(0)}_{ij}(x)+z^2 g^{(2)}_{ij}(x)+\ldots+z^d \ln (z^2)h^{(d)}_{ij}(x)+z^d g^{(d)}_{ij}
+\ldots~,
\label{gexpansion}
\end{equation}
and we have chosen to use a dimensionless radial coordinate $z$, such that the boundary is at $z=0$. In what follows, the number that appears here in parentheses as a superindex, indicating the order in the radial expansion, will be positioned at times as a subindex, if that gets less in the way of the other symbols.

Einstein's equation can be solved order by order in $z$ to determine all functions $g^{(p)}_{ij}(x)$ for $p<d$  in terms of $g^{(0)}_{ij}(x)$.  The function $h^{(d)}_{ij}(x)$ (not to be confused with the induced metric $h_{ij}(x)$) is also determined by $g^{(0)}_{ij}(x)$. It equals the metric variation of the holographic conformal anomaly \cite{Henningson:1998gx}, and is thus nonvanishing only for even $d$. The traceless and divergenceless piece of the function $g^{(d)}_{ij}(x)$ is indepedent of $g^{(0)}_{ij}(x)$, and sets the second boundary condition needed for the second-order Einstein equation. It is the normalizable mode of the bulk metric. In CFT language,  $g^{(0)}_{ij}(x)$ is the rigid background metric on which the field theory lives, while $g^{(d)}_{ij}(x)$ determines \cite{Balasubramanian:1999re,Emparan:1999pm,deHaro:2000vlm} the expectation value of the stress-energy tensor $T_{ij}(x)$, which is the operator dual to $G_{\mu\nu}(x,z)$. An important point is that bulk diffeomorphisms that preserve the FG gauge (\ref{fgmetric}) induce a conformal rescaling of $g^{(0)}_{ij}(x)$ \cite{Penrose:1986ca,Brown:1986nw,Imbimbo:1999bj}. This implies the well-known fact that a given bulk metric determines not a specific CFT metric, but only a conformal equivalence class  $\left[g^{(0)}_{ij}(x)\right]$  \cite{Witten:1998qj}.

As indicated in (\ref{gexpansion}), for the case of pure gravity the coefficients of the odd powers of $z$ vanish. For this reason, it is convenient to  rewrite the expansion in terms of the radial coordinate $\rho\equiv z^2$:
\begin{equation}
ds^2=
\frac{\ell^2}{4\rho^2}d\rho^2+\frac{1}{\rho}g_{ij}(x,\rho)dx^i dx^j~,
\label{fgmetricrho}
\end{equation}
with
\begin{equation}
g_{ij}(x,\rho)=g^{(0)}_{ij}(x)+\rho g^{(2)}_{ij}(x)+\ldots
+\rho^{d/2} \ln\rho\; h^{(d)}_{ij}(x)+\rho^{d/2} g^{(d)}_{ij}
+\ldots
\label{gexpansionrho}
\end{equation}

The FG expansion (\ref{fgmetricrho})-(\ref{gexpansionrho}) is the starting point for standard holographic renormalization. This entails expanding the action $I=I_{\mathrm{bulk}}+I_{\mathrm{YGH}}$ on shell, regularized with a radial cutoff at $\rho=\epsilon$, to identify the divergent terms \cite{deHaro:2000vlm}
\begin{equation}
I=\frac{1}{16\pi G_{\mathrm{N}}}{\displaystyle\int\limits_{\rho=\epsilon}}
d^{d}x\sqrt{-g^{(0)}}
\left(\epsilon^{-d/2}a_{(0)}+\epsilon^{-d/2+1}a_{(1)}+\ldots+\epsilon^{-1}a_{(d-2)}
+\ln\epsilon\; a_{(d)}+\mathcal{O}(\epsilon^0)\right)~,
\label{Idiv}
\end{equation}
and then eliminating these through minimal subtraction. The nontrivial part is that this subtraction can be reinterpreted as being due to the addition of local counterterms \emph{intrinsic} to the boundary, $I_{\mathrm{ren}}\equiv I+I_{\mathrm{ct}}$, with \cite{deHaro:2000vlm}
\begin{equation}
\begin{split}
I_{\mathrm{ct}}=\frac{1}{16\pi G_{\mathrm{N}}}{\displaystyle\int\limits_{\rho=\epsilon}}
d^{d}x\sqrt{-h}&\left[2(d-1)
+\frac{1}{d-2}\cR\right.
\\
&\left.+\frac{1}{(d-4)(d-2)^2}\left(\cR_{ij}\cR^{ij}-\frac{d}{4(d-1)}\cR^2\right)
+\ldots
-\epsilon^{d/2}\ln\epsilon\; a_{(d)}
\right]~.
\end{split}
\label{ct}
\end{equation}
It is understood here that, for a given $d$, one should only include the counterterms that are divergent, namely those where the first $(d-p)$ factor in the denominator is positive. The coefficient $a_{(d)}$, which is read off from (\ref{Idiv}), is related to both $h_{ij}^{(d)}$ and to the trace of $g_{ij}^{(d)}$ in (\ref{fgmetricrho}), and is nonvanishing only for even $d$. It sets the holographic conformal anomaly \cite{Witten:1998qj,Henningson:1998gx}, and admits a reformulation with the same structure as the term in (\ref{ct}) for which the first factor $(d-p)$ in the denominator would have vanished. E.g., for $d=2$, $\epsilon a_{(2)}=(1/2)\cR$, and for $d=4$, $\epsilon^2 a_{(4)}=(1/8)(\cR_{ij}\cR^{ij}-\cR^2/3)$.
It is worth emphasizing that the logarithimic counterterm is different from all the other terms in (\ref{ct}), due to its explicit dependence on the radial coordinate $\rho=\epsilon$. This lack of covariance is the origin of the anomaly.

The fact that (\ref{ct}) involves only curvatures intrinsic to the boundary guarantees that the addition of $I_{\mathrm{ct}}$ is compatible with the usual Dirichlet boundary condition $\delta g^{(0)}_{ij}=0$. Still, the brute-force computation of successive terms in (\ref{ct}) is tedious and highly theory-dependent. It is therefore interesting that, as mentioned in the Introduction, an alternative strategy is available for renormalization, which does not suffer from these drawbacks, and is thus more efficient for some purposes than traditional holographic renormalization. This is the Kounterterm method, to which we turn next.

\subsection{Kounterterms}\label{Ksubsec}

CFTs in even and odd spacetime dimension $d$ differ, due to the conformal anomaly. Correspondingly, ALAdS spacetimes with odd and even values of the bulk dimension $D=d+1$ differ too, as is  evident in the FG expansion reviewed in the previous subsection. The Kounterterm method reflects this distinction: for odd \cite{Aros:1999kt,Olea:2005gb} and even \cite{Mora:2004rx,Olea:2006vd} values of $d$, the required Kounterterms have a different structure and geometric interpretation. The detailed implementation of the method needs to be examined separately for the two cases, but we will begin here by writing out the defining expressions in parallel.

 The Kounterterm-renormalized gravitational action for an  ALAdS$_{d+1}$ bulk spacetime $\cM$ is given by \cite{Aros:1999kt,Mora:2004rx,Olea:2005gb,Olea:2006vd}
\begin{equation}
I_{\mathrm{ren}}\equiv I_{\mathrm{bulk}}+I_{\mathrm{Kt}}=\frac{1}{16\pi G_{\mathrm{N}}}
\left\{\;
{\displaystyle\int\limits_{\cM}}
d^{d+1}x\sqrt{-G}\left( R-2\Lambda\right)
+c_{d}%
{\displaystyle\int\limits_{\partial\cM}}
d^{d}x\sqrt{-h}\,B_{d}
\right\}~,
\label{iren}
\end{equation}
where the coefficient of the boundary action is
\begin{align}
c_{d}=
\begin{cases}
\frac{\left(  -1\right)^{\frac{d+1}{2}}
2{\ell}^{d-1}}{(d+1)\left(
d-1 \right)!}
\qquad\;\;\,  \text{for odd}\; d~, \\
\frac{\left(  -1\right)  ^{\frac{d}{2}}{\ell}^{d-2}}{2^{d-3}
d\left[  \left(\frac{d}{2}-1\right)  !\right]^{2}}
\qquad  \text{for even}\; d~,
\end{cases}
\label{cd}
\end{align}
and the Kounterterms can be expressed compactly as
\begin{align}
B_{d}=
\begin{cases}
\qquad
-(d+1)
{\displaystyle\int\limits_{0}^{1}}
ds\,\delta_{\left[d\right]  }^{\left[d\right]}
K\left(  \frac{1}{2}\cR ie-s^{2}KK\right)^{\frac{d-1}{2}}
\qquad\;\;\,  \ \text{for odd}\; d~,
\\
-d
{\displaystyle\int\limits_{0}^{1}}
ds
{\displaystyle\int\limits_{0}^{s}}
dt\,\delta_{\left[d-1\right]  }^{\left[d-1\right]  }K\left(  \frac{1}
{2}\cR ie-s^{2}KK+\frac{t^{2}}{{\ell}^{2}}\delta\delta\right)^{\frac{d}{2}-1}
\quad\;\text{for even} \; d~.
\end{cases}
\label{bd}
\end{align}
Here, and in the remainder of this paper, we use a shorthand notation to simplify tensorial
expressions. The generalized antisymmetric Kronecker delta,
defined as $\delta^{\left[  j_{1}\cdots j_{p}\right]}_{\left[ i_{1}\cdots i_{p}\right]  }
\equiv\det\left[  \delta_{i_{1}}^{j_{1}}\cdots\delta_{i_{p}}^{j_{p}}\right]  $, is indicated only
by the number of indices, i.e., $\delta_{\left[p\right]}^{\left[p\right]}$. Also, as the tensorial expressions appearing in Kounterterm Lagrangians are
fully contracted using Kronecker deltas, the indices of tensors are
omitted. For example, the rank $\binom{2}{2}$ intrinsic Riemann tensor $\cR_{j_{1}j_{2}}^{i_{1}i_{2}}$ is denoted as $\cR ie$,
the rank $\binom{1}{1}$
extrinsic curvature $K_{j_{1}}^{i_{1}}$ is denoted as $K$, etc. To avoid
ambiguity in the notation, traces of the tensors will be  explicitly indicated
 when needed (e.g., $\tr\left[  K\right]$). Also, uncontracted powers of
tensors (e.g., $K_{j_{1}}^{i_{1}}K_{j_{2}}^{i_{2}}$) are denoted as
powers of the symbol representing the tensor (e.g., $K^{2}$). As a simple illustration, we present here a Kronecker-delta identity \cite{Miskovic:2008ck} that will be useful below:
\begin{equation}
\delta_{\left[p\right]}^{\left[p\right]}\delta^q
\equiv
\delta^{\left[  j_{1}\cdots j_{p}\right]}_{\left[ i_{1}\cdots i_{p}\right]}
\delta^{i_1}_{j_1}\cdots\delta^{i_q}_{j_q}
=\frac{(d-p+q)!}{(d-p)!}
\delta_{\left[p-q\right]}^{\left[p-q\right]}~,
\label{kronecker}
\end{equation}
valid for $q\le p \le d$.
A final comment about our notation is that the integrals in (\ref{bd}) over the auxiliary parameters $s$ and $t$  are simply an efficient way to summarize the correct numerical coefficients for the various terms.

Action (\ref{iren}) reproduces the correct asymptotic charges (including the vacuum
energy) and thermodynamic properties of ALAdS manifolds (like
AdS black holes, for example). The presence of the extrinsic curvature in (\ref{iren}), and the absence of the YGH term (\ref{ygh}), imply that the variational problem defined by $I_{\mathrm{ren}}\equiv I_{\mathrm{bulk}}+I_{\mathrm{Kt}}$ is non-standard. The details were studied already in the original works \cite{Aros:1999kt,Mora:2004rx,Olea:2005gb,Olea:2006vd},
and tensorial expressions for the variation of the action in arbitrary dimension were presented in \cite{Miskovic:2008ck,Miskovic:2010ui}. An important property not elucidated in those early references is that, contrary to appearances, $I_{\mathrm{ren}}$ is in fact well-posed under the standard \emph{Dirichlet} boundary condition for the CFT metric, $\delta g^{(0)}=0$. We will illustrate the main ideas here by focusing for concreteness on the case with boundary dimension $d=4$ (bulk dimension $D=5$). The systematic analysis for arbitrary $d$ is postponed to Appendix \ref{D}.

Since we are dealing with Einstein gravity, the on-shell variation of the bulk action takes the well-known form
\begin{equation}
\Delta I_{\mathrm{bulk}}=-\frac{1}{16\pi G_{\mathrm{N}}}
{\displaystyle\int\limits_{\partial\cM}}
d^{4}x\sqrt{-h}\,
\left[(h^{-1}\Delta h)^i_k K^k_i + 2\Delta K^i_i\right]~.
\label{deltaeh}
\end{equation}
Notice that we are denoting infinitesimal variations with $\Delta$ instead of $\delta$, to avoid the risk of confusion with the Kronecker deltas employed in the Kounterterms.  Note also that the global sign of (\ref{deltaeh}) is opposite to the usual one, because for us, the radial coordinate $\rho$ \emph{decreases} toward the conformal boundary, and this impacts the sign of $K$. The $\Delta K^i_i$ term in (\ref{deltaeh}) would cancel against the corresponding part of $\Delta I_{\mathrm{YGH}}$ if (\ref{ygh}) were present; but, this not being the case, we must analyze how it combines with the variation of the Kounterterms.
{}From (\ref{iren})-(\ref{bd}), we see that for $d=4$, the normalization constant $c_4=\ell^2/8$, and after carrying out the integrals over the auxiliary parameters $s,t$, the Kounterterm action reads
\begin{equation}
I_{\mathrm{Kt}}=-\frac{\ell^2}{32\pi G_{\mathrm{N}}}
{\displaystyle\int\limits_{\partial\cM}}
d^{4}x\sqrt{-h}\,\delta^{[3]}_{[3]}K
\left(\frac{1}{4}\mathcal{R}ie-\frac{1}{3}K^2-\frac{1}{9\ell^2}\delta^2\right)
~.
\label{ikt}
\end{equation}
Its variation can be written as \cite{Miskovic:2010ui}
\begin{align}
\Delta I_{\mathrm{Kt}}=-\frac{\ell^2}{128\pi G_{\mathrm{N}}}
{\displaystyle\int\limits_{\partial\cM}}
d^{4}x\sqrt{-h}\,\delta^{[4]}_{[4]}
\left\{\left[(h^{-1}\Delta h)^{i_1}_k K^k_{j_1} + 2\Delta K^{i_1}_{j_1}\right]
\left( Rie\,\delta  + \frac{2}{3\ell^2}\delta^3 \right)
\right.~\nonumber\\
\left.
+\left[(h^{-1}\Delta h)^{i_1}_k
( K^k_{j_1}\delta^{i_2}_{j_2} - \delta^k_{j_1}K^{i_2}_{J_1})
+ 2\delta^{i_1}_{j_2}\Delta K^{i_2}_{j_2}\right]
\frac{1}{2}\left(\mathcal{R}ie-K^2+\frac{1}{\ell^2}\delta^2\right)
\right\}~,
\label{deltaikt}
\end{align}
where attention should be payed to the distinction between the boundary Riemann tensor $\mathcal{R}ie$ and (the boundary components of) its bulk counterpart $Rie$.

In the first line of (\ref{deltaikt}), it is convenient to use the on-shell relation with the bulk Weyl tensor $W$, to rewrite $Rie=W-2\delta^2/\ell^2$. The last term here combines then with the $\delta^3$ term in the first line of (\ref{deltaikt}), and knowing from (\ref{kronecker}) that
$\delta^{[4]}_{[4]}\delta^3=6\delta$, one can see that the result completely cancels (\ref{deltaeh}). We can similarly process the second line of (\ref{deltaikt}) by using first the Gauss-Codazzi equation
\begin{equation}
R^{ij}_{kl}=\cR^{ij}_{kl}-K^i_k K^j_l +K^i_l K^j_k~,
\end{equation}
which under the antisymmetrization imposed by $\delta^{[4]}_{[4]}$, implies that
$\cR ie = Rie +2K^2$. Next we can use again the relation $Rie=W-2\delta^2/\ell^2$.
Altogether, we are thus left with
\begin{align}
\Delta I_{\mathrm{ren}}&=-\frac{\ell^2}{128\pi G_{\mathrm{N}}}
{\displaystyle\int\limits_{\partial\cM}}
d^{4}x\sqrt{-h}\,\delta^{[4]}_{[4]}
\Bigg\{\left[(h^{-1}\Delta h)^{i_1}_k K^k_{j_1} + 2\Delta K^{i_1}_{j_1}\right]
W\,\delta
~\nonumber\\
&\left.
+\left[(h^{-1}\Delta h)^{i_1}_k
( K^k_{j_1}\delta^{i_2}_{j_2} - \delta^k_{j_1}K^{i_2}_{J_1})
+ 2\delta^{i_1}_{j_2}\Delta K^{i_2}_{j_2}\right]
\frac{1}{2}\left(W+K^2-\frac{1}{\ell^2}\delta^2\right)
\right\}~.
\label{deltairen}
\end{align}

If the boundary $\partial\cM$ were located at an arbitrary finite value of the radial coordinate $\rho$, the appearance of $\Delta K^{i}_{j}$ in  (\ref{deltairen})  would clearly signify an incompatibility with the choice of Dirichlet boundary conditions.  In spite of this, we will now demonstrate that compatibility is in fact achieved in the setting dual to a continuum CFT, where the boundary lies at $\rho\to 0$. For this purpose, one needs to examine the behavior of the objects appearing in (\ref{deltairen}) under the FG expansion (\ref{fgmetricrho})-(\ref{gexpansionrho}). We know of course that
\begin{equation}
\sqrt{-h}=\sqrt{-g_{(0)}}(\rho^{-2}+\ldots),
\qquad
(h^{-1}\Delta h)^{i}_k=(g^{-1}_{(0)}\Delta g_{(0)})^{i}_{k}+\ldots
\label{hfg}
\end{equation}
and solving the Einstein equation order by order in $\rho$, one finds that \cite{deHaro:2000vlm,Olea:2006vd,Anastasiou:2018the}
\begin{align}
K^i_j &= K^i_{(0)j} + \rho K^i_{(2)j}
+ \rho^2\ln\rho K^i_{(4h)j} + \rho^2 K^i_{(4)j} + \ldots ~,\\
W^{ij}_{kl} &= \rho W^{ij}_{(0)kl} +  \rho^2 W^{ij}_{(2)kl} + \ldots ~,
\nonumber
\end{align}
with
\begin{align}
K^i_{(0)j}=\frac{1}{\ell}\delta^i_j~,
\qquad
K^i_{(2)j}=-\frac{1}{\ell}g^i_{(2)j}
=\frac{\ell}{2}\left(\cR^i_{(0)j}-\frac{1}{6}\cR_{(0)}\delta^i_j\right),
\qquad
W^{ij}_{(0)kj}=0
\label{kwfg2}
\end{align}
(notice that in  $W_{(0)}$ we have taken a single trace over the boundary indices). The key point here is that in the intrinsic curvature, the normalizable mode of the metric, $g^{(4)}_{ij}$, enters only at order $\rho^2$, in $K^i_{(4)j}$. Additionally, the leading term of $K^i_j$ is a constant, so
$\Delta K^i_j=\rho \Delta K^i_{(2)j}
\propto\rho(g^{-1}_{(0)}\Delta g_{(0)})^i_j$, with corrections involving
$\Delta g^{(4)}_{kj}$ that appear at order $\rho^2$.

Armed with this information, we can work out the leading order contribution from each of the two lines in (\ref{deltairen}). In the first line, reading from (\ref{kronecker}) that
$\delta^{[4]}_{[4]}\delta=\delta^{[3]}_{[3]}$, we have
\begin{align}
\sqrt{-h}\delta^{[3]}_{[3]}\left[(h^{-1}\Delta h)^{i_1}_k K^k_{j_1} + 2\Delta K^{i_1}_{j_1}\right] W
&= \sqrt{-g_{(0)}}\rho^{-2}\delta^{[3]}_{[3]}
\left[(g_{(0)}^{-1}\Delta g_{(0)})^{i_1}_{k}\frac{1}{\ell}\delta^{k}_{j_1}\right]
W^{i_2 i_3}_{j_2 j_3}\nonumber\\
&=-\frac{4}{\ell}\sqrt{-g_{(0)}}\rho^{-2}(g_{(0)}^{-1}\Delta g_{(0)})^{i}_{j}
W^{jk}_{ik}\nonumber\\
&=-\frac{4}{\ell}\sqrt{-g_{(0)}}(g_{(0)}^{-1}\Delta g_{(0)})^{i}_{j}
W^{jk}_{(2)ik}~.
\label{firstline}
\end{align}
For the successive equalities, we first employed the initial relation in (\ref{kwfg2}) and the fact that $\Delta K^{i_1}_{j_1}$ is subleading, then applied the contractions with $\delta^{[3]}_{[3]}$, and finally took into account the last relation in (\ref{kwfg2}). In the second line of (\ref{deltairen}), we find
\begin{align}
\label{secondline}
\sqrt{-h}&\delta^{[4]}_{[4]}\left[(h^{-1}\Delta h)^{i_1}_k
( K^k_{j_1}\delta^{i_2}_{j_2} - \delta^k_{j_1}K^{i_2}_{J_1})
+ 2\delta^{i_1}_{j_2}\Delta K^{i_2}_{j_2}\right]
\left(W+K^2-\frac{1}{\ell^2}\delta^2\right)
\\
&=\sqrt{-g_{(0)}}\delta^{[4]}_{[4]}\left[(g_{(0)}^{-1}\Delta g_{(0)})^{i_1}_k
( K^k_{(2)j_1}\delta^{i_2}_{j_2} - \delta^k_{j_1}K^{i_2}_{(2)J_1})
+ 2\delta^{i_1}_{j_2}\Delta K^{i_2}_{(2)j_2}\right]
\left(W_{(0)}+2  K_{(0)} K_{(2)}\right)~.
\nonumber
\end{align}
 Here we have used (\ref{deltairen}) again to learn that the overall contribution of the brackets is of order $\rho$, as is the one from the final parentheses, thereby canceling the factor of $\rho^{-2}$ from $\sqrt{-h}$. We see that both (\ref{firstline}) and (\ref{secondline}) are finite, and purely proportional to $\Delta g_{(0)}$. For $\Delta K^{i_2}_{(2)j_2}$ this was established below (\ref{kwfg2}). Our conclusion then is that $I_{\mathrm{ren}}$ is fully compatible with the standard Dirichlet boundary condition $\Delta g_{(0)}=0$,
 just as we intended to show.

Aside from yielding a well-posed Dirichlet variational problem, the most important property of the Kounterterms $I_{\mathrm{Kt}}$ is that they serve the purpose of rendering the action finite
\cite{Aros:1999kt,Mora:2004rx,Olea:2005gb,Olea:2006vd}. Renormalizing via Kounterterms clearly seems very different from the intrinsic counterterm approach reviewed in the previous subsection. In spite of this, as mentioned in the Introduction, explicit agreement between the two methods has been proven in many different settings \cite{Miskovic:2006tm,Miskovic:2007mg,Anastasiou:2018the}.
In other words, upon writing out $I_{\mathrm{Kt}}$ in the FG expansion, it is found to match $I_{\mathrm{YGH}}+I_{\mathrm{ct}}$. In more detail,
$I_{\mathrm{Kt}}$ \emph{is found to perfectly cancel all power-law divergences, but it leaves logarithmic divergences untouched}. This will be crucial for our story in the following sections.

In \cite{Anastasiou:2018the}, equivalence between Kounterterms and counterterms was demonstrated for Einstein gravity in all boundary dimensions $d\le 8$ (bulk dimensions $D\le 9$). The matching holds without qualifications for the full class of ALAdS spacetimes with $d\le 4$ ($D\le 5$). For
 the calculation to be manageable in higher dimensions, attention was restricted to the subclass of ALAdS spacetimes for which
 the boundary components of the bulk Weyl tensor have the falloff of the normalizable mode, i.e., $W^{ij}_{kl} \sim O(\rho^{d/2})$. As shown in Appendix \ref{C}, for
 $5\le d\le 8$ this is achieved if the CFT metric is conformally flat, $\cW^{ij}_{kl}=0$. For this reason, we refer to the bulk spacetimes in question as \emph{asymptotically conformally flat} (ACF). Upon expressing $Rie$ and $\cR ie$ in the Kounterterms in terms of $W$ (just as we did in going from
 (\ref{deltaikt}) to (\ref{deltairen})), the ACF condition guarantees that all instances of $W$ are subleading, and can therefore be dropped. This greatly facilitates the proof of the matching between $I_{\mathrm{Kt}}$ and $I_{\mathrm{YGH}}+I_{\mathrm{ct}}$. In \cite{Anastasiou:2018the} this matching was established up to terms of eighth-order in derivatives, which first arise in $d=9$ or $10$ ($D=10$ or $11$). For the same range $5\le d\le 8$, the ACF condition is also relevant for the proof of the compatibility of $I_{\mathrm{ren}}$ with the Dirichlet boundary condition, given in Appendix \ref{D}.

\subsection{Renormalized EE for Odd CFT Dimension} \label{oddsubsec}

Using the Kounterterm method reviewed in the previous subsection, in \cite{Anastasiou:2018rla} a subset of us carried out the renormalization of the entanglement entropy for odd CFT dimension $d$. In that case, the bulk dimension $D=d+1$ is even, and the Kounterterms have a topological origin. More specifically, in the generalized Gauss-Bonnet (Chern) theorem, $B_d$ as given in (\ref{bd}) is precisely the boundary contribution to the Euler characteristic $\chi$:
\begin{equation}
{\displaystyle\int\limits_{\cM}}
d^{d+1}x\sqrt{-G}\,\cE_{d+1}
-{\displaystyle\int\limits_{\partial\cM}}
d^{d}x\sqrt{-h}\,B_{d}
= (4\pi)^{\frac{d+1}{2}}({\scriptstyle{\frac{d+1}{2}}})!\,\chi(\cM)~,
\label{gaussbonnet}
\end{equation}
where $\sqrt{-G}\,\cE_{d+1}$ is the Euler density,
\begin{equation}
\mathcal{E}_{d+1}={2^{-\frac{d+1}{2}}}\delta_{\left[  d+1\right]
}^{\left[ d+1\right]  }\left(  Rie\right)^{\frac{d+1}{2}}~.
\label{euler}
\end{equation}
Due to (\ref{gaussbonnet}), Kounterterm renormalization for even $D$ can be equivalently implemented via the bulk term involving the Euler density \cite{Aros:1999kt,Olea:2005gb}.

It was shown in \cite{Anastasiou:2018rla} that, on the squashed cone
$\cM^{(\alpha)}\equiv\cM(n)/{\bf Z}_n|_{n=1/\alpha}$ relevant for the LM replica trick reviewed in Section \ref{lmsubsec},
the renormalized action (\ref{iren}) can be
separated into a bulk contribution and a contribution coming from the
codimension-2 brane that becomes the minimal surface $\Sigma$ in the tensionless $\alpha\to 1$ limit. This decomposition is achieved by employing the
separation of the Euler density in its regular and codimension-2 conically
singular parts, demonstrated by Fursaev, Patrushev and Solodukhin (FPS) \cite{Fursaev:1995ef,Fursaev:2013fta}, and also by relating $\cE_{d+1}$ to $B_d$ using (\ref{gaussbonnet}). One finds that
\begin{align}
{\displaystyle\int\limits_{{\mathcal{M}}^{\left(  \alpha\right)  }}}%
d^{d+1}x\sqrt{G}\,R^{\left(  \alpha\right)  } &  ={\displaystyle\int
\limits_{{\mathcal{M}}^{\left(  \alpha\right)  }\backslash\Sigma}}%
d^{d+1}x\sqrt{G}\,R+4\pi\left(  1-\alpha\right)  {\displaystyle\int
\limits_{\Sigma}}d^{d-1}y\sqrt{\gamma}~,\label{ehsplit}\\
{\displaystyle\int\limits_{\partial{\mathcal{M}}^{{\left(  \alpha\right)  }%
}}}d^{d}x\sqrt{-h}\,B_{d}^{\left(  \alpha\right)  } &  ={\displaystyle\int
\limits_{\partial{\mathcal{M}}^{\left(  \alpha\right)  }\backslash
\partial\Sigma}}d^{d}x\sqrt{-h}\,B_{d}+2\pi\left(  d+1\right)  \left(
1-\alpha\right)  {\displaystyle\int\limits_{\partial\Sigma}}d^{d-2}%
x\sqrt{\tilde{\gamma}}\,B_{d-2}~,\label{B_2n-1_split}%
\end{align}
where $\tilde{\gamma}$ is the induced metric on
 $\partial\Sigma$, the boundary of the minimal surface, which is conformally equivalent to the entangling surface $\p A$ in the CFT.
The decomposition of $B_d$ will be obtained again in the following Section, by a route different than the one employed in \cite{Anastasiou:2018rla}. As a consequence of
(\ref{ehsplit}) and (\ref{B_2n-1_split}), one has
\begin{equation}
I_{\mathrm{ren}}\left[ \cM^{\left(  \alpha\right)  }\right]
=
I_{\mathrm{ren}}\left[ \cM^{\left(  \alpha\right)  }\backslash\Sigma\right]
+\frac{\left(  1-\alpha\right)  }{4G_{\mathrm{N}}}
\left(\mathrm{Area}\left[  \Sigma\right]
+{\scriptstyle{\left(\frac{d+1}{2}\right)}}c_d
\int_{\partial\Sigma}d^{d-2}x \sqrt{\tilde{\gamma}}\,B_{d-2}\right)~.
\label{ialpha}
\end{equation}

Given (\ref{ialpha}), the computation of
$-\left. \partial_{\alpha}I_{\mathrm{ren}}\left[\cM^{\left(
\alpha\right)  }\right]  \right\vert _{\alpha=1}$ becomes trivial, and
using (\ref{lm2}) one gets
\begin{equation}
S_{\mathrm{ren}}=\frac{1}{4G_{\mathrm{N}}}
\left(\mathrm{Area}\left[  \Sigma\right]
+{\scriptstyle{\left(\frac{d+1}{2}\right)}}c_d
\int_{\partial\Sigma}d^{d-2}x \sqrt{\tilde{\gamma}}\,B_{d-2}\right)~.
\label{srenodd}
\end{equation}
We note that in taking the derivative with respect to the conical parameter
$\alpha$, the position of the brane is considered fixed, as the action has
to be evaluated on-shell before computing said derivative.
In \cite{Anastasiou:2018rla} the
finiteness of $S_{\mathrm{ren}}$ was demonstrated, and it was shown that
this renormalized holographic entropy corresponds as expected to $S_{\mathrm{univ}}$, the universal part of the EE. As explained in the Introduction,
for a spherical entangling surface in an odd-dimensional CFT one finds
\cite{Casini:2011kv}  that
$S_{\mathrm{univ}}=(-1)^{\frac{d-1}{2}}2\pi F$,
where $F$ is the finite part of the free energy $-\ln Z$ associated with the Euclidean partition function $Z$ for the CFT on the $d$-sphere. The $F$-theorem \cite{Myers:2010tj,Jafferis:2011zi,Klebanov:2011gs} stipulates that $F$ decreases under renormalization group flows, and is therefore directly analogous to the central charge $a$ of even-dimensional CFTs.

An interesting geometric interpretation of (\ref{srenodd}) and (\ref{iren}) was
established later in \cite{Anastasiou:2018mfk}, by demonstrating explicitly that $4G_{\mathrm{N}} S_{\mathrm{ren}}$ is the \emph{renormalized area} of the extremal surface $\Sigma$, and $I_{\mathrm{ren}}\left[\cM\right] $ is proportional
to the \emph{renormalized volume} of the manifold $\cM$, in the context of
conformal calculus \cite{ALBIN2009140,Alexakis2010,Yang2008}. These results are tied to the fact that, when examining the renormalized Einstein-AdS gravity action in the LM replica trick, the coupling of
the cosmic brane which sources the conical singularity is through an action
which is of Nambu-Goto form, but where instead of considering the area functional
of the brane, one considers its renormalized area.

\section{Renormalized EE for Even CFT Dimension} \label{evensec}

We now turn to the central objective of this paper: studying the renormalization of entanglement entropy (EE) in
even-dimensional CFTs through extrinsic counterterms.
For this purpose, we will follow the logic of \cite{Anastasiou:2018rla}, reviewed in the previous subsection. Starting with the action (\ref{iren}), the key step is again to determine its decomposition
into a regular and a conically singular part, by evaluating said action on
the orbifold $\cM^{\left(\alpha\right)}$. Given such a decomposition, we
will be able to extract $S_{\mathrm{ren}}$ using the replica formula (\ref{lm2}).

The decomposition of the Einstein action, known from \cite{Fursaev:2013fta}, is given by (\ref{ehsplit}).   In the following subsections, we will
show that the Kounterterms (\ref{bd})
do likewise split into a regular and a codimension-2  conically
singular part. In order to do this, we must evaluate
 $B_{d}$ on the boundary of the orbifold,
$\partial\cM^{\left(\alpha\right)}$. For achieving the desired split, we will first define
in Section \ref{bpqsubsec} certain
 boundary terms $b_{pq}$ from which $B_{d}$ is constructed, both for odd and even CFT dimension $d$. Then, in Section \ref{bpqsplitsubsec}
we will motivate the decomposition of each $b_{pq}$ into its regular and conically
singular parts. {}From this we will first recover the split for the $B_{d}$ in the case of odd $d$,
given in (\ref{B_2n-1_split}) and previously reported in \cite{Anastasiou:2018rla}, and we then determine the corresponding split for even $d$.
With this second result in hand, we will arrive in Section \ref{srensubsec} at an explicit formula for $S_{\mathrm{ren}}$ for even $d$.

\subsection{Decomposition of $B_d$ (for arbitrary $d$) into individual terms}
\label{bpqsubsec}

Considering the basic structures present in (\ref{bd}), we will find it convenient to define a set of boundary terms as follows:
\begin{equation}
b_{pq}\equiv
{\displaystyle\int\limits_{\partial \cM}}
d^{d}x\sqrt{-h}\,\delta_{\left[  2p+1\right]  }^{\left[  2p+1\right]  }\left(
\cR ie\right)  ^{q}\left(  K\right)  ^{2\left(  p-q\right)  +1}~.
\label{b_pq}%
\end{equation}
These boundary terms are evaluated at the asymptotic boundary of a
$(d+1)-$dimensional ALAdS manifold $\cM$, considering the standard foliation
along the holographic radial coordinate. Our notation in (\ref{b_pq}) was explained below (\ref{bd}): $\cR ie$ is the
Riemann tensor of the induced metric $h$ at the boundary, and $K$ is its
extrinsic curvature tensor. Considering certain linear combinations of the $b_{pq}$, we
can construct the Kounterterms $B_{d}$ that
renormalize the Einstein-AdS action for even-dimensional and odd-dimensional
bulk manifolds respectively. We now turn to determining these combinations explicitly.

\subsubsection{Construction of $B_{d}$ for odd $d$}

From (\ref{bd}), we know that in odd-dimensional CFTs
\begin{equation}
{\displaystyle\int\limits_{\partial \cM}}
d^d x\sqrt{-h}\,B_{d}=
-(d+1){\displaystyle\int\limits_{\partial \cM}}
d^{d}x\sqrt{-h}\,\delta_{\left[d\right]  }^{\left[d\right]}
{\displaystyle\int\limits_{0}^{1}}
ds\,
K\left(  \frac{1}{2}\cR ie-s^{2}KK\right)  ^{\frac{d-1}{2}}
\label{bdodd}
\end{equation}
Now, opening up the parenthesis, we obtain%
\begin{align}
{\displaystyle\int\limits_{\partial \cM}}
d^d x\sqrt{-h}\,B_{d}
&=-(d+1)
{\displaystyle\int\limits_{\partial\cM}}
d^{d}x\sqrt{-h}\,\delta_{\left[d\right]  }^{\left[d\right]  }%
{\displaystyle\int\limits_{0}^{1}}
ds
{\displaystyle\sum\limits_{q=0}^{\frac{d-1}{2}}}
\binom{\frac{d-1}{2}}{q}\frac{\left(-1\right)^{\left(\frac{d-1}{2}-q\right)}}{2^q}
s^{d-2q-1}
\left(\cR ie\right)^{q}
\left(K\right)^{d-2q-1}
\nonumber \\
&=
{\displaystyle\sum\limits_{q=0}^{\frac{d-1}{2}}}
\binom{\frac{d-1}{2}}{q}
\frac{\left(-1\right)^{\left(\frac{d+1}{2}-q\right)}}{2^q(d-2q)}
(d+1)
{\displaystyle\int\limits_{\partial \cM}}
d^{d}x\sqrt{-h}\,\delta_{\left[d\right]}^{\left[d\right]}
\left(\cR ie\right)^{q}
\left(K\right)^{d-2q-1}~.
\end{align}
Comparing with (\ref{b_pq}), we see that we
can write the Kounterterms for odd $d$ as%
\begin{equation}
{\displaystyle\int\limits_{\partial \cM}}
d^d x\sqrt{-h}\,B_{d}
=
{\displaystyle\sum\limits_{q=0}^{\frac{d-1}{2}}}
C_{dq}\,b_{\frac{d-1}{2},q}~,
\label{B_2n-1_In_b}%
\end{equation}
where
\begin{equation}
C_{dq}=\frac{\left(-1\right)^{\left(\frac{d+1}{2}-q\right)}
\left(\frac{d+1}{2}\right)!}{2^{q-1}\left(
\frac{d-1}{2}-q\right)!q!\left(d-2q\right)}~.
\label{C_nq}
\end{equation}

\subsubsection{Construction of $B_{d}$ for even $d$}
Starting from (\ref{bd}) for even $d$, one can similarly carry out the
integrals over the auxiliary parameters $s,t$ to write out $B_d$ in terms of the
$b_{pq}$. The result is \cite{Kofinas:2007ns}
\begin{equation}
{\displaystyle\int\limits_{\partial \cM}}
d^d x\sqrt{-h}\,B_{d}=
{\displaystyle\sum\limits_{p=0}^{\frac{d}{2}}}
{\displaystyle\sum\limits_{q=0}^{p}}
C_{dp}^{\left(  1\right)  }C_{dpq}^{\left(  2\right)  }b_{pq},
\label{B_2n_In_b}
\end{equation}
where
\begin{align}
C_{dp}^{\left( 1\right)}
&=
-\left(\frac{d}{2}\right)!\frac{\left(d-2p-3\right)!!}{{\ell
}^{d-2p-2}}~,
\nonumber\\
C_{dpq}^{\left(2\right)}
&=\frac{\left(-1\right)^{p-q}}{\left(
p-q\right)  !q!}\frac{2^{\frac{d}{2}-\left(p+q+1\right)  }}{\left(\frac{d}{2}-q\right)
}
\label{C_npq}%
\end{align}
and $d \geq 2$.

\subsection{Splitting of $b_{pq}$ into its regular and conically singular parts}
\label{bpqsplitsubsec}

Having written the Kounterterms in terms of the $b_{pq}$, we
now proceed to decompose the $b_{pq}$ into their regular and conically singular parts.
First we have, from our proposed generalization of the FPS relation
\cite{Anastasiou:2018rla,Anastasiou:2018mfk}, that for even-dimensional bulk
manifolds (odd $d$), the integral of the Euler density (\ref{euler}) decomposes as
\begin{equation}
{\displaystyle\int\limits_{\cM^{\left(  \alpha\right)  }}}
d^{d+1} x\sqrt{-G}\,\mathcal{E}_{d+1}^{\left(  \alpha\right)  }=
{\displaystyle\int\limits_{\cM}}
d^{d+1}x\sqrt{-G}\,\mathcal{E}_{d+1}
+\left(  1-\alpha\right)2\pi(d+1)
{\displaystyle\int\limits_{\Sigma}}
d^{d-1}y\sqrt{\gamma}\,\mathcal{E}_{d-1}~.
\label{Euler-split}%
\end{equation}
In this equation $\cM$ refers to the regular part of the
$\cM^{\left(  \alpha\right)  }$ orbifold, i.e.,
$\cM=\cM^{\left(\alpha\right)  }\backslash\Sigma$.\footnote{We emphasize that this decomposition was proven 
by Fursaev and Solodukhin in \cite{Fursaev:1995ef} for conically-singular manifolds with a continuous $U(1)$ 
isometry and by Fursaev, Patrushev and Solodukhin in \cite{Fursaev:2013fta} for 
the particular case of the Gauss-Bonnet term in four-dimmensional squashed cones (conically-singular manifolds with a 
discrete symmetry instead of the U(1) isometry).}

Next, we note that, as observed by FPS in
\cite{Fursaev:1995ef,Fursaev:2013fta}, the expression resulting from the
decomposition of curvature invariants in their regular and conically-singular
parts is independent of the dimension of the ambient spacetime. Therefore, the
decomposition of the Euler density generalizes automatically to the Lovelock
densities $\cL_{2p}$, defined as
\begin{equation}
\cL_{2p}=\frac{1}{2^{p}}\delta_{\left[  2p\right]  }^{\left[  2p\right]
}\left(Rie\right)^{p}%
\end{equation}
for $2p<D$, where $D=d+1$ is the dimension of the ambient spacetime. Thus, we have that
\begin{equation}
{\displaystyle\int\limits_{\cM^{\left(  \alpha\right)  }}}
d^{d+1}x\sqrt{-G}\,\cL_{2p}^{\left(  \alpha\right)  }
=
{\displaystyle\int\limits_{\cM}}
d^{d+1}x\sqrt{-G}\,\cL_{2p}
+\left(  1-\alpha\right)4\pi p
{\displaystyle\int\limits_{\Sigma}}
d^{d-1}y\sqrt{\gamma}\,\cL_{2p-2}~.
\end{equation}
This relation can be rewritten as
\begin{equation}
{\displaystyle\int\limits_{\cM^{\left(  \alpha\right)  }}}
d^{d+1}x\sqrt{-G}\,\delta_{\left[  2p\right]  }^{\left[  2p\right]  }\left(
Rie^{\left(  \alpha\right)  }\right)  ^{p}
=
{\displaystyle\int\limits_{\cM}}
d^{d+1}x\sqrt{-G}\,\delta_{\left[  2p\right]  }^{\left[  2p\right]  }\left(
Rie\right)  ^{p}+\left(  1-\alpha\right)8\pi p
{\displaystyle\int\limits_{\Sigma}}
d^{d-1}y\sqrt{\gamma}\,\delta_{\left[  2p-2\right]  }^{\left[  2p-2\right]
}\left(\mathcal{R}ie\right)^{p-1}~.
\end{equation}

Finally, by considering that the minimal surface $\Sigma$ is normal to the
spacetime boundary, we have that the conical singularity does not affect the
radial extrinsic curvature, as discussed by Taylor and Woodhead in
\cite{Taylor:2016aoi}. Therefore, by analogy with the relation for the Lovelock densities, it is natural to propose a dimensional continuation of this decomposition at the boundary, such that
\begin{align}
{\displaystyle\int\limits_{\partial \cM^{\left(  \alpha\right)  }}}
d^{d}x\sqrt{-h}\,\delta_{\left[  2p+1\right]  }^{\left[  2p+1\right]  }\left(
Rie^{\left(  \alpha\right)  }\right)  ^{q}\left(  K\right)  ^{2\left(
p-q\right)  +1}  =%
{\displaystyle\int\limits_{\partial \cM}}
d^{d}x\sqrt{-h}\,\delta_{\left[  2p+1\right]  }^{\left[  2p+1\right]  }
Rie^{q}\left(  K\right)^{2\left(  p-q\right)
+1}
\nonumber\\
\qquad\qquad\qquad\qquad\qquad
+\left(  1-\alpha\right)8\pi q
{\displaystyle\int\limits_{\partial\Sigma}}
d^{d-2}y\sqrt{\tilde{\gamma}}\delta_{\left[  2p-1\right]  }^{\left[
2p-1\right]  }\left(  \mathcal{R}ie\right)  ^{q-1}\left(  k\right)  ^{2\left(
p-q\right)  +1},
\end{align}
where $K_{ij}$ is the radial extrinsic curvature of the spacetime boundary
$\partial M$ and $k_{ab}$ is the radial extrinsic curvature of the border
$\partial\Sigma$ of the RT surface, which has an induced metric denoted by
$\tilde{\gamma}$.

The preceding discussion motivates a decomposition formula for the $b_{pq}$ boundary terms. Namely, we propose that
\begin{equation}
b_{pq}^{\left(  \alpha\right)  }
=b_{pq}
+\left(
1-\alpha\right)8\pi q \, b_{p-1,q-1}^{\partial\Sigma}
\label{b_pq_split}
\end{equation}
for $p\geq q\geq1$, $b_{pq}^{\left(  \alpha\right)  }=b_{pq}$ for $q=0$
and $b_{p-1,q-1}^{\partial\Sigma}$ of course refers to the analog of (\ref{b_pq})
constructed from the geometric quantities $\cR ie$ and $k$ relevant to $\partial\Sigma$.
Ultimately, it is the results to be found below, on the cancellation of divergences and the recovery of the correct $a$ anomaly coefficient, that lend credence to this proposed decomposition.

\subsection{Splitting of $B_{d}$ for both odd and even $d$}
\label{bdsplitsubsec}

We now proceed to derive the decomposition formulas for the Kounterterms in both odd and even CFT dimension, by considering the way in which they are constructed from the $b_{pq}$, which decompose according to (\ref{b_pq_split}).

For odd $d$, $B_{d}$ is written in terms of the
$b_{pq}$ as shown in (\ref{B_2n-1_In_b}). Then, its decomposition is given by
\begin{equation}%
{\displaystyle\int\limits_{\partial\cM^{\left(  \alpha\right)  }}}
d^d x \sqrt{-h}\,B_{d}^{\left(  \alpha\right)  }
=
{\displaystyle\int\limits_{\partial \cM}}
d^d x\sqrt{-h}\,B_{d}
+4\pi\left(  1-\alpha\right)
{\displaystyle\sum\limits_{q=1}^{\frac{d-1}{2}}}
2qC_{dq}\,b_{\frac{d-3}{2},q-1}^{\partial\Sigma}~.
\end{equation}
Now, it can be seen from (\ref{C_nq})
that $2qC_{dq}=\frac{d+1}{2}C_{d-2,q-1}$,  so we have that
\begin{align}
{\displaystyle\int\limits_{\partial \cM^{\left(  \alpha\right)  }}}
d^dx\sqrt{-h}\,B_{d}^{\left(  \alpha\right)  }
& =
{\displaystyle\int\limits_{\partial \cM}}
d^dx\sqrt{-h}\,B_{d}
+2\pi (d+1)\left(  1-\alpha\right)
{\displaystyle\sum\limits_{q=1}^{\frac{d-1}{2}}}
C_{d-2,q-1}\,b_{\frac{d-3}{2},q-1}^{\partial\Sigma}\nonumber\\
&  =
{\displaystyle\int\limits_{\partial \cM}}
d^d x\sqrt{-h}\,B_{d}
+2\pi (d+1)\left(  1-\alpha\right)
{\displaystyle\sum\limits_{q=0}^{\frac{d-3}{2}}}
C_{d-2,q}b_{\frac{d-3}{2},q}^{\partial\Sigma}~,
\end{align}
and therefore,
\begin{equation}
{\displaystyle\int\limits_{\partial \cM^{\left(  \alpha\right)  }}}
d^d x\sqrt{-h}\,B_{d}^{\left(  \alpha\right)  }
=
{\displaystyle\int\limits_{\partial \cM}}
d^d x\sqrt{-h}\,B_{d}
+2\pi (d+1)\left(  1-\alpha\right)
{\displaystyle\int\limits_{\partial\Sigma}}
d^{d-2} y\sqrt{\tilde{\gamma}}\,B_{d-2}~.
\end{equation}
We thus recover here the splitting relation of (\ref{B_2n-1_split}), as conjectured in
\cite{Anastasiou:2018rla}.

Next we turn to the case of even $d$, where $B_{d}$ is written in terms of the
$b_{pq}$ as given in (\ref{B_2n_In_b}). Using again (\ref{b_pq_split}), its decomposition is given by
\begin{equation}
{\displaystyle\int\limits_{\partial \cM^{\left(  \alpha\right)  }}}
d^d x\sqrt{-h}\,B_{d}^{\left(  \alpha\right)  }
=
{\displaystyle\int\limits_{\partial \cM}}
d^d x\sqrt{-h}\,B_{d}
+8\pi\left(  1-\alpha\right)
{\displaystyle\sum\limits_{p=1}^{\frac{d}{2}-1}}
{\displaystyle\sum\limits_{q=1}^{p}}
C_{dp}^{\left(  1\right)  }C_{dpq}^{\left(  2\right)  }qb_{p-1,q-1}%
^{\partial\Sigma}~.
\end{equation}
Now, it can be seen from (\ref{C_npq}) that
$2qC_{dp}^{\left(  1\right)  }C_{dpq}^{\left(  2\right)
}=\frac{d}{2}C_{d-2,p-1}^{\left(  1\right)  }C_{d-2,p-1,q-1}^{\left(  2\right)}$, so we have
\begin{align}
{\displaystyle\int\limits_{\partial\cM^{\left(  \alpha\right)  }}}
d^d x\sqrt{-h}\,B_{d}^{\left(  \alpha\right)  }
&  =
{\displaystyle\int\limits_{\partial \cM}}
d^d x\sqrt{-h}\,B_{d}
+2\pi d\left(  1-\alpha\right)
{\displaystyle\sum\limits_{p=1}^{\frac{d}{2}-1}}
{\displaystyle\sum\limits_{q=1}^{p}}
C_{d-2,p-1}^{\left(  1\right)  }C_{d-2,p-1,q-1}^{\left(  2\right)
}b_{p-1,q-1}^{\partial\Sigma}
\nonumber\\
&  =
{\displaystyle\int\limits_{\partial \cM}}
d^d x\sqrt{-h}\,B_{d}
+2\pi d\left(  1-\alpha\right)
{\displaystyle\sum\limits_{p=0}^{\frac{d}{2}-2}}
{\displaystyle\sum\limits_{q=0}^{p}}
C_{d-2,p}^{\left(  1\right)  }C_{d-2,p,q}^{\left(  2\right)  }b_{p,q}%
^{\partial\Sigma}~,
\end{align}
and therefore
\begin{equation}
{\displaystyle\int\limits_{\partial \cM^{\left(  \alpha\right)  }}}
d^dx\sqrt{-h}\,B_{d}^{\left(  \alpha\right)  }
=
{\displaystyle\int\limits_{\partial \cM}}
d^dx\sqrt{-h}\,B_{d}
+2\pi d\left(  1-\alpha\right)
{\displaystyle\int\limits_{\partial\Sigma}}
d^{d-2}y\sqrt{\tilde{\gamma}}\,B_{d-2}~.
\label{B_2n_split}
\end{equation}
This finally gives us the desired splitting formula for the Kounterterms in odd-dimensional bulks (even-dimensional CFTs).  As expected, the codimension-2
$B_{d-2}$ is evaluated on the boundary $\partial\Sigma$ of the minimal surface,

\subsection{Explicit expression for $S_{\mathrm{ren}}$}
\label{srensubsec}

Having obtained the decomposition of $B_{d}$, we now find the explicit
expression for the renormalized EE of even-dimensional holographic CFTs.
Considering the expression for the renormalized Einstein-AdS action (\ref{iren}) evaluated
on the orbifold $\cM^{\left(  \alpha\right)  }$,
and using  the splitting formulas (\ref{ehsplit}) and
(\ref{B_2n_split}), we have that the renormalized on-shell action
decomposes as
\begin{equation}
I_{\mathrm{ren}}\left[\cM^{\left(  \alpha\right)  }\right]
=I_{\mathrm{ren}}\left[\cM^{\left(  \alpha\right)  }\backslash\Sigma\right]
+\frac{\left(  1-\alpha\right)  }{4G_{\mathrm{N}}}
\left({\mathrm{Area}}\left[\Sigma\right]
+\left({\scriptstyle\frac{d}{2}}\right)c_d
{\displaystyle\int\limits_{\partial\Sigma}}
d^{d-2}y\sqrt{\tilde{\gamma}}\,B_{d-2}\right)~.
\label{I_odd_decomp}
\end{equation}
With this, using the replica formula of (\ref{lm2}), we finally obtain
\begin{equation}
S_{\mathrm{ren}}=\frac{1}{4G_{\mathrm{N}}}
\left(\mathrm{Area}\left[  \Sigma\right]
+{\scriptstyle{\left(\frac{d}{2}\right)}}c_d
\int_{\partial\Sigma}d^{d-2}y \sqrt{\tilde{\gamma}}\,B_{d-2}\right)~,
\label{sreneven}
\end{equation}
which is the main result of this paper. The first term here is of course the Ryu-Takayangi entropy (\ref{rt}). The second term gives an explicit formula for
the Kounterterms $S_{\mathrm{Kt}}$ that are needed to renormalize it. The nontrivial observation is that, thanks to the self-replicating nature of the splitting formula (\ref{B_2n_split}), the Kounterterms for EE have exactly the same structure as the Kounterterms for the bulk action, in 2 dimensions less. We thus find that $S_{\mathrm{Kt}}$ indeed inherits all of the attractive features of
$I_{\mathrm{Kt}}$, described in Section \ref{Ksubsec}.

\section{Verification of divergence cancellation in $S_{\mathrm{ren}}$ to
next-to-leading order}
\label{verifsec}

Having obtained the formula (\ref{sreneven}) for the renormalized EE in even-dimensional CFTs, we proceed to evaluate
its two constituent parts: the usual
Ryu-Takayanagi piece \cite{Ryu:2006ef,Ryu:2006bv}
\begin{equation}
S_{\mathrm{RT}}=\frac{\mathrm{Area}\left[  \Sigma\right]  }{4G_{\mathrm{N}}}~,
\label{rt2}
\end{equation}
and the newly derived Kounterterms
\begin{equation}
S_{\mathrm{Kt}}=\frac{d\,c_d}{8G_{\mathrm{N}}}%
{\displaystyle\int\limits_{\partial\Sigma}}
d^{d-2}y\sqrt{\tilde{\gamma}}\,B_{d-2}~.
\label{sktbd-2}
\end{equation}
In this evaluation, we consider the explicit embedding of the minimal surface
$\Sigma$ in the ambient manifold $\cM$, as given in
\cite{Anastasiou:2018rla}.

\subsection{Evaluation of $S_{\mathrm{Kt}}$} \label{sktsubsec}

Using  (\ref{cd}) and (\ref{bd}) for the case of even $d$, we have
\begin{align}
S_{\mathrm{Kt}}
&  =\frac{\left(-1\right)^{\frac{d}{2}+1}{\ell}^{d-2}(d-2)}{2^{d}\,G_{\mathrm{N}}\left[
\left(\frac{d}{2}-1\right) !\right] ^{2}}
{\displaystyle\int\limits_{\partial\Sigma}}
d^{d-2}y
{\displaystyle\int\limits_{0}^{1}}
ds
{\displaystyle\int\limits_{0}^{s}}
dt\sqrt{\widetilde{\gamma}}\,
\delta_{\left[d-3\right]  }^{\left[d-3\right]}\,k
\left(  \frac{1}{2}\mathcal{R}ie-s^{2}kk+\frac{t^{2}}{{\ell
}^{2}}\delta\delta\right)^{\frac{d}{2}-2}~.
\label{B2n-2}
\end{align}
In this expression, $\mathcal{R}ie$ is the Riemann tensor of the induced
metric $\widetilde{\gamma}$ on the border of the minimal surface
$\partial\Sigma$, and $k$ is the radial extrinsic curvature of $\partial
\Sigma$.

We now consider a Fefferman-Graham-like decomposition of the relevant
codimension-2 tensors, given previously in \cite{Anastasiou:2018rla}, based on the works by Hung, Myers and Smolkin \cite{Hung:2011nu} and by Schwimmer and Theisen
\cite{Schwimmer:2008yh}. By definition, the induced metric $\gamma$ on the
minimal surface $\Sigma$ is given by $\gamma_{\alpha\beta}=\frac{\partial x^{\mu}%
}{\partial Y^{\alpha}}\frac{\partial x^{\nu}}{\partial Y^{\beta}}G_{\mu\nu}$. Then,
when $G_{\mu\nu}$ is of the usual FG form (\ref{fgmetricrho}),
$\gamma_{\alpha\beta}$ is
given by
\begin{align}
ds_{\gamma}^{2} &  =\gamma_{\alpha\beta}\,dY^{\alpha}dY^{\beta}=\frac{{\ell}^{2}}{4\rho^{2}%
}\left(  1+\frac{\rho{\ell}^{2}\kappa_{a}^{\widehat{i}a}\kappa_{b}%
^{\widehat{i}b}}{\left(d-2\right)  ^{2}}+...\right)  d\rho^{2}%
+\widetilde{\gamma}_{ab}\,dy^{a}dy^{b},\nonumber\\
\widetilde{\gamma}_{ab} &  =\frac{\sigma_{ab}}{\rho};
\qquad
\sigma_{ab}=\sigma
_{ab}^{\left(  0\right)  }+\rho\sigma_{ab}^{\left(  2\right)  }%
+...
\label{General_embedding}%
\end{align}
In this expression, $\widetilde{\gamma}_{ab}$ is the induced metric on
$\partial\Sigma$ (which is codimension-3 with respect to the bulk) and $\kappa
_{ab}^{\widehat{i}}$ is the extrinsic curvature of $\partial\Sigma$ along the
$\widehat{i}$ direction (orthogonal to the radial direction). In turn,
$\widetilde{\gamma}_{ab}$ has its own FG-like expansion, such that $\sigma
_{ab}^{\left(  0\right)  }$ and $\sigma_{ab}^{\left(  2\right)  }$ are its
leading and next-to-leading order coefficients. This means then that
$\sigma_{ab}^{\left(0\right)  }$
is the induced metric on the entangling surface $\p A$ in the CFT.

To simplify the expression of $B_{d-2}$, we consider the following relations:%
\begin{align}
\mathcal{R}ie &  =\mathcal{W}+4\mathcal{S}\delta~,
\quad \tr\left[  \mathcal{W}\right]=0~;
\nonumber\\
\mathcal{W}&=\rho W^{\left(
0\right)  }+...~,
\quad \mathcal{S} =\rho S^{\left(  0\right)  }+...~,
\label{RelationsB2n-2}
\\
k&=k^{\left(  0\right)
}+\rho k^{\left(  2\right)  }+...~,
\nonumber\\
k^{\left(  0\right)  } &  =\frac{\delta}{{\ell}}~,
\quad k^{\left(  2\right)
}=-\frac{1}{{\ell}}\left[  \sigma^{\left(  2\right)  }+\frac{{\ell}^{2}%
\kappa_{a}^{\widehat{i}a}\kappa_{b}^{\widehat{i}b}}{2\left(d-2\right)
^{2}}\delta\right]~.
\nonumber
\end{align}
Here, $\mathcal{W}$ and $\mathcal{S}$ are the Weyl and Schouten tensors of
$\widetilde{\gamma}$
(i.e., $\cS\equiv \frac{1}{3}(\mathcal{R}ic-\frac{1}{8}\cR\delta)$),
$W^{\left(  0\right)  }$ and $S^{\left(  0\right)  }$
are the Weyl and Schouten tensors of $\sigma^{\left(  0\right)  }$, and
$k^{\left(  0\right)  }$ and $k^{\left(  2\right)  }$ are the leading and
next-to-leading FG coefficients of the radial extrinsic curvature $k$ of
$\partial\Sigma$. By definition, $k_{ab}=\frac{-1}{2\sqrt{\gamma_{\rho\rho}}%
}\partial_{\rho}\widetilde{\gamma}_{ab}$~. The first equation in (\ref{RelationsB2n-2}) is understood to hold inside $B_{d-2}$~, due to the antisymmetrization  enforced by $\delta^{[d-3]}_{[d-3]}$
(otherwise there are 4 terms involving $\cS\delta$ with distinct choices of indices).

From here on, we proceed with the simplification of $S_{\mathrm{Kt}}$ in Appendix
\ref{A}. Then, from (\ref{Answer}) we have that
\begin{align}
S_{\mathrm{Kt}} &  =-\frac{{\ell}}{4G_{\mathrm{N}}}
{\displaystyle\int\limits_{\partial\Sigma}}
d^{d-2}y\,\frac{\sqrt{\sigma^{\left(  0\right)  }}}{\left(d-2\right)
\rho^{\left(d-2\right)  /2}}\Bigg(  1+
\label{AnswerBody}
\\
&  \qquad\qquad
\left.  \rho\left[  -\frac{1}{2}\tr\left[  \sigma^{\left(  2\right)
}\right]  -\frac{{\ell}^{2}\kappa_{a}^{\widehat{i}a}\kappa_{b}^{\widehat{i}b}%
}{2\left(  d-2\right)  }-\frac{{\ell}^{2}}{2\left(  d-4\right)  }%
\mathcal{R}^{\left(  0\right)  }\right]  +...\right)~,
\nonumber
\end{align}
where $\mathcal{R}^{\left(  0\right)  }$ is the Ricci scalar of $\sigma
^{\left(  0\right)  }$.

\subsection{Evaluation of $S_{\mathrm{RT}}$}\label{rtsubsec}

Now we evaluate the usual Ryu-Takayanagi expression (\ref{rt2}) for the EE, and expand it
in powers of the holographic coordinate $\rho$. We have
\begin{equation}
S_{\mathrm{RT}}=\frac{1}{4G_{\mathrm{N}}}
{\displaystyle\int\limits_{\partial\Sigma}}
d^{d-2}y
{\displaystyle\int\limits_{\rho}^{\rho_{\max}}}
d\rho^{\prime}\sqrt{\gamma}~,
\label{srteq}
\end{equation}
where
\begin{equation}
\sqrt{\gamma}=\frac{{\ell}\sqrt{\sigma^{\left(  0\right)  }}}{2\rho^{\prime\frac{d}{2}}}\left(  1+\rho^{\prime}\left[  \frac{{\ell}^{2}\kappa_{a}^{\widehat{i}%
a}\kappa_{b}^{\widehat{i}b}}{2\left(d-2\right)  ^{2}}+\frac{1}{2}\tr\left[
\sigma^{\left(  2\right)  }\right]  \right]  +...\right)  .
\end{equation}
Then, after carrying out the radial integral, we get
\begin{equation}
S_{\mathrm{RT}}=C_{1}+\frac{1}{4G}
{\displaystyle\int\limits_{\partial\Sigma}}
d^{d-2}y{\ell}\frac{\sqrt{\sigma^{\left(  0\right)  }}}{\left(d-2\right)
\rho^{\left(d-2\right)  /2}}\left(  1+  \rho\left[  \frac{\left(d-2\right)  }{2\left(  d-4\right)
}\tr\left[  \sigma^{\left(  2\right)  }\right]  +\frac{{\ell}^{2}\kappa
_{a}^{\widehat{i}a}\kappa_{b}^{\widehat{i}b}}{2\left(d-4\right)  \left(
d-2\right)  }\right]  +...\right)~,
\label{thiseq}
\end{equation}
where $C_{1}$ is a finite term (which for even $d$ is non-universal).

As discussed in the Introduction, in the case of even-dimensional CFTs,
the universal part $S_{\mathrm{univ}}$ of the
EE is logarithmically divergent. This divergence arises from the integration of the
$\frac{1}{\rho}$ power in the expansion of $\sqrt{\gamma}$ from the
Ryu-Takayanagi part. This term however does not appear explicitly in the
general-dimensional expansion given in (\ref{thiseq}), because it enters at higher order.
We will see this term in Section IV, when we use $S_{\mathrm{ren}}$ to compute the
type A conformal anomaly of the CFT.

\subsection{Evaluation of $S_{\mathrm{RT}}+S_{\mathrm{Kt}}$}
\label{cancellationsubsec}

We now evaluate the sum of the two parts that constitute $S_{\mathrm{ren}}$ in the
general-dimensional case. Combining (\ref{AnswerBody}) and (\ref{thiseq}), we have
\begin{align}
S_{\mathrm{ren}}
&  =
C_{1}+\frac{1}{4G_{\mathrm{N}}}
{\displaystyle\int\limits_{\partial\Sigma}}
d^{d-2}y\frac{{\ell}\sqrt{\sigma^{\left(  0\right)  }}}{\left(d-2\right)
\rho^{\left(d-2\right)  /2}}\rho\left(  \left(  \frac{\left(d-2\right)
}{2\left(d-4\right)  }+\frac{1}{2}\right)  \tr\left[  \sigma^{\left(
2\right)  }\right]  \right.  \nonumber\\
&  +\left.  \left(  \frac{1}{2\left(d-4\right)  \left(d-2\right)  }%
+\frac{1}{2\left(d-2\right)  }\right)  {\ell}^{2}\kappa_{a}^{\widehat{i}%
a}\kappa_{b}^{\widehat{i}b}+\frac{{\ell}^{2}}{2\left(d-4\right)
}\mathcal{R}^{\left(  0\right)  }\right)  +\ldots~,
\label{sreneq}
\end{align}
where ``$\ldots$'' denotes higher-order terms.
We note that the leading-order
divergence in $S_{\mathrm{RT}}$ is cancelled out directly. This is the only
power-law divergence that appears in the case of AdS$_{5}$/CFT$_{4}$, so for
4D CFTs, we have shown here that the renormalization procedure based on extrinsic counterterms works in full generality.

To check the cancellation of the next-to-leading order divergence in $S_{\mathrm{RT}}$, we define the ``difference'' EE 
$S_{\mathrm{diff}}$ as the leading term in (\ref{sreneq}); i.e.,
\begin{align}
S_{\mathrm{diff}}
 \equiv \frac{1}{4G_{\mathrm{N}}}
{\displaystyle\int\limits_{\partial\Sigma}}
d^{d-2}y\frac{{\ell}\sqrt{\sigma^{\left(  0\right)  }}}{\left(d-2\right)
\left(d-4\right)  \rho^{\left(d-4\right)  /2}}\left(  \left(
d-3\right)  \tr\left[  \sigma^{\left(  2\right)  }\right]  +  \frac{\left(  d-3\right)  }{2\left(d-2\right)  }{\ell}%
^{2}\kappa_{a}^{\widehat{i}a}\kappa_{b}^{\widehat{i}b}+\frac{l^{2}}%
{2}\mathcal{R}^{\left(  0\right)  } \right).
\end{align}
At first sight, $S_{\mathrm{diff}}$
represents an $O\left(  \rho^{-\left(
d-4\right)  /2}\right)  $ divergence in the renormalized EE. For our renormalization procedure to be successful, these polynomial divergences
 ought not to be present. We will show in what follows that indeed $S_{\mathrm{diff}}=0$, restricting attention for simplicity to the case of
spherical entangling surfaces for CFTs in flat spacetime. As explained in the Introduction, this is the setting that makes contact with the central charge $a$ of the CFT, that is the subject of the $a$-theorem \cite{Cardy:1988cwa,Komargodski:2011xv,Komargodski:2011vj}. We will
extract the value of $a$ for our class of theories in Section \ref{anomalysec}.

\subsubsection{Simplification of $S_{\mathrm{diff}}$}

Considering that
\begin{equation}
\sigma_{ab}^{\left(  2\right)  }=-{\ell}^{2}S_{ab}-\frac{{\ell}^{2}}{\left(
d-2\right)  }\kappa_{c}^{\widehat{i}c}\kappa_{ab}^{\widehat{i}}~,
\label{Sigma2}
\end{equation}
where $S_{ab}$ is the Schouten tensor of the CFT metric $g^{\left(  0\right)
}$ evaluated with codimension-3 indices, and that, as derived in Appendix \ref{B},
\begin{align}
\tr\left[  \sigma^{\left(  2\right)  }\right]   &  =
\label{TrSig2}
\\
&  -\frac{{\ell}^{2}}{\left(
d-2\right)  \left(d-1\right)  }\left(  R_{\left(  i\right)  \left(
i\right)  }+  \frac{\left(d-2\right)  }{2}\kappa_{d}^{\widehat{i}d}\kappa
_{a}^{\widehat{i}a}-\frac{d}{2}\left(  R_{\left(  i\right)  \left(  j\right)
\left(  i\right)  \left(  j\right)  }-\mathcal{R}^{\left(  0\right)  }%
-\kappa_{d}^{\widehat{i}a}\kappa_{a}^{\widehat{i}d}\right)  \right)~,
\nonumber%
\end{align}
we have that
\begin{align}
S_{\mathrm{diff}}  =&\frac{1}{4G_{\mathrm{N}}}
{\displaystyle\int\limits_{\partial\Sigma}}
d^{d-2}y\frac{{\ell}^{3}\sqrt{\sigma^{\left(  0\right)  }}}{2\left(
d-4\right)  \left(d-2\right)  \rho^{\left(d-4\right)  /2}}\left(
\frac{2}{\left(d-2\right)  \left(d-1\right)  }\mathcal{R}^{\left(
0\right)  }\right.
\nonumber\\
& \qquad \left. -\frac{2\left(d-3\right)  }{\left(d-2\right)  \left(
d-1\right)  }R_{\left(  i\right)  \left(  i\right)  }+\frac{d\left(
d-3\right)  }{\left(d-2\right)  \left(d-1\right)  }R_{\left(  i\right)
\left(  j\right)  \left(  i\right)  \left(  j\right)  }\right.
\nonumber\\
&  \qquad\left. -\frac{d\left(d-3\right)  }{\left(d-2\right)  \left(
d-1\right)  }\kappa_{d}^{\widehat{i}a}\kappa_{a}^{\widehat{i}d}+\frac{\left(
d-3\right)  }{\left(d-2\right)  \left(d-1\right)  }\kappa_{d}%
^{\widehat{i}d}\kappa_{a}^{\widehat{i}a}\right)~.
\label{toreferto}
\end{align}
In this expression, $R_{\left(  i\right)  \left(  i\right)  }$ is the partial
trace of the Ricci tensor of $g^{\left(  0\right)  }$, along the directions of
the codimension-3 foliation that are orthogonal to the radial coordinate. $R_{\left(
i\right)  \left(  j\right)  \left(  i\right)  \left(  j\right)  }$ is a
similar partial trace of the Riemann tensor of $g^{\left(  0\right)  }$.

Using the fact, demonstrated in Appendix \ref{B}, that
\begin{align}
\delta_{a}^{c}\delta_{b}^{d}\left(  W^{\left(  0\right)  }\right)  _{cd}^{ab}
=&
\frac{2}{\left(d-2\right)  \left(d-1\right)  }\mathcal{R}^{\left(
0\right)  }-\frac{2\left(d-3\right)  }{\left(d-2\right)  \left(
d-1\right)  }R_{\left(  i\right)  \left(  i\right)  }
\label{TrWeyl}
\\
&  +\frac{d\left(d-3\right)  }{\left(d-2\right)  \left(d-1\right)
}R_{\left(  i\right)  \left(  j\right)  \left(  i\right)  \left(  j\right)
}+\frac{2}{\left(d-2\right)  \left( d-1\right)  }\left(  \kappa
_{b}^{\widehat{i}a}\kappa_{a}^{\widehat{i}b}-\kappa_{a}^{\widehat{i}a}%
\kappa_{b}^{\widehat{i}b}\right)~,
\nonumber%
\end{align}
 we can write (\ref{toreferto}) as
\begin{equation}
S_{\mathrm{diff}}  =\frac{1}{4G_{\mathrm{N}}}
{\displaystyle\int\limits_{\partial\Sigma}}
d^{d-2}y\frac{{\ell}^{3}\sqrt{\sigma^{\left(  0\right)  }}}{2\left(
d-4\right)  \left(d-2\right)  \rho^{\left(d-4\right)  /2}}\left(
\delta_{a}^{c}\delta_{b}^{d}\left(  W^{\left(  0\right)  }\right)  _{cd}%
^{ab}  -  \left(  \kappa_{d}^{\widehat{i}a}\kappa_{a}^{\widehat{i}d}-\frac
{1}{\left(  d-2\right)  }\kappa_{d}^{\widehat{i}d}\kappa_{a}^{\widehat{i}%
a}\right)  \right)~.
\label{sdiffeq}
\end{equation}
In this expression, $\delta_{a}^{c}\delta_{b}^{d}\left(  W^{\left(  0\right)
}\right)  _{cd}^{ab}$ is a partial trace of the Weyl tensor of $g^{\left(
0\right)  }$, along the directions that comprise the worldvolume of
$\partial\Sigma$.

We now proceed to show explicitly that $S_{\mathrm{diff}}$ is identically zero, specializing for simplicity to the case of a spherical entangling surface
in a CFT on flat spacetime which is in its ground state, dual to global AdS.

\subsubsection{Vanishing of $S_{\mathrm{diff}}$ for CFTs in flat spacetime and
spherical entangling surfaces}

In flat spacetime, $W^{\left(  0\right)  }=0$, so
the first term in (\ref{sdiffeq}) is zero. Here we will examine the second term. We define%
\begin{equation}
T_{2}\equiv\left(  \kappa_{d}^{\widehat{i}a}\kappa_{a}^{\widehat{i}d}-\frac
{1}{\left(  d-2\right)  }\kappa_{d}^{\widehat{i}d}\kappa_{a}^{\widehat{i}%
a}\right)~,
\label{T2}%
\end{equation}
and we proceed to verify that $T_{2}=0$ in our case of interest.

For a CFT in its
ground state (which is dual to global AdS), the induced metric $\gamma$ on the
Ryu-Takayanagi surface $\Sigma$ associated with a spherical entangling surface $\p A$ of radius $L$ is given (as discussed e.g. in \cite{Anastasiou:2018rla}) by
\begin{align}
ds_{\gamma}^{2} &  \equiv
\gamma_{ab}dy^{a}dy^{b}
=
\frac{{\ell}^{2}}{4\rho^{2}}\left(  1+\frac{{\ell}^{2}%
\rho}{L^{2}-{\ell}^{2}\rho  }\right)  d\rho^{2}+\frac{
L^{2}-{\ell}^{2}\rho  }{\rho}\,d\Omega_{d-2}^{2}~,\nonumber\\
d\Omega_{d-2}^{2} &  =d\theta_{1}^{2}+\sin^{2}\theta_{1}d\theta_{2}%
^{2}+\ldots+\sin^{2}\theta_{1}\cdots\sin^{2}\theta_{d-3}d\theta_{d-2}%
^{2}~.
\label{Spherical_Metric}
\end{align}
Near the conformal boundary we know that
\begin{equation}
\frac{{\ell}^{2}\rho}{ L^{2}-l^{2}\rho  }=\frac{{\ell}^{2}%
}{L^{2}}\rho+\frac{{\ell}^{4}}{L^{4}}\rho^{2}+....~.
\end{equation}
Thus, comparing the $\rho$ component of the induced metric $\gamma$ with the
expression for the general FG-like embedding given in
(\ref{General_embedding}), we see that, for spherical $\p A$,
\begin{equation}
\kappa_{d}^{\widehat{i}d}\kappa_{a}^{\widehat{i}a}=\frac{\left(  d-2\right)
^{2}}{L^{2}}~,
\label{Trace1}%
\end{equation}
and
\begin{equation}
\sigma_{ab}^{\left(  0\right)  }dy^{a}dy^{b}=L^{2}d\Omega_{d-2}^{2}~,
\end{equation}
which means that the induced metric on the entangling surface is the metric of
a $\left(d-2\right)-$sphere with radius $L$. Therefore $\mathcal{R}%
^{\left(  0\right)  }$, which is the Ricci scalar of $\sigma^{\left(
0\right)  }$, is given by
\begin{equation}
\mathcal{R}^{\left(  0\right)  }=\frac{\left(  d-2\right)  \left(
d-3\right)  }{L^{2}}.
\end{equation}

Now, to compute $\kappa_{d}^{\widehat{i}a}\kappa_{a}^{\widehat{i}d}$, we
notice that, because the CFT is on a Minkowski background,%
\begin{align}
\delta_{a}^{c}\delta_{b}^{d}\left(  W^{\left(  0\right)  }\right)  _{cd}^{ab}
&  =0~,\nonumber\\
R_{\left(  i\right)  \left(  i\right)  } &  =0~,\\
R_{\left(  i\right)  \left(  j\right)  \left(  i\right)  \left(  j\right)  }
&  =0~.
\nonumber
\end{align}
Then, from the expression for $\delta_{a}^{c}\delta_{b}^{d}\left(  W^{\left(
0\right)  }\right)  _{cd}^{ab}$ given in Appendix \ref{B}, we have that%
\begin{equation}
\kappa_{b}^{\widehat{i}a}\kappa_{a}^{\widehat{i}b}=\frac{\left(d-2\right)
}{L^{2}}~.
\label{Trace2}%
\end{equation}

Finally, we compute $T_{2}$ by substituting (\ref{Trace1}) and
 (\ref{Trace2}) into (\ref{T2}), finding that $T_{2}=0$. Therefore,%
\begin{equation}
S_{\mathrm{diff}}=0~,
\end{equation}
as we intended to show.

With this result, we have explicitly shown that $S_{\mathrm{ren}}$ is divergence-free up to
next-to-leading order, considering spherical entangling surfaces in the ground state of a CFT in
flat spacetime (dual to global AdS). Previously, we had also demonstrated in  (\ref{sreneq})
that the leading-order divergence vanishes in full generality.

\section{Type A anomaly computation from EE} \label{anomalysec}

In the previous section, we showed that the leading-order divergence  in the Ryu-Takayanagi entanglement entropy
$S_{\mathrm{RT}}$ (which is the only power-law divergence for $d\le 4$) is cancelled in full generality by the addition of the
$S_{\mathrm{Kt}}$ counterterm (\ref{B2n-2}), constructed from the $B_{d-2}$ Kounterterm
evaluated on the boundary $\partial\Sigma$ of the minimal surface, as indicated in
(\ref{sktbd-2}). We  also
proved that the next-to-leading order divergence (present for the first time in $d=6$) is cancelled in the case of
spherical entangling surfaces for CFTs in flat spacetime.

Since $S_{\mathrm{Kt}}$  has been derived from the action Kounterterm $I_{\mathrm{Kt}}$, in any dimension we expect it to cancels all of the
power-law divergences. On the other hand, it is easy to see that $S_{\mathrm{Kt}}$ (just like $I_{\mathrm{Kt}}$) does \emph{not} produce logarithmic divergences,
and therefore, the standard logarithmic divergence coming from $S_{\mathrm{RT}}$
will survive, yielding the usual universal term of the EE for
even-dimensional CFTs (recall our discussion around (\ref{Sdiv})). Thus,
\begin{equation}
S_{\mathrm{ren}}=S_{\mathrm{univ}}+C~,
\label{srensplit}
\end{equation}
where $C$ stands for finite terms that are in general non-universal and
renormalization-scheme-dependent.

We will now specialize to spherical entangling surfaces in
CFTs on flat spacetime. This is one of the settings of greatest physical interest, where the result is related, as explained in the Introduction, to the conformal, or Weyl, or trace anomaly of the CFT. In  more detail: as a consequence of the anomaly, the expectation value of the trace of the energy-momentum tensor is generally nonvanishing for even  CFT dimension $d$, and takes the form \cite{Duff:1977ay,Myers:2010tj,Myers:2010xs}
\begin{equation}
\left\langle T_{i}^{i}\right\rangle _{\mathrm{CFT}}
=
\left( -1\right)^{\frac{d}{2}+1}2\mathrm{A} E_d
+{\displaystyle\sum\limits_{m=1}^M} \mathrm{B}_{m}I_{m}
+B^{\prime}\nabla_{i}J^{i}~,
\label{tii}
\end{equation}
The A anomaly coefficient multiplies
$E_{d}\equiv\frac{\mathcal{E}_{d}}{\left(
4\pi\right)  ^{d/2}\left(  \frac{d}{2}\right)  !}$~,
where $\mathcal{E}_{d}$ is
the Euler density, defined as in (\ref{euler}) but using the  Riemann tensor built from the CFT metric $g_{(0)}$. $E_d$ is normalized such that on a $d$-sphere
$\int d^d x\sqrt{g_{(0)}}\,E_d=2$.
The B$_{m}$ anomaly coefficients multiply the  various conformal invariants $I_{m}$ in the spacetime of the CFT, whose number $M$ increases with $d$, starting with $M=0,1$ for $d=2,4$, respectively. The last term in (\ref{tii}) is scheme-dependent and can be
neglected. Common alternative names for the type A and B coefficients, or central charges, are $a$ and $c$ ($c_m$ for $d>4$). The statement of monotonicity under RG flow alluded to in the Introduction, or $a$ theorem \cite{Cardy:1988cwa,Komargodski:2011xv,Komargodski:2011vj}, refers to the type A anomaly coefficient, and in the normalization used in (\ref{tii}), we have A$=a$, and $B_1=16\pi^2 c$ in $d=4$. The B$_m$ or $c_m$ central charges are in general not monotonic. An important point is that  CFTs dual to Einstein gravity are special, in that they have only \emph{one} independent central charge, e.g., $a=c$ in $d=4$. More generic holographic CFTs are dual to higher-curvature theories of gravity.

As explained one paragraph below (\ref{Sdiv}), the relation between the conformal anomaly and the entanglement entropy for a spherical entangling surface in flat spacetime is \cite{Casini:2011kv,Solodukhin:2008dh}
\begin{equation}
    S_{\mathrm{univ}}=(-1)^{\frac{d}{2}+1}4a\ln(L/\varepsilon)~,
\label{suniveven}
\end{equation}
where $L$ is the radius of the sphere. Care should be taken to relate the UV cutoff $\varepsilon$ seen in this expression, which has dimensions of length, to the dimensionless UV cutoff $\epsilon$ that we have used throughout this paper, starting in (\ref{Idiv}). The latter is a lower bound on the FG coordinate $\rho$ defined in (\ref{fgmetricrho}), from which we see that $\rho$ scales like length \emph{squared} under the AdS isometry that is dual to a rigid rescaling of the CFT coordinates $x^i$ (it is $z\equiv\sqrt{\rho}$ in (\ref{fgmetric}) that scales like a length). The relation between the two cutoffs is therefore $\epsilon=\varepsilon^2/L^2$, where the factor of $L$ has been introduced on dimensional grounds.\footnote{Another natural possibility would be to use the AdS curvature radius $\ell$ instead of $L$, but this simply amounts to a shift in the constant $C$ in (\ref{srensplit}), and does not modify the coefficient of the logarithm in (\ref{suniveven2}).} Under this translation, (\ref{suniveven}) becomes
\begin{equation}
    S_{\mathrm{univ}}=(-1)^{\frac{d}{2}}\,2a\ln\epsilon~.
\label{suniveven2}
\end{equation}

We now proceed to compute $S_{\mathrm{univ}}$ holographically. Restating (\ref{srteq}),
\begin{equation}
S_{\mathrm{RT}}=\frac{1}{4G_{\mathrm{N}}}%
{\displaystyle\int\limits_{\partial\Sigma_{\epsilon}}}
d^{d}y
{\displaystyle\int\limits_{\epsilon}^{\rho_{\max}}}
d\rho\sqrt{\gamma},
\end{equation}
and knowing that for a spherical entangling surface the induced metric $\gamma$ on the
corresponding minimal surface is given by (\ref{Spherical_Metric}), we have
\begin{align}
S_{\mathrm{RT}} &  =\frac{\mathrm{Vol}\left(\mathbf{S}^{d-2}\right)  }{4G_{\mathrm{N}}}
{\displaystyle\int\limits_{\epsilon}^{\rho_{\max}}}
d\rho\left[  \frac{{\ell}^{2}}{4\rho^{2}}\left(  1+\frac{{\ell}^{2}\rho
}{L^{2}-{\ell}^{2}\rho  }\right)
\left(  \frac{L^{2}-{\ell}^{2}\rho  }{\rho}\right)  ^{d-2 }\right]
^{1/2}
\nonumber\\
&  =\frac{\mathrm{Vol}\left( \mathbf{S}^{d-2}\right)  L^{d-2  }{\ell}}{8G_{\mathrm{N}}}%
{\displaystyle\int\limits_{\epsilon}^{\rho_{\max}}}
d\rho\frac{1}{\rho^{\frac{d}{2}}}\left(  1-\frac{{\ell}^{2}}{L^{2}}\rho\right)
^{\frac{d-3}{2}}~.
\end{align}

In the preceding equation, we see that the logarithmic divergence will come from the $O\left(  \rho^{\frac{d}{2}-1}\right)  $
term in the Taylor expansion of
\begin{equation}
f(  \rho)  =\left(  1-\frac{{\ell}^{2}}{L^{2}}\rho\right)
^{\frac{ d-3}{2}}
\end{equation}
at small $\rho$. Writing out
\begin{equation}
f(\rho)
=
{\displaystyle\sum\limits_{i=0}^{\infty}}
\rho^{i}\left(  -1\right)  ^{i}\frac{\left(  \frac{{\ell}^{2}}{L^{2}}\right)
^{i}}{i!}
{\displaystyle\prod\limits_{j=0}^{i-1}}
\left(  \frac{\left(d-3\right)  }{2}-j\right)~,
\end{equation}
the $O\left(  \rho^{\frac{d}{2}-1}\right)  $ term is found to be given by%
\begin{equation}
f(  \rho)
=
\ldots+\rho^{\frac{d}{2}-1}\left(  -1\right)  ^{\frac{d}{2}-1}\frac{\left(
\frac{{\ell}^{2}}{L^{2}}\right)  ^{\frac{d}{2}-1}}{\left(\frac{d}{2}-1\right)  !}
{\displaystyle\prod\limits_{j=0}^{\frac{d}{2}-2}}
\left(  \frac{\left(  d-3\right)  }{2}-j\right)  +\cdots.
\end{equation}
Thus, we find that
\begin{align}
S_{\mathrm{univ}} &  =\frac{\mathrm{Vol}\left(  \mathbf{S}^{d-2}\right)  L^{d-2
}{\ell}}{8G_{\mathrm{N}}}
{\displaystyle\int\limits_{\epsilon}}
d\rho\,\frac{1}{\rho}\left(  -1\right)  ^{\frac{d}{2}-1}\frac{\left(  \frac{{\ell}^{2}%
}{R^{2}}\right)  ^{\frac{d}{2}-1}}{\left(  \frac{d}{2}-1\right)  !}%
{\displaystyle\prod\limits_{j=0}^{\frac{d}{2}-2}}
\left(  \frac{d-3  }{2}-j\right)
\nonumber\\
&  =\left(  -1\right)  ^{\frac{d}{2}}\ln \epsilon\,
\frac{{\ell}^{d-1  }}{8G_{\mathrm{N}}\left( \frac{d}{2}-1\right)  !}\left[
\mathrm{Vol}\left(  \mathbf{S}^{d-2}\right)
{\displaystyle\prod\limits_{j=0}^{\frac{d}{2}-2}}
\left(  \frac{d-3  }{2}-j\right)  \right]
\nonumber\\
&  =\left(  -1\right)  ^{\frac{d}{2}}\ln \epsilon\,
\frac{{\ell}^{d-1  }}{8G_{\mathrm{N}}\left(  \frac{d}{2}-1\right)  !}\mathrm{Vol}\left(
\mathbf{S}^{d-2}\right)  \left[  \frac{\Gamma\left(  \frac{d-1}{2}\right)  }%
{\Gamma\left(  \frac{1}{2}\right)  }\right]~.
\end{align}
For our last step, we use
\begin{equation}
\mathrm{Vol}\left(\mathbf{S}^{d-2}\right)  =
\frac{\left(  d-1\right)
\pi^{\frac{d-1  }{2}}}{\Gamma\left(  \frac{d+1}{2}\right)  }~,
\quad
\Gamma\left(
\frac{1}{2}\right)  =\sqrt{\pi}~,
\quad
\Gamma\left(  \frac{d+1}{2}\right)    =
\frac{ d!\sqrt{\pi}}{\left(\frac{d}{2}\right)!2^{d}}~,
\end{equation}
to obtain
\begin{equation}
S_{\mathrm{univ}}=\left(  -1\right)  ^{\frac{d}{2}}\,2\ln\epsilon
\left[
\frac{{\ell}^{d-1  }\pi^{\frac{d}{2}-1  }}{8G_{\mathrm{N}}\left(
\frac{d}{2}-1\right)  !}\right]~.
\label{S_univ}
\end{equation}

Finally, by comparing  (\ref{S_univ}) against (\ref{suniveven2}), we learn that
\begin{equation}
a=\frac{{\ell}^{d-1  }\pi^{\frac{d}{2}-1}}{8G_{\mathrm{N}}\left(
\frac{d}{2}-1\right)  !}~,
\label{A-coeff}%
\end{equation}
in agreement with the standard results (see, e.g.,
\cite{Imbimbo:1999bj,Schwimmer:2008yh}). For example, in
the $d=4$ case,
 $a=\frac{\pi{\ell}}{8G_{\mathrm{N}}}$. With the aid of the
$G_{\mathrm{N}}=\pi l^{3}/2N^{2}$ entry of the AdS/CFT dictionary
for $\mathcal{N}=4$ SYM in the large $N$ limit \cite{Gubser:1998bc}, this translates into  $a=N^{2}/4$, which is indeed the correct $a$ central charge \cite{Henningson:1998gx}.

\section{Conclusions} \label{conclusionsec}

In this work, we have adapted the Kounterterm procedure \cite{Aros:1999kt,Mora:2004rx,Olea:2005gb,Olea:2006vd} to renormalize
the  entanglement entropy (EE) of even-dimensional CFTs dual to
Einstein gravity in asymptotically locally AdS (ALAdS)  manifolds. Along the way, we have elucidated two important points regarding the applicability of the
Kounterterm prescription and its relation to the standard method of holographic
renormalization \cite{Henningson:1998gx,deHaro:2000vlm,Skenderis:2002wp}.  First, as shown explicitly
in Section \ref{Ksubsec} and Appendix \ref{D}, the Kounterterm-renormalized action has a
well-defined variational principle, that is fully consistent with the usual boundary condition for holography, a Dirichlet condition for
the CFT metric $g_{\left(  0\right)  }$. Second, we have emphasized that for asymptotically conformally flat bulk spacetimes of dimension
$ D\leq 9$ (ACF, see Section \ref{Ksubsec}), the two methods of renormalization, via counterterms or Kounterterms, are
essentially equivalent  \cite{Miskovic:2006tm, Miskovic:2007mg,Anastasiou:2018the}. As shown in Appendix \ref{C}, for bulk dimension $D\leq5$ the ACF
condition holds automatically for all ALAdS spacetimes, with no restriction on the CFT metric $g_{\left(  0\right)  }$,
while for $6\leq D\leq9$, it requires that
$g_{\left(  0\right)  }$ be conformally flat.

For odd-dimensional bulk spacetimes (dual to even-dimensional CFTs), there is one important difference between standard holographic renormalization and the Kounterterm method. Whereas the former procedure cancels all divergences by expressly  constructing the required counterterms one by one,
in the Kounterterm prescription, as seen in (\ref{iren}), there is a \emph{single} object that is added to the action, fully specified from the beginning, involving $B_d$ defined in (\ref{bd}). Remarkably, this procedure, where the only freedom is in
fixing the overall constant $c_d$ in front of $B_d$, turns out to be enough to cancel all of the power-law divergences. However, \emph{the logarithmic divergence in the action
arising from the volume integral is not cancelled}. As shown by Graham in \cite{Graham:1999pm}, for Einstein-AdS gravity, this logarithmic term is the volume anomaly
of the bulk manifold, which is universal and proportional to the conformal
anomaly of the CFT \cite{Henningson:1998gx}. Thus, the absence of a logarithmic Kounterterm has the effect of isolating this universal contribution from the
bulk when computing the renormalized action, instead of cancelling it. Most
importantly, despite maintaining the logarithmic divergence, the general
variation of the resulting action is finite for ACF manifolds, therefore
leading to finite conserved charges, with the corresponding Noether
prepotential being proportional to the electric part of the bulk Weyl tensor
\cite{Jatkar:2014npa}. It is also worth emphasizing that, as seen in (\ref{ct}),
even in the standard method of holographic renormalization the logarithmic
counterterm is different from the other terms, as it cannot be written
covariantly with respect to the boundary metric $h$.

The application of the Kounterterm procedure for renormalizing EE
 uses the result of Lewkowycz and
Maldacena \cite{Lewkowycz:2013nqa}, which expresses the EE of holographic CFTs in terms of the
on-shell gravity action $I$ of the bulk dual. The renormalized EE, $S_{\mathrm{ren}}$,
is obtained when one starts from the renormalized gravity action $I_{\mathrm{ren}}$ instead of $I$. This was worked out in \cite{Taylor:2016aoi} for the case of standard holographic renormalization. The
Kounterterm renormalization of holographic EE in the case of odd-dimensional
CFTs was developed in \cite{Anastasiou:2018rla}. In this work we focused on the case of
even-dimensional CFTs, where the main novelty is the existence of the
conformal anomaly. In Section \ref{evensec}
we derived $S_{\mathrm{ren}}$, by evaluating the Kounterterm-renormalized
gravity action on the usual replica orbifold, considering the self-replicating
property (\ref{B_2n_split}) of the $B_{d}$ boundary term. In more detail, this property entails that
$B_{d}$, when evaluated on a conically singular manifold, decomposes into a
regular part plus a codimension-2 version of itself (i.e., $B_{d-2}$) located
at the conical singularity, in direct analogy to the decomposition of the
Euler density \cite{Fursaev:2013fta,Fursaev:1995ef}. We emphasize that the self-replicating property
is sensitive to the particular coefficients that appear when expressing
$B_{d}$ as in (\ref{B_2n-1_In_b}) and (\ref{B_2n_In_b}), in terms of the $b_{pq}$ terms introduced in (\ref{b_pq}),
and it is therefore \emph{not} general for an arbitrary boundary term constructed out
of Riemann tensors and extrinsic curvatures of the boundary.

After obtaining the Kounterterm expression  for $S_{\text{ren}}$ in even-dimensional
CFTs, we proceeded to verify by explicit computation that
indeed the power-law divergences coming from the RT part of the EE are
cancelled by the $B_{d-2}$ term, while at the same time leaving the universal
(logarithmically-divergent) part unchanged. In Section \ref{cancellationsubsec}, the cancellation of divergences
was verified explicitly up to next-to-leading order in the holographic coordinate $\rho$
for spherical entangling regions of CFTs in conformally flat manifolds. More generally, the cancellation of all power-law divergences is expected to be inherited directly from the corresponding cancellation in the Kounterterm-renormalized bulk action. This implies again that the Kounterterm method conveniently isolates the universal part of the EE, $S_{\mathrm{univ}}$, defined below (\ref{Sdiv}). This is in fact true both for even and odd $d$, the latter case having been established in \cite{Anastasiou:2018rla}. The interesting difference between the two cases is the presence of the logarithmic divergence for even $d$. For spherical entangling surfaces in flat spacetime, in Section \ref{anomalysec} we arrived at a concrete result for $S_{\mathrm{univ}}$, given in (\ref{S_univ}). By comparing with  (\ref{suniveven2}), we were able to extract the Type A anomaly coefficient for our CFT, shown in (\ref{A-coeff}).
Its value agrees with expectations \cite{Myers:2010tj,Myers:2010xs,Casini:2011kv}. As  is well-known, this $a$ (or A) central charge
 is a $c$-function candidate \cite{Cardy:1988cwa}, meaning that it
is conjectured to decrease between any two conformal fixed points connected by
a renormalization group flow. This was proven
by Komargodski and Schwimmer for the $d=4$ case \cite{Komargodski:2011xv,Komargodski:2011vj}, and important evidence exists for
the higher even-dimensional cases \cite{Myers:2010tj,Myers:2010xs,Liu:2012eea,Myers:2012ed,Cordova:2015fha}.

To summarize, combining our results with those in \cite{Anastasiou:2018rla}, we have demonstrated that the Kounterterm method is a powerful tool to renormalize holographic entanglement entropy. The result (\ref{sreneven}) efficiently computes $S_{\mathrm{univ}}$ in both even and odd dimensions, and directly inherits all of the virtues of the corresponding Kounterterm-renormalized bulk action, including compactness and uniform applicability across dimensions and across different theories of gravity. The main drawbacks of the method are its current limitation to the pure gravity setting, and to ACF bulk spacetimes when $D\ge 6$.

In the near future, we will study the
renormalization of the Renyi entropy  for
even-dimensional CFTs dual to Einstein-AdS gravity. We will also examine the relation
between the notions of renormalized volume and area of extremal surfaces in
odd-dimensional ALAdS manifolds and their corresponding on-shell Einstein
action and modular entropies, in the same spirit as in \cite{Anastasiou:2018the}. Finally, we
will consider higher-curvature theories of gravity beyond the Einstein case,
like for example those of Lovelock class \cite{Lovelock:1971yv,Lovelock:1972vz}, where we can show that the
Kounterterm procedure gives a direct answer for the renormalized holographic
EE of their dual CFT (the Gauss-Bonnet case has been addressed in \cite{Taylor:2016aoi} using standard holographic renormalization).

\section*{Acknowledgements}

It is a pleasure to thank C\'esar Arias, Stefan Theisen and David Vergara for useful discussions. The work of GA, IJA and RO was funded in part by FONDECYT Grants No.~1170765 \emph{Boundary dynamics in anti-de Sitter gravity and gauge/gravity duality}, No.~3180620 \emph{Entanglement Entropy and AdS gravity} 
and No.~3190314 \emph{Holographic Complexity from anti-de Sitter gravity}.
The work of AG was partially supported by Mexico's National Council of Science and Technology (CONACyT) grant A1-S-22886 and DGAPA-UNAM grant IN107520.

\appendix

\section{Simplification of $S_{\mathrm{Kt}}$}\label{A}

We proceed to simplify $S_{\mathrm{Kt}}$. Using (\ref{B2n-2}) and the relations given in (\ref{RelationsB2n-2}), we obtain
\begin{align}
S_{\mathrm{Kt}}
&  =\frac{\left(-1\right)^{\frac{d}{2}+1}{\ell}^{d-2}(d-2)}{2^{d}\,G_{\mathrm{N}}\left[
\left(\frac{d}{2}-1\right) !\right] ^{2}}\times
\nonumber\\
&\quad\times {\displaystyle\int\limits_{\partial\Sigma}}
d^{d-2}y
{\displaystyle\int\limits_{0}^{1}}
ds
{\displaystyle\int\limits_{0}^{s}}
dt\sqrt{\widetilde{\gamma}}\,\delta_{\left[d-3\right]  }^{\left[
d-3\right]  }k\left(  \frac{1}{2}\mathcal{W}+  2\mathcal{S}\delta-s^{2}kk+t^{2}k^{\left(  0\right)  }k^{\left(
0\right) }\right)^{\frac{d}{2}-2}~.
\end{align}
Now we use the fact that, up to linear order in $\rho$, $kk$ is given by
\begin{equation}
kk=k^{\left(  0\right)  }k^{\left(  0\right)  }+2\rho k^{\left(  0\right)
}k^{\left(  2\right)  }+...~,
\end{equation}
and that the Weyl tensor $\mathcal{W}$ and the Schouten tensor $\mathcal{S}$
of the induced metric $\widetilde{\gamma}$ at $\partial\Sigma$ can be written
in terms of the corresponding tensors of $\sigma^{\left(  0\right)  }$ at the
entangling surface $\p A$ as $\mathcal{W}$ $=\rho W^{\left(  0\right)  }+\ldots$ and
$\mathcal{S}=\rho S^{\left(  0\right)} +\ldots$~. Then, we have that%
\begin{equation}
S_{\mathrm{Kt}}=
\frac{\left(-1\right)^{\frac{d}{2}+1}{\ell}^{d-2}(d-2)}{2^{d}\,G_{\mathrm{N}}\left[
\left(\frac{d}{2}-1\right) !\right] ^{2}}
{\displaystyle\int\limits_{\partial\Sigma}}
d^{d-2}y\,I~,
\label{I_def}%
\end{equation}
where%
\begin{equation}
I \equiv
{\displaystyle\int\limits_{0}^{1}}
ds\!
{\displaystyle\int\limits_{0}^{s}}
dt\sqrt{\widetilde{\gamma}}\delta_{\left[d-3\right]  }^{\left[
d-3\right]  }
\left(  k^{\left(  0\right)  }+\rho k^{\left(  2\right)
}+\ldots\right)
\left[ \left( t^{2}-s^{2}\right)  k^{\left(  0\right)
}k^{\left(  0\right)  }+  2\rho\left(  {\ell}S^{\left(  0\right)  }-s^{2}k^{\left(  2\right)
}\right)  k^{\left(  0\right)  }+\ldots\right]  ^{\frac{d}{2}-2}.
\end{equation}
In this expression we have have neglected the contribution of $W^{\left(
0\right)  }$, as the fact that $\tr\left[  W^{\left(  0\right)  }\right]  =0$
implies that it does not contribute to $I$ at next-to-leading order in $\rho$.

We now  proceed to simplify $I$ explicitly, keeping only the terms up to
next-to-leading order. In particular, expanding the brackets to the power of
$\left(
\frac{d}{2}-2\right)  $, we have that
\begin{align}
I  & =
{\displaystyle\int\limits_{0}^{1}}
ds
{\displaystyle\int\limits_{0}^{s}}
dt\sqrt{\widetilde{\gamma}}\,\delta_{\left[d-3\right]  }^{\left[
d-3\right]  }
\left(  k^{\left(  0\right)  }+\rho k^{\left(  2\right)
}+...\right)
\left[ \left(  t^{2}-s^{2}\right)  \left(  k^{\left(  0\right)
}\right)^{d-4}+\right.
\nonumber\\
& \qquad \left.  +\rho\left( d-4\right)  \left(t^{2}-s^{2}\right)^{\frac{d}{2}-3}\left(
k^{\left(  0\right)  }\right) ^{d-5}\left(  {\ell}S^{\left(  0\right)
}-s^{2}k^{\left(  2\right)  }\right)  +...\right]~.
\end{align}
After multiplying and collecting terms, we find
\begin{align}
I  &  =
{\displaystyle\int\limits_{0}^{1}}
ds
{\displaystyle\int\limits_{0}^{t}}
dt\sqrt{\widetilde{\gamma}}\delta_{\left[d-3\right]  }^{\left[
d-3\right]  }\left(  \left(  t^{2}-s^{2}\right)  ^{\frac{d}{2}-2}\left(  k^{\left(
0\right)  }\right)  ^{d-3}+\right.
\\
& + \rho\left(  \left(  t^{2}-s^{2}\right)  ^{\frac{d}{2}-2}\left(  k^{\left(  0\right)
}\right)  ^{d-4}k^{\left(  2\right)  }+\right. \left.  \left.  \left(d-4\right)  \left(  t^{2}-s^{2}\right)
^{\frac{d}{2}-3}\left(  k^{\left(  0\right)  }\right)  ^{d-4}\left(  {\ell}S^{\left(
0\right)  }-s^{2}k^{\left(  2\right)  }\right)  \right)  +\ldots\right)~.
\nonumber
\end{align}
We now consider that $k^{\left(  0\right)  }=\frac{\delta}{{\ell}}$, and
therefore we have that
\begin{align}
I  &  =
{\displaystyle\int\limits_{0}^{1}}
ds
{\displaystyle\int\limits_{0}^{s}}
dt\frac{\sqrt{\widetilde{\gamma}}}{{\ell}^{\left(d-3\right)  }}%
\delta_{\left[d-3\right]  }^{\left[d-3\right]  }\left(  \left(
t^{2}-s^{2}\right)  ^{\frac{d}{2}-2}\delta^{d-3}+\right.
\\
&  +\rho\left(  {\ell}\left(  t^{2}-s^{2}\right)  ^{\frac{d}{2}-2}\delta^{d-4}k^{\left(
2\right)  }+  \left.  {\ell}^{2}\left( d-4\right)  \left(  t^{2}-s^{2}\right)
^{\frac{d}{2}-3}\delta^{d-4}\left(  S^{\left(  0\right)  }-\frac{s^{2}}{{\ell}%
}k^{\left(  2\right)  }\right)  \right)  +\ldots\right)~.
\nonumber
\end{align}
Now we use the contraction property (\ref{kronecker}) of the antisymmetrized Kronecker delta, which
states that
\begin{equation}
\delta_{\left[  m\right]  }^{\left[  m\right]  }\delta^{k}=\frac{\left(
N-m+k\right)  !}{\left(  N-m\right)  !}\delta_{\left[  m-k\right]  }^{\left[
m-k\right]  }~,
\end{equation}
where for our purposes $N=d-2$ is the dimension of the $\p\Sigma$ submanifold.
Substituting this into the expression for $I$, we have that%
\begin{align}
I  &  =
{\displaystyle\int\limits_{0}^{1}}
ds
{\displaystyle\int\limits_{0}^{s}}
dt\frac{\sqrt{\widetilde{\gamma}}}{{\ell}^{\left(d-3\right)  }}\Bigg(
\left(d-2\right)  !\left(  t^{2}-s^{2}\right)  ^{\frac{d}{2}-2}
+\rho\left(  {\ell}\left(  t^{2}-s^{2}\right)  ^{\frac{d}{2}-2}\left(d-3\right)
!\tr\left[  k^{\left(  2\right)  }\right]  \right. \nonumber\\
& \qquad +\left.  \left.  {\ell}^{2}\left(d-4\right)  \left(  t^{2}-s^{2}\right)
^{\frac{d}{2}-3}\left(d-3\right)  !\left(\tr\left[  S^{\left(  0\right)  }\right]
-\frac{s^{2}}{{\ell}}\tr\left[  k^{\left(  2\right)  }\right]  \right)
\right)  +\ldots\right)~,
\end{align}
and then
\begin{align}
I  &  =\frac{\sqrt{\widetilde{\gamma}}}{{\ell}^{\left(d-3\right)  }}\left(
\left(d-2\right)  !\left[
{\displaystyle\int\limits_{0}^{1}}
ds {\displaystyle\int\limits_{0}^{s}}
dt\left(  t^{2}-s^{2}\right)  ^{\frac{d}{2}-2}\right]  +\right. \nonumber\\
&  + \rho\left(  {\ell}\left(d-3\right)  !\left(  \left[
{\displaystyle\int\limits_{0}^{1}}
ds
{\displaystyle\int\limits_{0}^{s}}
dt\left(  t^{2}-s^{2}\right)  ^{\frac{d}{2}-2}\right]  -\right.  \right. \left.  \left.  \left(d-4\right)  \left[
{\displaystyle\int\limits_{0}^{1}}
ds
{\displaystyle\int\limits_{0}^{s}}
dt\left(  t^{2}-s^{2}\right)  ^{\frac{d}{2}-3}s^{2}\right]  \right)\tr\left[
k^{\left(  2\right)  }\right]\right.
\nonumber\\
&  \qquad\quad\left.  \left. + {\ell}^{2}\left(d-4\right)  \left[
{\displaystyle\int\limits_{0}^{1}}
ds
{\displaystyle\int\limits_{0}^{s}}
dt\left( t^{2}-s^{2}\right)  ^{\frac{d}{2}-3}\right]  \left(d-3\right)  !\tr\left[
S^{\left(  0\right)  }\right]  \right)  +\ldots\right)~.
\label{previous}
\end{align}
Next we carry out the parametric integrals:
\begin{align}
{\displaystyle\int\limits_{0}^{1}}
ds
{\displaystyle\int\limits_{0}^{s}}
dt\left(  t^{2}-s^{2}\right)  ^{\frac{d}{2}-2}
&  =
\frac{\left(  -1\right)
^{\frac{d}{2}-2}2^{2\left(\frac{d}{2}-2\right)  }\left[  \left(\frac{d}{2}-2\right)  !\right]  ^{2}%
}{\left(  d-2\right)  !}~,
\nonumber\\
{\displaystyle\int\limits_{0}^{1}}
ds
{\displaystyle\int\limits_{0}^{s}}
dt\left(  t^{2}-s^{2}\right)  ^{\frac{d}{2}-3}
&  =
\frac{\left(  -1\right)
^{\frac{d}{2}-3}2^{2\left(\frac{d}{2}-3\right)  }\left[  \left(\frac{d}{2}-3\right)  !\right]^{2}%
}{\left(d-4\right)  !}~,
\\
{\displaystyle\int\limits_{0}^{1}}
ds
{\displaystyle\int\limits_{0}^{s}}
dt\left(  t^{2}-s^{2}\right)  ^{\frac{d}{2}-3}s^{2}
&  =
\frac{\left(  -1\right)
^{\frac{d}{2}-3}2^{2\left(\frac{d}{2}-3\right)  }\left(\frac{d}{2}-3\right)  !
\left(\frac{d}{2}-2\right)
!}{\left(d-4\right)  !\left(\frac{d}{2}-1\right)  }~.
\nonumber
\end{align}
Then, replacing these expressions in (\ref{previous}), we find that
\begin{align}
I  &  =\frac{\sqrt{\widetilde{\gamma}}}{{\ell}^{\left(d-3\right)  }}\left(
\left(d-2\right)  !\frac{\left(  -1\right)  ^{\frac{d}{2}-2}2^{2\left(\frac{d}{2}-2\right)
}\left[  \left(\frac{d}{2}-2\right)  !\right]  ^{2}}{\left(d-2\right)  !}+\right.
\nonumber\\
&  \rho\left(  {\ell}\left(d-3\right)  !\left(  \frac{\left(  -1\right)
^{\frac{d}{2}-2}2^{d-4  }\left[  \left(\frac{d}{2}-2\right)  !\right]  ^{2}%
}{\left(d-2\right)  !}-  \left(d-4\right)  \frac{\left(  -1\right)
^{\frac{d}{2}-3}2^{2\left(\frac{d}{2}-3\right)  }\left(\frac{d}{2}-3\right)  !\left(  \frac{d}{2}-2\right)
!}{\left(d-4\right)  !\left(\frac{d}{2}-1\right)  }\right)  \tr\left[  k^{\left(
2\right)  }\right] \right.
\nonumber\\
&  +\left.  \left.  {\ell}^{2}\left(d-4\right)  \frac{\left(  -1\right)
^{\frac{d}{2}-3}2^{2\left(\frac{d}{2}-3\right)  }\left[  \left(\frac{d}{2}-3\right)  !\right]  ^{2}%
}{\left(d-4\right)  !}\left(d-3\right)  !\tr\left[  S^{\left(  0\right)
}\right]  \right)  +\ldots\right)  .
\end{align}

Plugging this result into (\ref{I_def}), we obtain
\begin{align}
S_{\mathrm{Kt}}
&  =
\frac{{\ell}\left(d-2\right)  \left[  \left(\frac{d}{2}-3\right)
!\right]  ^{2}}{2^{5}G_{\mathrm{N}}\left[  \left(\frac{d}{2}-1\right)  !\right]  ^{2}}
{\displaystyle\int\limits_{\partial\Sigma}}
d^{d-2}y\sqrt{\widetilde{\gamma}}\left(  -2\left(\frac{d}{2}-2\right)  ^{2}\right.
\nonumber\\
&  \left. + \rho\left(  -{\ell}2\left(\frac{d}{2}-2\right)  ^{2}\tr\left[  k^{\left(
2\right)  }\right]  +{\ell}^{2}\left(\frac{d}{2}-2\right)  \left(d-3\right)
\tr\left[  S^{\left(  0\right)  }\right]  \right)  +\ldots\right)  ,
\end{align}
which simplifies to
\begin{align}
S_{\mathrm{Kt}}  &  =\frac{1}{4G_{\mathrm{N}}}
{\displaystyle\int\limits_{\partial\Sigma}}
d^{d-2}y{\ell}\sqrt{\widetilde{\gamma}}\left(  \frac{-1}{\left(d-2\right)
}\right. \nonumber\\
&  +\rho{\ell}\left.  \left(  -\frac{1}{\left(d-2\right)  }\tr\left[
k^{\left(  2\right)  }\right]  +\frac{\left(d-3\right)  }{\left(
d-4\right)  \left(d-2\right)  }{\ell}\tr\left[  S^{\left(  0\right)
}\right]  \right)  +\ldots\right)~.
\end{align}

Now, from (\ref{RelationsB2n-2}), and the definition (\ref{schouten})
of the Schouten tensor, we have that
\begin{align}
\tr\left[  k^{\left(  2\right)  }\right]
&  =
-\frac{1}{{\ell}}\left[
\tr\left[  \sigma^{\left(  2\right)  }\right]  +\frac{{\ell}^{2}\kappa
_{a}^{\widehat{i}a}\kappa_{b}^{\widehat{i}b}}{2\left(d-2\right)  }\right]~,
\nonumber\\
\tr\left[  S^{\left(  0\right)  }\right]
&  =
\frac{\mathcal{R}^{\left(
0\right)  }}{2\left(d-3\right)  }~,
\nonumber\\
\sqrt{\widetilde{\gamma}}
&  =
\frac{\sqrt{\sigma^{\left(  0\right)  }}}%
{\rho^{\left(d-2\right)  /2}}\left(  1+\frac{\rho}{2}\tr\left[
\sigma^{\left(  2\right)  }\right]  +...\right)  .
\end{align}
Then, replacing these relations into the expression for $S_{\mathrm{Kt}}$, we have that
\begin{align}
S_{\mathrm{Kt}}
&  =
\frac{1}{4G_{\mathrm{N}}}
{\displaystyle\int\limits_{\partial\Sigma}}
d^{d-2}y{\ell}\frac{\sqrt{\sigma^{\left(  0\right)  }}}{\rho^{\left(
d-2\right)  /2}}\left(  1+\frac{\rho}{2}tr\left[  \sigma^{\left(  2\right)
}\right]  +...\right)  \left(  \frac{-1}{\left(d-2\right)  }\right.
\nonumber\\
&  \left.  +\rho\left(  \frac{1}{\left(d-2\right)  }\tr\left[  \sigma
^{\left(  2\right)  }\right]  +\frac{{\ell}^{2}\kappa_{a}^{\widehat{i}a}%
\kappa_{b}^{\widehat{i}b}}{2\left(d-2\right)  ^{2}}+\frac{{\ell}%
^{2}\mathcal{R}^{\left(  0\right)  }}{2\left(d-4\right)  \left(
d-2\right)  }\right)  +\ldots\right)~.
\end{align}
Expanding the product we find
\begin{align}
S_{\mathrm{Kt}}
&  =
\frac{1}{4G_{\mathrm{N}}}
{\displaystyle\int\limits_{\partial\Sigma}}
d^{d-2}y{\ell}\frac{\sqrt{\sigma^{\left(  0\right)  }}}{\rho^{\left(
d-2\right)  /2}}\left(  \frac{-1}{\left(d-2\right)  }\right.
\nonumber\\
&  +\rho\left(  \frac{1}{2\left(d-2\right)  }\tr\left[  \sigma^{\left(
2\right)  }\right]  +\frac{1}{2\left(d-4\right)  \left(d-2\right)
}{\ell}^{2}\mathcal{R}^{\left(  0\right)  }+ \left.  \frac{{\ell}^{2}\kappa_{a}^{\widehat{i}a}\kappa
_{b}^{\widehat{i}b}}{2\left(d-2\right)  ^{2}}\right)  +\ldots\right)~,
\end{align}
which simplifies to
\begin{align}
S_{\mathrm{Kt}}
&  \label{Answer} =
\\
&  -\frac{1}{4G_{\mathrm{N}}}{\displaystyle\int\limits_{\partial\Sigma}}
d^{d-2}y{\ell}\frac{\sqrt{\sigma^{\left(  0\right)  }}}{\left(d-2\right)
\rho^{\left(d-2\right)  /2}}\left(  1+  \rho\left(  -\frac{1}{2}\tr\left[  \sigma^{\left(  2\right)
}\right]  -\frac{{\ell}^{2}\kappa_{a}^{\widehat{i}a}\kappa_{b}^{\widehat{i}b}%
}{2\left(d-2\right)  }-\frac{{\ell}^{2}}{2\left(d-4\right)  }%
\mathcal{R}^{\left(  0\right)  }\right)  +\ldots\right)~.
\nonumber
\end{align}
This is our final expression for $S_{\mathrm{Kt}}$, which is used in
Eq.~(\ref{AnswerBody}) of the main text.

\section{Relations between boundary curvature tensors}\label{B}

We proceed to derive some useful relations between the codimension-1 and
codimension-3 boundary curvature tensors. For this, we consider an orthogonal
codimension-$p$ foliation of $d$-dimensional spacetime. The completeness relation
for the metric states that
\begin{equation}
g^{\mu\nu}=\left(  \left(  n_{\left(  i\right)  }\right)  ^{\mu}\left(
n_{\left(  i\right)  }\right)  ^{\nu}+e_{a}^{\mu}e_{b}^{\nu}\sigma
^{ab}\right)~,
\end{equation}
such that $\left(  n_{\left(  i\right)  }\right)  ^{\mu}$ are the unit vectors
pointing along the foliation directions $\left(  i\right)  =1,\ldots,p$,
$e_{a}^{\mu}$ are the frame vectors tangent to the foliation sheets along the
$a=p+1,...,d$ directions, and $\sigma_{ab}$ is the induced metric on the sheets.
By choosing the worldvolume of the sheets to be parametrized by coordinates
$y^{a}$, such that they are a subset of the bulk coordinates $x^{\mu}$, we
have the following orthogonality relations:
\begin{equation}
e_{a}^{\mu}=\frac{\partial x^{\mu}}{\partial y^{a}}=\delta_{a}^{\mu},\text{
}\left(  n_{\left(  i\right)  }\right)  ^{a}=0~.
\end{equation}
Then, for this foliation, the Gauss-Codazzi decomposition of the Riemann
tensor is given by
\begin{equation}
R_{cd}^{ab}=\mathcal{R}_{cd}^{ab}-\kappa_{c}^{\widehat{i}a}\kappa
_{d}^{\widehat{i}b}+\kappa_{d}^{\widehat{i}a}\kappa_{c}^{\widehat{i}b}~,
\label{Gauss-Codazzi}%
\end{equation}
where $\mathcal{R}_{cd}^{ab}$ is the intrinsic Riemann tensor on the sheets
and $\kappa_{ab}^{\widehat{i}}$ is the extrinsic curvature along the
$\widehat{i}$-th direction.

Stating from eq.(\ref{Gauss-Codazzi}), we proceed to derive the decomposition
of other useful curvature tensors. In particular, for the Ricci tensor, we
have that
\begin{equation}
R_{ab}=g^{\lambda\sigma}R_{\lambda a\sigma b}=\left(  \left(  n_{\left(
i\right)  }\right)  ^{\lambda}\left(  n_{\left(  i\right)  }\right)  ^{\sigma
}+e_{c}^{\lambda}e_{d}^{\sigma}\sigma^{cd}\right)  R_{\lambda a\sigma b}~,
\end{equation}
and therefore,
\begin{align}
R_{ab}  &  =R_{\left(  i\right)  a\left(  i\right)  b}+\sigma^{cd}\left(
\mathcal{R}_{cadb}-\kappa_{cd}^{\widehat{i}}\kappa_{ab}^{\widehat{i}}%
+\kappa_{cb}^{\widehat{i}}\kappa_{ad}^{\widehat{i}}\right)  ,\nonumber\\
&  =R_{\left(  i\right)  a\left(  i\right)  b}+\mathcal{R}_{ab}-\kappa
_{d}^{\widehat{i}d}\kappa_{ab}^{\widehat{i}}+\kappa_{ad}^{\widehat{i}}%
\kappa_{b}^{\widehat{i}d}~.
\end{align}

For the Ricci scalar, we have
\begin{align}
R  &  =g^{\lambda\sigma}g^{\mu\nu}R_{\lambda\mu\sigma\nu}\nonumber\\
&  =\left(  \left(  n_{\left(  i\right)  }\right)  ^{\lambda}\left(
n_{\left(  i\right)  }\right)  ^{\sigma}+e_{c}^{\lambda}e_{d}^{\sigma}%
\sigma^{cd}\right)  \left(  \left(  n_{\left(  j\right)  }\right)  ^{\mu
}\left(  n_{\left(  j\right)  }\right)  ^{\nu}+e_{a}^{\mu}e_{b}^{\nu}%
\sigma^{ab}\right)  R_{\lambda\mu\sigma\nu}~,
\end{align}
and therefore,
\begin{equation}
R=R_{\left(  i\right)  \left(  j\right)  \left(  i\right)  \left(  j\right)
}+2\left(  n_{\left(  i\right)  }\right)  ^{\lambda}\left(  n_{\left(
i\right)  }\right)  ^{\sigma}e_{a}^{\mu}e_{b}^{\nu}\sigma^{ab}R_{\lambda
\mu\sigma\nu}+\sigma^{cd}\sigma^{ab}R_{cadb}~.
\end{equation}
Now, considering that%
\begin{equation}
e_{a}^{\mu}e_{b}^{\nu}\sigma^{ab}=g^{\mu\nu}-\left(  n_{\left(  j\right)
}\right)  ^{\mu}\left(  n_{\left(  j\right)  }\right)  ^{\nu}~,
\end{equation}
we have
\begin{equation}
\left(  n_{\left(  i\right)  }\right)  ^{\lambda}\left(  n_{\left(  i\right)
}\right)  ^{\sigma}\left(  g^{\mu\nu}-\left(  n_{\left(  j\right)  }\right)
^{\mu}\left(  n_{\left(  j\right)  }\right)  ^{\nu}\right)  R_{\lambda
\mu\sigma\nu}=R_{\left(  i\right)  \left(  i\right)  }-R_{\left(  i\right)
\left(  j\right)  \left(  i\right)  \left(  j\right)  }~,%
\end{equation}
and therefore,
\begin{equation}
R
=-R_{\left(  i\right)  \left(  j\right)  \left(  i\right)  \left(
j\right)  }+2R_{\left(  i\right)  \left(  i\right)  }+R_{ab}^{ab}~.
\end{equation}
Finally, using the Gauss-Codazzi decomposition of Eq.~(\ref{Gauss-Codazzi}), we
have that
\begin{equation}
R=-R_{\left(  i\right)  \left(  j\right)  \left(  i\right)  \left(  j\right)
}+2R_{\left(  i\right)  \left(  i\right)  }+\mathcal{R}-\kappa_{a}%
^{\widehat{i}a}\kappa_{b}^{\widehat{i}b}+\kappa_{b}^{\widehat{i}a}\kappa
_{a}^{\widehat{i}b}~.
\end{equation}

For the decomposition of the Schouten tensor, we consider its definition as%
\begin{equation}
S_{\nu}^{\mu}=\frac{1}{\left(d-2\right)  }\left(  R_{\nu}^{\mu}-\frac
{R}{2\left( d-1\right)  }\delta_{\nu}^{\mu}\right)~,
\label{schouten}
\end{equation}
such that its trace is given by
\begin{equation}
S=\frac{1}{2\left(d-1\right)  }R~.
\end{equation}
Then, we have that the bulk Schouten tensor with indices along the sheet
directions is given by
\begin{align}
S_{ab}  &  =\frac{1}{\left( d-2\right)  }\left(  \left(  R_{\left(
i\right)  a\left(  i\right)  b}+\mathcal{R}_{ab}-\kappa_{d}^{\widehat{i}%
d}\kappa_{ab}^{\widehat{i}}+\kappa_{ad}^{\widehat{i}}\kappa_{b}^{\widehat{i}%
d}\right)  \right. \nonumber\\
&  \left.  -\frac{\left(  -R_{\left(  i\right)  \left(  j\right)  \left(
i\right)  \left(  j\right)  }+2R_{\left(  i\right)  \left(  i\right)
}+\mathcal{R}-\kappa_{a}^{\widehat{i}a}\kappa_{b}^{\widehat{i}b}+\kappa
_{b}^{\widehat{i}a}\kappa_{a}^{\widehat{i}b}\right)  }{2d-2
}\sigma_{ab}\right)~.
\end{align}
Thus, the partial trace of the Schouten tensor along the sheet directions is%
\begin{align}
S_{a}^{a}  &  =\frac{1}{d-2  }\left(  \left(  R_{\left(
i\right)  \left(  i\right)  a}^{a}+\mathcal{R}-\kappa_{d}^{\widehat{i}d}%
\kappa_{a}^{\widehat{i}a}+\kappa_{d}^{\widehat{i}a}\kappa_{a}^{\widehat{i}%
d}\right)  \right. \nonumber\\
&  \left.  -\left(  -R_{\left(  i\right)  \left(  j\right)  \left(  i\right)
\left(  j\right)  }+2R_{\left(  i\right)  \left(  i\right)  }+\mathcal{R}%
-\kappa_{a}^{\widehat{i}a}\kappa_{b}^{\widehat{i}b}+\kappa_{b}^{\widehat{i}%
a}\kappa_{a}^{\widehat{i}b}\right)  \frac{d-2  }{2d-2 }\right)~,
\end{align}
and considering that
\begin{equation}
R_{\left(  i\right)  \left(  i\right)  a}^{a}=R_{\left(  i\right)  \left(
i\right)  }-R_{\left(  i\right)  \left(  j\right)  \left(  i\right)  \left(
j\right)  }~,
\end{equation}
we obtain
\begin{equation}
S_{a}^{a}=\frac{1}{\left(d-2\right)  \left( d-1\right)  }\left(
R_{\left(  i\right)  \left(  i\right)  }-\frac{d}{2}\left(  R_{\left(
i\right)  \left(  j\right)  \left(  i\right)  \left(  j\right)  }%
-\mathcal{R}+\kappa_{d}^{\widehat{i}d}\kappa_{a}^{\widehat{i}a}-\kappa
_{d}^{\widehat{i}a}\kappa_{a}^{\widehat{i}d}\right)  \right)  . \label{Saa}%
\end{equation}

Now that we have obtained the Gauss-Codazzi decompositions of the Riemann,
Ricci and Schouten tensors, as well as that of the Ricci scalar, we proceed to
compute $\tr\left[  \sigma^{\left(  2\right)  }\right]  $ and $\delta_{a}%
^{c}\delta_{b}^{d}W_{cd}^{ab}$, which are needed in Eqs.~(\ref{TrSig2}) and
(\ref{TrWeyl}) of the main text. In this context, the bulk quantities refer
to the tensors of the codimension-1 boundary which corresponds to the CFT
spacetime, and the sheet quantities refer to the codimension-3 tensors that
are defined on the entangling surface.

We first consider the definition of $\sigma_{ab}^{\left(  2\right)  }$, as
given in (\ref{Sigma2}).Using the result for $S_{a}^{a}$ given in
(\ref{Saa}), we have
\begin{align}
\tr\left[  \sigma^{\left(  2\right)  }\right]
=&
\\
&  -\frac{{\ell}^{2}}{d-2  }\left(  \frac{1}{d-1 }R_{\left(  i\right)  \left(  i\right)  }-\frac{d}{2d-2  }\left(  R_{\left(  i\right)  \left(  j\right)  \left(  i\right)
\left(  j\right)  }- \mathcal{R}^{\left(  0\right)  }+\kappa_{d}^{\widehat
{i}d}\kappa_{a}^{\widehat{i}a}-\kappa_{d}^{\widehat{i}a}\kappa_{a}%
^{\widehat{i}d}\right)  +\kappa_{d}^{\widehat{i}d}\kappa_{a}^{\widehat{i}%
a}\right)  ,
\nonumber
\end{align}
which simplifies to
\begin{align}
\tr\left[  \sigma^{\left(  2\right)  }\right]
=&
\\
&  -\frac{{\ell}^{2}}{\left(d-2\right)  \left( d-1\right)  }\left(
R_{\left(  i\right)  \left(  i\right)  }+\frac{d-2  }{2}%
\kappa_{d}^{\widehat{i}d}\kappa_{a}^{\widehat{i}a}- \frac{d}{2}\left(
R_{\left(  i\right)  \left(  j\right)  \left(  i\right)  \left(  j\right)
}-\mathcal{R}^{\left(  0\right)  }-\kappa_{d}^{\widehat{i}a}\kappa
_{a}^{\widehat{i}d}\right)  \right)~.
\nonumber
\end{align}

Now, we proceed to compute $\delta_{a}^{c}\delta_{b}^{d}W_{cd}^{ab}$, which is
the partial trace of the Weyl tensor of $g^{\left(  0\right)  }$, along the
codimension-3 directions. We first consider the definition of the Weyl tensor
in terms of the Riemann and Schouten tensors, given by
\begin{equation}
W_{\rho\sigma}^{\mu\nu}=R_{\rho\sigma}^{\mu\nu}-4S_{\left[  \rho\right.
}^{\left[  \mu\right.  }\delta_{\left.  \sigma\right]  }^{\left.  \nu\right]
}~.
\end{equation}
Then, we have that
\begin{align}
\delta_{a}^{c}\delta_{b}^{d}W_{cd}^{ab}  &  =\delta_{a}^{c}\delta_{b}%
^{d}\left(  R_{cd}^{ab}-4S_{\left[  c\right.  }^{\left[  a\right.  }%
\delta_{\left.  d\right]  }^{\left.  b\right]  }\right)
\nonumber\\
&  =\delta_{a}^{c}\delta_{b}^{d}\left(  R_{cd}^{ab}-S_{c}^{a}\delta_{d}%
^{b}+S_{d}^{a}\delta_{c}^{b}+S_{c}^{b}\delta_{d}^{a}-S_{d}^{b}\delta_{c}%
^{a}\right)
\nonumber\\
&  =R_{ab}^{ab}-S_{a}^{a}\left(d-2\right)  +S_{a}^{a}+S_{a}^{a}-S_{a}%
^{a}\left( d-2\right)
\nonumber\\
&  =R_{ab}^{ab}-2\left( d-3\right)  S_{a}^{a}.
\end{align}
Finally, using the Gauss-Codazzi decompositions found above, we obtain%
\begin{align}
\delta_{a}^{c}\delta_{b}^{d}W_{cd}^{ab}
=&
\left(  \mathcal{R}^{\left(
0\right)  }-\kappa_{a}^{\widehat{i}a}\kappa_{b}^{\widehat{i}b}+\kappa
_{b}^{\widehat{i}a}\kappa_{a}^{\widehat{i}b}\right)
\\
&  -\left(d-3\right)  \left(  \frac{2}{\left(  d-2\right)  \left(
d-1\right)  }\left(  R_{\left(  i\right)  \left(  i\right)  }-\right.
\right.  \left.  \left.  d\left(  R_{\left(  i\right)  \left(
j\right)  \left(  i\right)  \left(  j\right)  }-\mathcal{R}^{\left(  0\right)
}+\kappa_{d}^{\widehat{i}d}\kappa_{a}^{\widehat{i}a}-\kappa_{d}^{\widehat{i}%
a}\kappa_{a}^{\widehat{i}d}\right)  \right)  \right)  ,
\nonumber
\end{align}
which simplifies to
\begin{align}
\delta_{a}^{c}\delta_{b}^{d}W_{cd}^{ab}
=&
\frac{2}{\left(d-2\right)
\left(  d-1\right)  }\mathcal{R}^{\left(  0\right)  }-\frac{2\left(
d-3\right)  }{\left( d-2\right)  \left( d-1\right)  }R_{\left(  i\right)
\left(  i\right)  }
\\
&  +\frac{d\left(d-3\right)  }{\left(d-2\right)  \left(d-1\right)
}R_{\left(  i\right)  \left(  j\right)  \left(  i\right)  \left(  j\right)
}+\frac{2}{\left(d-2\right)  \left(d-1\right)  }\left(  \kappa
_{b}^{\widehat{i}a}\kappa_{a}^{\widehat{i}b}-\kappa_{a}^{\widehat{i}a}%
\kappa_{b}^{\widehat{i}b}\right)~.
\nonumber
\end{align}
We use this expression for $\delta_{a}^{c}\delta_{b}^{d}W_{cd}^{ab}$ in
Eq.~(\ref{TrWeyl}) of the main text.

\section{Asymptotic conformal flatness}\label{C}

The simplifying condition used in \cite{Anastasiou:2018the} to show the equivalence
between the Kounterterms and the standard counterterms of holographic
renormalization is that, in the evaluation of the $B_{d}$ boundary terms, the
bulk Weyl tensor with boundary indices has a negligible contribution. This is
the working definition of the Asymptotically Conformally Flat (ACF) condition,
and it requires the fall-off of the bulk Weyl tensor with boundary indices to
be the same than the normalizable mode, i.e. $W_{kl}^{ij}\sim O\left(
\rho^{\frac{d}{2}}\right)  $, where $d=D-1$ is the dimension of the boundary.

Let us consider first the case of an even-$D$ bulk manifold, with $D=2n$. We know from (\ref{bd})
that the boundary term is given by%
\begin{equation}
B_{2n-1}=-2n\sqrt{-h}%
{\displaystyle\int\limits_{0}^{1}}
dt\delta_{\left[  2n-1\right]  }^{\left[  2n-1\right]  }K\left(  \frac{1}%
{2}\mathcal{R}ie-t^{2}KK\right)  ^{n-1}.
\end{equation}
Also, considering the Gauss-Codazzi decomposition of the Riemann tensor, we have%
\begin{equation}
R_{kl}^{ij}=\mathcal{R}_{kl}^{ij}+K_{k}^{i}K_{l}^{j}-K_{l}^{i}K_{k}^{j}~.
\end{equation}
Now, for the on-shell bulk Weyl tensor, considering Einstein spacetimes, we have%
\begin{equation}
W_{kl}^{ij}=R_{kl}^{ij}+\frac{1}{\ell^{2}}\delta_{\left[  kl\right]
}^{\left[  ij\right]  }~.
\end{equation}
Thus, we can write%
\begin{equation}
B_{2n-1}=-2n\sqrt{-h}
{\displaystyle\int\limits_{0}^{1}}
dt\delta_{\left[  2n-1\right]  }^{\left[  2n-1\right]  }K\left(  \frac{1}%
{2}W+\left(  1-t^{2}\right)  KK-\frac{1}{\ell^{2}}\delta\delta\right)  ^{n-1}.
\end{equation}
Then, the ACF condition corresponds to considering that $W_{kl}^{ij}\sim
O\left(  \rho^{\frac{2n-1}{2}}\right)  $, such that $W_{kl}^{ij}$ can be
neglected in the above expression. To see that this is the case, we consider
the lowest order in $\rho$ contribution that can come from $W_{kl}^{ij}$.
Taking into account the Fefferman-Graham decomposition of $K$ as $K=\frac
{1}{\ell}\delta+\rho\ell S+O\left(  \rho^{2}\right)  $, the lowest order
contribution will be proportional to $\tr_{\partial M}\left[  W\right]  $. We
note that this is not the full trace of the bulk Weyl, which is zero as usual,
but it is an double trace over the boundary indices only. We proceed to show
that this is indeed zero. We have that
\begin{equation}
\tr_{\partial M}\left[  W\right]  =W_{ij}^{ij}=\left(  W_{\mu\nu}^{\mu\nu
}-W_{\rho\rho}^{\rho\rho}-2W_{i\rho}^{i\rho}\right)  =-2W_{i\rho}^{i\rho
}=-2\epsilon_{i}^{i}~, \label{electric_W}%
\end{equation}
where $\epsilon_{j}^{i}$ is the electric part of the (bulk) Weyl tensor.

Now,
$\epsilon_{i}^{i}$ can be shown to be zero as follows:%
\begin{equation}
W_{\mu\nu}^{\mu\nu}=2W_{i\rho}^{i\rho}+W_{ij}^{ij}+W_{\rho\rho}^{\rho\rho
}=2\epsilon_{i}^{i}+W_{ij}^{ij}=0~,
\end{equation}
and also
\begin{align}
W_{\mu j}^{\mu i}  &  =W_{ij}^{ij}+W_{i\rho}^{i\rho}=W_{ij}^{ij}+\epsilon
_{i}^{i}=0~,
\nonumber\\
W_{ij}^{ij}  &  =-\epsilon_{i}^{i}~%
\end{align}
and therefore, $\epsilon_{i}^{i}=0$. Thus,%
\begin{equation}
\tr_{\partial M}\left[  W\right]  =0. \label{tr_dM_W}%
\end{equation}

We see then that, if $W_{kl}^{ij}\sim O\left(  \rho^{\frac{2n-1}{2}}\right)  $, the lowest-order contribution to $\int B_{2n-1}$ due to the bulk Weyl, which naively would
be of $O\left(  1\right)  $, will in fact vanish, and all higher-order contributions
will fall-off to zero at the conformal bondary. Therefore, indeed one can
neglect the contribution of $W_{kl}^{ij}$ in the computation of $B_{2n-1}$~,
provided that the ACF condition is satisfied. As shown in
\cite{Anastasiou:2018the}, when $W_{kl}^{ij}$ is neglected in the above
computation, the standard holographic renormalization counterterms are
recovered up to sixth order in derivatives of the metric (which is sufficient
to have full agreement up to $D=8$).

Analogously, in odd-$D$ bulk manifolds, $D=2n+1$, the boundary term
defined in (\ref{bd}) is given by
\begin{equation}
B_{2n}=-2n\sqrt{-h}
{\displaystyle\int\limits_{0}^{1}}
ds
{\displaystyle\int\limits_{0}^{s}}
dt\,\delta_{\left[  2n\right]  }^{\left[  2n\right]  }K\delta\left(  \frac{1}%
{2}\mathcal{R}ie-s^{2}KK+\frac{t^{2}}{\ell^{2}}\delta\delta\right)  ^{n-1}~,
\end{equation}
and considering the previous Gauss-Codazzi decomposition we can again write
\begin{equation}
B_{2n}=-2n\sqrt{-h}
{\displaystyle\int\limits_{0}^{1}}
ds
{\displaystyle\int\limits_{0}^{s}}
dt\,\delta_{\left[  2n\right]  }^{\left[  2n\right]  }K\delta\left(  \frac{1}%
{2}W+\left(  1-s^{2}\right)  KK-\frac{\left(  1-t^{2}\right)  }{\ell^{2}%
}\delta\delta\right)  ^{n-1}~.
\end{equation}
Then, the ACF condition corresponds to considering that $W_{kl}^{ij}\sim
O\left(  \rho^{n}\right)  $, such that $W_{kl}^{ij}$ can be neglected in the
above expression. To see that this is true, we proceed like in the previous
case and again consider the lowest order in $\rho$ contribution that can come
from $W_{kl}^{ij}$. Taking into account the Fefferman-Graham decomposition of
$K$, the lowest order contribution will again be proportional to $\tr_{\partial
M}\left[  W\right]  =0$. Thus, if $W_{kl}^{ij}\sim O\left(  \rho^{n}\right)
$, all the contributions due to $W_{kl}^{ij}$ will fall-off faster than the
normalizable mode and will not modify the computation of $B_{2n}$ at the
conforml boundary. As shown in \cite{Anastasiou:2018the}, when $W_{kl}^{ij}$
is neglected in the above computation, the standard Holographic
Renormalization counterterms are recovered up to sixth order in derivatives of
the metric (which is sufficient to have full agreement up to $D=9$, excluding
the logarithmic term, which is not cancelled in the case of the Kounterterms).

We now proceed to show that the Conformally Flat boundary condition (CF
condition) implies the ACF condition, at least up to $D=9$, such that CF$=$ACF
up to that dimension. If this is the case, then the applicability of the
Kounterterms is restricted only to manifolds with $CF$ boundary up to $D=9$
(while being fully general up to $D=5$). In order to show this, we use the FG
expansion of $\mathcal{W}\left[  h\right]  $, given by $\mathcal{W}\left[
h\right]  =\rho W_{\left(  0\right)  }+\rho^{2}W_{\left(  2\right)  }+\ldots$
Then, we consider that the bulk Weyl with boundary indices $W_{kl}^{ij}$ being
of order $O\left(  \rho^{\frac{d}{2}}\right)  $ is equivalent to $W_{\left(
0\right)  }=W_{\left(  2\right)  }=\ldots=0$ up to the $O\left(  \rho
^{\frac{d}{2}}\right)  $ coefficient. Thus, if $W_{\left(  0\right)  }=0$
implies that $W_{\left(  2\right)  }=\ldots=0$, then the CF condition is
equivalent to the ACF condition up to the corresponding order.
For example, if
$W_{\left(  0\right)  }=0$ implies $W_{\left(  2\right)  }=0$, then CF$=$ACF
up to $D=9$. Thus, we verify that $W_{\left(  0\right)  }=0$ implies
$W_{\left(  2\right)  }=0$ by considering the FG expansion of the boundary
tensors.

{}From (\ref{fgmetricrho}), we know that the boundary metric $h_{ij}\left(  \rho,x^{i}\right)  $
has an FG expansion given by
\begin{equation}
h_{ij}\left(  \rho,x^{i}\right)  =\frac{g_{\left(  0\right)  ij}\left(
x^{i}\right)  +\rho g_{\left(  2\right)  ij}\left(  x^{i}\right)  +\rho
^{2}g_{\left(  4\right)  ij}\left(  x^{i}\right)  +\ldots}{\rho}~,
\end{equation}
such that
\begin{equation}
\left(  g_{\left(  4\right)  }\right)  _{j}^{i}=\frac{\ell^{4}}{4\left(
d-4\right)  }\left[  \left(  d-4\right)  S_{k}^{i}S_{j}^{k}-B_{j}^{i}\right]~,
\end{equation}
where $S_{j}^{i}$ is the Schouten tensor and $B_{j}^{i}$ is the Bach tensor
(both being tensors of $g_{\left(  0\right)  }$). Also,%
\begin{align}
B_{k}^{j}  &  =\nabla^{i}C_{ik}^{j}+S_{n}^{m}\left(  W_{\left(  0\right)
}\right)  _{mk}^{nj},\\
B_{ij}  &  =\nabla^{k}\nabla_{k}S_{ij}-\nabla_{k}\nabla_{j}S_{i}^{k}%
-S^{km}\left(  W_{\left(  0\right)  }\right)  _{kijm}~,
\end{align}
where $C_{ik}^{j}$ is the Cotton tensor of $g_{\left(  0\right)  }$ and the
covariant derivative $\nabla_{i}$ is with respect to $g_{\left(  0\right)  }$
as well. Then we have%
\begin{equation}
C_{l}^{jk}=\frac{1}{\ell^{2}\left(  d-3\right)  }\nabla^{i}\left(  W_{\left(
0\right)  }\right)  _{li}^{jk}~,
\end{equation}
where $\left(  W_{\left(  0\right)  }\right)  _{li}^{jk}$ is the Weyl tensor
of $g_{\left(  0\right)  }$. Now, the Weyl tensor of $h_{ij}$ has an FG
expansion given by%
\begin{equation}
\mathcal{W}_{kl}^{ij}=\rho\left(  W_{\left(  0\right)  }\right)  _{kl}%
^{ij}+\rho^{2}\left(  W_{\left(  2\right)  }\right)  _{kl}^{ij}+\ldots~,
\end{equation}
where%
\begin{equation}
\left(  W_{\left(  0\right)  }\right)  _{kl}^{ij}=R_{kl}^{ij}-4\delta_{\lbrack
k}^{[i}S_{l]}^{j]}~,%
\end{equation}
and $R_{kl}^{ij}$ is the Riemann tensor of $g_{\left(  0\right)  }$. Also%
\begin{align}
\left(  W_{\left(  2\right)  }\right)  _{kl}^{ij}
=&
\ell^{2}\left(
\frac{1}{2}\left(  \nabla_{k}C_{l}^{ij}-\nabla_{l}C_{k}^{ij}+S_{m}^{j}%
R_{kl}^{im}-S_{m}^{i}R_{kl}^{jm}\right)  -\left(  S_{k}^{i}S_{l}^{j}-S_{l}%
^{i}S_{k}^{j}\right)  \right.
\nonumber\\
&  +\left(  -\delta_{k}^{i}\left(  S^{2}\right)  _{l}^{j}+\delta_{l}%
^{i}\left(  S^{2}\right)  _{k}^{j}+\delta_{k}^{j}\left(  S^{2}\right)
_{l}^{i}-\delta\left(  S^{2}\right)  _{k}^{i}\right)
\nonumber\\
&  \left.  +\frac{2}{\ell^{4}}\left(  \delta_{k}^{i}\left(  g_{\left(
4\right)  }\right)  _{l}^{j}-\delta_{l}^{i}\left(  g_{\left(  4\right)
}\right)  _{k}^{j}-\delta_{k}^{j}\left(  g_{\left(  4\right)  }\right)
_{l}^{i}+\delta_{l}^{j}\left(  g_{\left(  4\right)  }\right)  _{k}^{i}\right)
\right)~.
\end{align}

Now, to show that $W_{\left(  0\right)  }=0$ implies $W_{\left(  2\right)
}=0$, we first simplify the different terms under the condition that the Weyl
tensor of $g_{\left(  0\right)  }$ is zero (i.e., that $g_{\left(  0\right)
}$ is conformally flat). We then have that%
\begin{align}
C_{l}^{jk}  &  =\frac{1}{d-3}\nabla^{i}\left(  W_{\left(  0\right)  }\right)
_{li}^{jk}=0~,\nonumber\\
\nabla^{i}C_{ik}^{j}  &  =\frac{1}{d-3}\nabla_{l}\nabla^{i}\left(  W_{\left(
0\right)  }\right)  _{ik}^{jl}=0~,\nonumber\\
B_{k}^{j}  &  =\nabla^{i}C_{ik}^{j}+S_{n}^{m}\left(  W_{\left(  0\right)
}\right)  _{mk}^{nj}=0~.
\end{align}
We also have that
\begin{align}
S_{m}^{j}R_{kl}^{im}  &  =S_{m}^{j}\left(  \left(  W_{\left(  0\right)
}\right)  _{kl}^{im}+4\delta_{\lbrack k}^{[i}S_{l]}^{m]}\right)  =4S_{m}%
^{j}\delta_{\lbrack k}^{[i}S_{l]}^{m]}~,\nonumber\\
\left(  g_{\left(  4\right)  }\right)  _{j}^{i}  &  =\frac{\ell^{4}}{4}%
S_{k}^{i}S_{j}^{k}~.
\end{align}
Therefore, $W_{\left(  2\right)  }$ simplifies to%
\begin{align}
\left(  W_{\left(  2\right)  }\right)  _{kl}^{ij}
=&
\ell^{2}\left(
\frac{1}{2}\left(  4S_{m}^{j}\delta_{\lbrack k}^{[i}S_{l]}^{m]}-4S_{m}%
^{i}\delta_{\lbrack k}^{[j}S_{l]}^{m]}\right)  -\left(  S_{k}^{i}S_{l}%
^{j}-S_{l}^{i}S_{k}^{j}\right)  \right. \nonumber\\
&  +\left(  -\delta_{k}^{i}\left(  S^{2}\right)  _{l}^{j}+\delta_{l}%
^{i}\left(  S^{2}\right)  _{k}^{j}+\delta_{k}^{j}\left(  S^{2}\right)
_{l}^{i}-\delta_{l}^{j}\left(  S^{2}\right)  _{k}^{i}\right) \nonumber\\
&  \left.  +\frac{2}{\ell^{4}}\left(  \delta_{k}^{i}\frac{\ell^{4}}{4}\left(
S^{2}\right)  _{l}^{j}-\delta_{l}^{i}\frac{\ell^{4}}{4}\left(  S^{2}\right)
_{k}^{j}-\delta_{k}^{j}\frac{\ell^{4}}{4}\left(  S^{2}\right)  _{l}^{i}%
+\delta_{l}^{j}\frac{\ell^{4}}{4}\left(  S^{2}\right)  _{k}^{i}\right)
\right)~,
\end{align}
i.e.,
\begin{align}
\left(  W_{\left(  2\right)  }\right)  _{kl}^{ij}
=&
\ell^{2}\left(
\frac{1}{2}\left(  4S_{m}^{j}\delta_{\lbrack k}^{[i}S_{l]}^{m]}-4S_{m}%
^{i}\delta_{\lbrack k}^{[j}S_{l]}^{m]}\right)  -\left(  S_{k}^{i}S_{l}%
^{j}-S_{l}^{i}S_{k}^{j}\right)  \right. \nonumber\\
&  \left.  +\frac{1}{2}\left(  -\delta_{k}^{i}\left(  S^{2}\right)  _{l}%
^{j}+\delta_{l}^{i}\left(  S^{2}\right)  _{k}^{j}+\delta_{k}^{j}\left(
S^{2}\right)  _{l}^{i}-\delta_{l}^{j}\left(  S^{2}\right)  _{k}^{i}\right)
\right)~.
\end{align}
Using that
\begin{align}
4S_{m}^{j}\delta_{\lbrack k}^{[i}S_{l]}^{m]}  &  =S_{m}^{j}\left(  \delta
_{k}^{i}S_{l}^{m}-\delta_{k}^{m}S_{l}^{i}-\delta_{l}^{i}S_{k}^{m}+\delta
_{l}^{m}S_{k}^{i}\right)
\nonumber\\
&  =-S_{k}^{j}S_{l}^{i}+S_{l}^{j}S_{k}^{i}-\delta_{l}^{i}\left(  S^{2}\right)
_{k}^{j}+\delta_{k}^{i}\left(  S^{2}\right)  _{l}^{j}~,
\nonumber\\
-4S_{m}^{i}\delta_{\lbrack k}^{[j}S_{l]}^{m]}  &  =-S_{k}^{j}S_{l}^{i}%
+S_{l}^{j}S_{k}^{i}+\delta_{l}^{j}\left(  S^{2}\right)  _{k}^{i}-\delta
_{k}^{j}\left(  S^{2}\right)  _{l}^{i}~,
\end{align}
we obtain
\begin{align}
\left(  W_{\left(  2\right)  }\right)  _{kl}^{ij}  &  =\ell^{2}\left(  \left(
-S_{k}^{j}S_{l}^{i}+S_{l}^{j}S_{k}^{i}\right)  -\frac{1}{2}\left(  \delta
_{l}^{i}\left(  S^{2}\right)  _{k}^{j}-\delta_{k}^{i}\left(  S^{2}\right)
_{l}^{j}-\delta_{l}^{j}\left(  S^{2}\right)  _{k}^{i}+\delta_{k}^{j}\left(
S^{2}\right)  _{l}^{i}\right)  \right.
\nonumber\\
&  \left.  -\left(  S_{k}^{i}S_{l}^{j}-S_{l}^{i}S_{k}^{j}\right)  +\frac{1}%
{2}\left(  -\delta_{k}^{i}\left(  S^{2}\right)  _{l}^{j}+\delta_{l}^{i}\left(
S^{2}\right)  _{k}^{j}+\delta_{k}^{j}\left(  S^{2}\right)  _{l}^{i}-\delta
_{l}^{j}\left(  S^{2}\right)  _{k}^{i}\right)  \right)~,
\end{align}
and thus%
\begin{equation}
\left(  W_{\left(  2\right)  }\right)  _{kl}^{ij}=0.
\end{equation}

Therefore, we have shown that requiring $W_{\left(  0\right)  }=0$ implies
that $W_{\left(  2\right)  }=0$ as well. Thus, up to $D=9$ (included), having
a conformally flat boundary manifold implies that one can neglect the full
bulk Weyl with boundary indices in the evaluation of the extrinsic boundary
counterterms, as in this case, $W_{kl}^{ij}\sim O\left(  \rho^{\frac{d}{2}%
}\right)  $, which corresponds to the Asymptotically Conformally Flat (ACF)
condition by definition. In other words, up to $D=9$, the ACF condition for
the bulk is equivalent to having a conformally flat boundary.

Note that in the general case, for $W_{\left(  0\right)  }\neq0$, $\left(
W_{\left(  2\right)  }\right)  _{kl}^{ij}$ can be further simplified and
rewritten in a more compact form. In particular, we have that%
\begin{align}
\left(  W_{\left(  2\right)  }\right)  _{kl}^{ij}  &  =\ell^{2}\left(
\frac{1}{2}\left(  \nabla_{k}C_{l}^{ij}-\nabla_{l}C_{k}^{ij}+S_{m}^{j}\left(
W_{\left(  0\right)  }\right)  _{kl}^{im}-S_{m}^{i}\left(  W_{\left(
0\right)  }\right)  _{kl}^{jm}\right)  \right. \nonumber\\
&  \left.  -\frac{1}{2\left(  d-4\right)  }\left(  \delta_{k}^{i}B_{l}%
^{j}-\delta_{l}^{i}B_{k}^{j}-\delta_{k}^{j}B_{l}^{i}+\delta_{l}^{j}B_{k}%
^{i}\right)  \right)~,
\end{align}
or equivalently,%
\begin{equation}
\left(  W_{\left(  2\right)  }\right)  _{kl}^{ij}=\ell^{2}\left(
\nabla_{\lbrack k}C_{l]}^{ij}+S_{m}^{[j}\left(  W_{\left(  0\right)  }\right)
_{kl}^{i]m}-\frac{2}{\left(  d-4\right)  }\delta_{\lbrack k}^{[i}B_{l]}%
^{j]}\right)~.
\end{equation}

\section{The Kounterterm variational principle}\label{D}

The purpose of this appendix is to show that, for the Kounterterm-renormalized
Einstein-AdS action, an arbitrary variation is consistent with a Dirichlet
condition at the conformal boundary. This means that, at the conformal
boundary, the on-shell variation can be written as a total variation of
$g_{\left(  0\right)  }$ (i.e., as proportional to $\delta g_{\left(
0\right)  }$).

We consider the usual Einstein-Hilbert action, given by%
\begin{equation}
I_{EH}=\frac{1}{16\pi G}%
{\displaystyle\int\limits_{M}}
d^{D}x\sqrt{- G}\left(  R-2\Lambda\right)~,
\end{equation}
where%
\begin{equation}
\Lambda=-\frac{\left(  D-1\right)  \left(  D-2\right)  }{2\ell^{2}}~.
\end{equation}
Using the generalized Kronecker deltas, this action can be rewritten as%
\begin{equation}
I_{EH}=\frac{1}{16\pi G_{\mathrm{N}}}%
{\displaystyle\int\limits_{M}}
d^{D}x\sqrt{- G}\delta_{\left[  \mu_{1}\mu_{2}\right]  }^{\left[
\nu_{1}\nu_{2}\right]  }\left(  \frac{1}{2}R_{\nu_{1}\nu_{2}}^{\mu_{1}\mu_{2}%
}+\frac{\left(  D-2\right)  }{D\ell^{2}}\delta_{\nu_{1}}^{\mu_{1}}\delta
_{\nu_{2}}^{\mu_{2}}\right)~.
\end{equation}
Now, by performing an arbitrary variation, we have that%
\begin{align}
\delta\left[  \frac{1}{2}\delta_{\left[  \mu_{1}\mu_{2}\right]  }^{\left[
\nu_{1}\nu_{2}\right]  }\sqrt{- G} G^{\mu_{2}\sigma
}R_{\sigma\nu_{1}\nu_{2}}^{\mu_{1}}\right]   &  =\frac{1}{2}\delta_{\left[
\mu_{1}\mu_{2}\right]  }^{\left[  \nu_{1}\nu_{2}\right]  }\left(
 G^{\mu_{2}\sigma}R_{\sigma\nu_{1}\nu_{2}}^{\mu_{1}}\delta
\sqrt{- G}\right. \nonumber\\
&  \left.  +\sqrt{- G}R_{\sigma\nu_{1}\nu_{2}}^{\mu_{1}}%
\delta G^{\mu_{2}\sigma}+\sqrt{- G} G^{\mu
_{2}\sigma}\delta R_{\sigma\nu_{1}\nu_{2}}^{\mu_{1}}\right)  ,\nonumber\\
\delta\left[  \sqrt{- G}\frac{\left(  D-2\right)  }{D\ell^{2}}%
\delta_{\left[  \mu_{1}\mu_{2}\right]  }^{\left[  \nu_{1}\nu_{2}\right]
}\delta_{\nu_{1}}^{\mu_{1}}\delta_{\nu_{2}}^{\mu_{2}}\right]   &
=-2\Lambda\delta\sqrt{- G}~,
\end{align}
and considering that%
\begin{equation}
\delta\sqrt{- G}=\frac{1}{2}\sqrt{- G} G%
^{\mu\lambda}\delta G_{\mu\lambda}=-\frac{1}{2}\sqrt{- G%
} G_{\mu\lambda}\delta G^{\mu\lambda}~,
\end{equation}
we have that%
\begin{align}
\delta I_{EH}  &  =\frac{1}{16\pi G_{\mathrm{N}}}
{\displaystyle\int\limits_{M}}
d^{D}x\sqrt{- G}\left[  \frac{1}{2}\delta_{\left[  \mu_{1}\mu
_{2}\right]  }^{\left[  \nu_{1}\nu_{2}\right]  }\left(  -\frac{1}%
{2} G_{\mu\lambda} G^{\mu_{2}\sigma}R_{\sigma\nu_{1}\nu_{2}%
}^{\mu_{1}}\delta G^{\mu\lambda}+R_{\sigma\nu_{1}\nu_{2}}^{\mu_{1}%
}\delta G^{\mu_{2}\sigma}\right.  \right.
\nonumber\\
&  \left.  \left.  + G^{\mu_{2}\sigma}\delta R_{\sigma\nu_{1}\nu_{2}%
}^{\mu_{1}}\right)  +\Lambda G_{\mu\lambda}\delta G%
^{\mu\lambda}\right]
\nonumber\\
&  =\frac{1}{16\pi G_{\mathrm{N}}}%
{\displaystyle\int\limits_{M}}
d^{D}x\sqrt{- G}\left[  \underset{G_{\mu\nu}=0\text{ on-shell}%
}{\underbrace{\left(  R_{\mu\lambda}-\frac{1}{2}\left(  R-2\Lambda\right)
 G_{\mu\lambda}\right)  }}\delta G^{\mu\lambda}+\frac{1}%
{2}\delta_{\left[  \mu_{1}\mu_{2}\right]  }^{\left[  \nu_{1}\nu_{2}\right]
} G^{\mu_{2}\sigma}\delta R_{\sigma\nu_{1}\nu_{2}}^{\mu_{1}}\right]
~.
\end{align}

Now we simplify%
\begin{align}
\frac{1}{2}\delta_{\left[  \mu_{1}\mu_{2}\right]  }^{\left[  \nu_{1}\nu
_{2}\right]  } G^{\mu_{2}\sigma}\delta R_{\sigma\nu_{1}\nu_{2}}%
^{\mu_{1}}  &  =\frac{1}{2}\delta_{\left[  \mu_{1}\mu_{2}\right]  }^{\left[
\nu_{1}\nu_{2}\right]  } G^{\mu_{2}\sigma}\left(  \nabla_{\nu_{1}%
}\left(  \delta\Gamma_{\sigma\nu_{2}}^{\mu_{1}}\right)  -\nabla_{\nu_{2}%
}\left(  \delta\Gamma_{\sigma\nu_{1}}^{\mu_{1}}\right)  \right) \nonumber\\
&  = G^{\mu_{2}\sigma}\left(  \nabla_{\mu_{1}}\left(  \delta
\Gamma_{\sigma\mu_{2}}^{\mu_{1}}\right)  -\nabla_{\mu_{2}}\left(  \delta
\Gamma_{\sigma\mu_{1}}^{\mu_{1}}\right)  \right) \nonumber\\
&  =\nabla_{\mu_{1}}\left(   G^{\mu_{2}\sigma}\delta\Gamma_{\sigma
\mu_{2}}^{\mu_{1}}- G^{\mu_{1}\sigma}\delta\Gamma_{\sigma\mu_{2}%
}^{\mu_{2}}\right)  .
\end{align}
Thus,
\begin{equation}
\delta I_{EH}=\frac{1}{16\pi G_{\mathrm{N}}}%
{\displaystyle\int\limits_{M}}
d^{D}x\sqrt{- G}\nabla_{\mu_{1}}\left(   G^{\mu_{2}\sigma
}\delta\Gamma_{\sigma\mu_{2}}^{\mu_{1}}- G^{\mu_{1}\sigma}%
\delta\Gamma_{\sigma\mu_{2}}^{\mu_{2}}\right)~,
\end{equation}
and using Gauss' theorem we have%
\begin{equation}
{\displaystyle\int\limits_{M}}
d^{D}x\sqrt{- G}\nabla_{\mu}\left[  \cdot\right]  \rightarrow%
{\displaystyle\int\limits_{\partial M}}
d^{D-1}x\sqrt{-h}n_{\mu}\left[  \cdot\right]~.
\end{equation}
In our case, $n_{\mu}=\left(  -N,0,\cdots,0\right)  $ is the covariant normal
vector of the sheets along the usual radial foliation (Gauss-Normal foliation
along the holographic $\rho$ coordinate of the FG gauge), where the vector
points in the outward radial direction as usual and the minus sign appears
because $\rho$ increases towards the bulk. Then, $N$ is the (radial) lapse
function. Also, in terms of the extrinsic curvature of our foliation, we have that%
\begin{align}
\Gamma_{ij}^{\rho}  &  =\frac{1}{N}K_{ij},\nonumber\\
\Gamma_{\rho j}^{i}  &  =-NK_{j}^{i},\nonumber\\
 G^{\rho\rho}  &  =\frac{1}{N^{2}}.
\end{align}
Then, we can write
\begin{align}
\delta I_{EH}  &  =\frac{1}{16\pi G_{\mathrm{N}}}%
{\displaystyle\int\limits_{\partial M}}
d^{D-1}x\sqrt{-h}n_{\rho}\left[  h^{ij}\delta\Gamma_{ij}^{\rho}- G%
^{\rho\rho}\delta\Gamma_{\rho i}^{i}\right] \nonumber\\
&  =-\frac{1}{16\pi G_{\mathrm{N}}}%
{\displaystyle\int\limits_{\partial M}}
d^{D-1}x\sqrt{-h}\left[  h^{ij}\delta K_{ij}+\delta K_{i}^{i}\right]~.
\end{align}

Finally, considering that%
\begin{equation}
h^{ij}\delta K_{ij}=h^{ij}\delta\left(  h_{il}K_{j}^{l}\right)  =h^{ij}\delta
h_{il}K_{j}^{l}+\delta K_{i}^{i}~,
\end{equation}
we obtain
\begin{equation}
\delta I_{EH}=-\frac{1}{16\pi G_{\mathrm{N}}}%
{\displaystyle\int\limits_{\partial M}}
d^{D-1}x\sqrt{-h}\left[  \left(  h^{-1}\delta h\right)  _{j}^{i}K_{i}%
^{j}+2\delta K_{i}^{i}\right]
\equiv
{\displaystyle\int\limits_{\partial M}}
d^{D-1}x\left(  -\Theta_{EH}\right)~.
\end{equation}
The EH boundary term, which will be used in what follows, is then defined as%
\begin{equation}
\Theta_{EH}=\frac{1}{16\pi G_{\mathrm{N}}}\sqrt{-h}\left[  \left(  h^{-1}\delta h\right)
_{j}^{i}K_{i}^{j}+2\delta K_{i}^{i}\right]~.
\end{equation}
As stated in (\ref{deltaeh}), this is the standard form of the variation
for the on-shell bulk EH action, and usually it requires the addition of the
YGH term in order to implement a Dirichlet variational principle, thus
cancelling the $\delta K$ part.

We now proceed to consider the variation of the Kounterterm-renormalized
action, in order to check its finiteness and compatibility with a Dirichlet
condition at the conformal boundary (for $g_{\left(  0\right)  }$). We have that%
\begin{equation}
\delta I_{EH}^{ren}=\delta I_{EH}+\delta I_{B_{D-1}}~,
\end{equation}
where the form of $I_{B_{D-1}}$ depends on whether the manifold is even or odd
dimensional. In what follows, we consider both cases separately.

\subsection*{Variational principle for even bulk dimension (odd CFT
dimension)}

We will focus first on the case when $D=2n$, where
\begin{equation}
I_{EH}^{ren}=I_{EH}+c_{2n-1}%
{\displaystyle\int\limits_{\partial M}}
d^{2n-1}xB_{2n-1}~,
\end{equation}
with
\begin{equation}
c_{2n-1}=\frac{\left(  -1\right)  ^{n}\ell^{2n-2}}{n\left(  2n-2\right)  !}~.
\end{equation}
Then, the variation of the renormalized action is given by%
\begin{equation}
\delta I_{EH}^{ren}=
{\displaystyle\int\limits_{\partial M}}
d^{2n-1}x\left(  -\Theta_{EH}\right)  +\frac{c_{2n-1}}{16\pi G}%
{\displaystyle\int\limits_{\partial M}}
d^{2n-1}x\delta B_{2n-1}~,
\end{equation}
where%
\begin{equation}
B_{2n-1}=-2n\sqrt{-h}%
{\displaystyle\int\limits_{0}^{1}}
dt\delta_{\left[  2n-1\right]  }^{\left[  2n-1\right]  }K\left(  \frac{1}%
{2}\mathcal{R}-t^{2}KK\right)  ^{n-1}~.
\end{equation}
One can expand the parentheses and write $B_{2n-1}$ as%
\begin{align}
B_{2n-1}  &  =\delta_{\left[  2n-1\right]  }^{\left[  2n-1\right]  }%
{\displaystyle\sum\limits_{i=0}^{n-1}}
\frac{\left(  -1\right)  ^{n-i}n!}{2^{i-1}i!\left(  n-i-1\right)  !\left(
2n-2i-1\right)  }A_{i}~,
\nonumber\\
A_{i}  &  =\sqrt{-h}K^{2n-2i-1}\mathcal{R}^{i}~.
\end{align}

Now, we consider the general variation of the different terms. In particular,
we have that
\begin{align}
\delta\sqrt{-h}  &  =\sqrt{-h}\frac{1}{2}\left(  h^{-1}\delta h\right)
_{s}^{s}~,
\nonumber\\
\delta\mathcal{R}_{j_{1}j_{2}}^{i_{1}i_{2}}  &  =\delta h^{i_{2}s}%
\mathcal{R}_{sj_{1}j_{2}}^{i_{1}}+h^{i_{2}s}\delta\mathcal{R}_{sj_{1}j_{2}%
}^{i_{1}}
\nonumber\\
&  =-\left(  h^{-1}\delta h\right)  _{s}^{i_{2}}\mathcal{R}_{j_{1}j_{2}%
}^{i_{1}s}+2h^{i_{2}s}D_{j_{1}}\delta\Gamma_{sj_{2}}^{i_{1}}\nonumber\\
&  =-\left(  h^{-1}\delta h\right)  _{s}^{i_{2}}\mathcal{R}_{j_{1}j_{2}%
}^{i_{1}s}+2h^{i_{2}s}D_{j_{1}}h^{i_{1}m}D_{s}\delta h_{mj_{2}}\nonumber\\
&  =-\left(  h^{-1}\delta h\right)  _{s}^{i_{2}}\mathcal{R}_{j_{1}j_{2}%
}^{i_{1}s}+2D^{i_{2}}D_{j_{1}}\left(  h^{-1}\delta h\right)  _{j_{2}}^{i_{1}}~,
\label{eq.Rie}%
\end{align}
where in the simplification of $\delta\mathcal{R}$ we have used the symmetry
properties of the different terms, and the fact that there is an overall
$\delta_{\left[  2n-1\right]  }^{\left[  2n-1\right]  }$ in front. Then, we
have that%
\begin{align}
\delta A_{i}  &  =\sqrt{-h}\left(  \frac{1}{2}\left(  h^{-1}\delta h\right)
_{s}^{s}K^{2n-2i-1}\mathcal{R}^{i}+\left(  2n-2i-1\right)  K^{2n-2i-2}%
\mathcal{R}^{i}\delta K\right. \nonumber\\
&  \left.  -iK^{2n-2i-1}\mathcal{R}^{i-1}\left(  h^{-1}\delta h\right)
_{s}^{i_{2}}\mathcal{R}_{j_{1}j_{2}}^{i_{1}s}+2iK^{2n-2i-1}\mathcal{R}%
^{i-1}D^{i_{2}}D_{j_{1}}\left(  h^{-1}\delta h\right)  _{j_{2}}^{i_{1}%
}\right)~.
\end{align}

We now define
\begin{equation}
\delta A_{DD,i}=2i\sqrt{-h}K^{2n-2i-1}\mathcal{R}^{i-1}D^{i_{2}}D_{j_{1}%
}\left(  h^{-1}\delta h\right)  _{j_{2}}^{i_{1}}%
\end{equation}
and%
\begin{equation}
\delta I_{DD}=\delta_{\left[  2n-1\right]  }^{\left[  2n-1\right]  }%
{\displaystyle\sum\limits_{i=0}^{n-1}}
\frac{\left(  -1\right)  ^{n-i}n!}{2^{i-1}i!\left(  n-i-1\right)  !\left(
2n-2i-1\right)  }\delta A_{DD,i}~.
\label{deltaIDD}%
\end{equation}
Integrating by parts, using the Bianchi identity for $\mathcal{R}ie$ and
considering that as the boundary manifold itself has no boundaries, total
derivative contributions vanish, it can be shown that $\delta I_{DD}=0$ and
therefore, the double derivative terms do not contribute to the arbitrary
variation. Then, we have%
\begin{align}
\delta A_{i}  &  =\sqrt{-h}\left(  \frac{1}{2}\left(  \left(  h^{-1}\delta
h\right)  _{s}^{s}K^{2n-2i-1}\mathcal{R}^{i}-2i\left(  h^{-1}\delta h\right)
_{s}^{i_{2}}K^{2n-2i-1}\mathcal{R}_{j_{1}j_{2}}^{i_{1}s}\mathcal{R}%
^{i-1}\right)  \right. \nonumber\\
&  \left.  +\left(  2n-2i-1\right)  K^{2n-2i-2}\mathcal{R}^{i}\delta K\right)
\end{align}
and%
\begin{equation}
\delta B_{2n-1}=\delta_{\left[  2n-1\right]  }^{\left[  2n-1\right]  }%
{\displaystyle\sum\limits_{i=0}^{n-1}}
\frac{\left(  -1\right)  ^{n-i}n!}{2^{i-1}i!\left(  n-i-1\right)  !\left(
2n-2i-1\right)  }\delta A_{i}~.
\end{equation}

In order to further simplify our expression for $\delta B_{2n-1}$, we consider
the following antisymmetric contraction:%
\begin{align}
\delta_{\left[  2n\right]  }^{\left[  2n\right]  }\left(  h^{-1}\delta
h\right)  K^{2n-2i-1}\mathcal{R}^{i}  &  =\delta_{\left[  2n-1\right]
}^{\left[  2n-1\right]  }\left(  \left(  h^{-1}\delta h\right)  _{s}%
^{s}K^{2n-2i-1}\mathcal{R}^{i}-2i\left(  h^{-1}\delta h\right)  _{s}^{i_{2}%
}K^{2n-2i-1}\mathcal{R}_{j_{1}j_{2}}^{i_{1}s}\mathcal{R}^{i-1}\right.
\nonumber\\
&  \left.  -\left(  2n-2i-1\right)  \left(  h^{-1}\delta h\right)  _{s}%
^{i_{1}}K_{j_{1}}^{s}K^{2n-2i-2}\mathcal{R}^{i}\right)~.
\end{align}
Now in turn, we also have that%
\begin{equation}
\delta_{\left[  2n\right]  }^{\left[  2n\right]  }\left(  h^{-1}\delta
h\right)  K^{2n-2i-1}\mathcal{R}^{i}=0
\end{equation}
because the manifold is $\left(  2n-1\right)  -$dimensional, and therefore,
the indices of the generalized Kronecker delta are necessarily over-saturated,
such that some indices will be repeated. Thus, considering an overall
$\delta_{\left[  2n-1\right]  }^{\left[  2n-1\right]  }$ in front, we have that%
\begin{equation}
\left(  h^{-1}\delta h\right)  _{s}^{s}K^{2n-2i-1}\mathcal{R}^{i}-2i\left(
h^{-1}\delta h\right)  _{s}^{i_{2}}K^{2n-2i-1}\mathcal{R}_{j_{1}j_{2}}%
^{i_{1}s}\mathcal{R}^{i-1}=\left(  2n-2i-1\right)  \left(  h^{-1}\delta
h\right)  _{s}^{i_{1}}K_{j_{1}}^{s}K^{2n-2i-2}\mathcal{R}^{i}~,
\end{equation}
and therefore,%
\begin{equation}
\delta A_{i}=\frac{\sqrt{-h}}{2}\left(  2n-2i-1\right)  K^{2n-2i-2}%
\mathcal{R}^{i}\left(  \left(  h^{-1}\delta h\right)  _{s}^{i_{1}}K_{j_{1}%
}^{s}+2\delta K\right)~.
\end{equation}
Finally, using the Gauss-Codazzi relation for $\mathcal{R}ie$ and the on-shell
Einstein condition for the bulk Weyl tensor $W$, we have that%
\begin{align}
\mathcal{R}  &  =\left(  W+2K^{2}-\frac{2}{\ell^{2}}\delta^{2}\right)
,\nonumber\\
\mathcal{R}^{i}  &  =%
{\displaystyle\sum\limits_{j=0}^{i}}
\frac{i!}{j!\left(  i-j\right)  !}W^{j}2^{i-j}%
{\displaystyle\sum\limits_{m=0}^{i-j}}
\frac{\left(  i-j\right)  !\left(  -1\right)  ^{i-j-m}}{m!\left(
i-j-m\right)  !\ell^{2\left(  i-j-m\right)  }}K^{2m}\delta^{2\left(
i-j-m\right)  }~.
\end{align}

Thus, defining%
\begin{equation}
J_{j_{1}}^{i_{1}}=\left(  \left(  h^{-1}\delta h\right)  _{s}^{i_{1}}K_{j_{1}%
}^{s}+2\delta K_{j_{1}}^{i_{1}}\right)~,
\end{equation}
we have%
\begin{align}
\delta B_{2n-1}  &  =\sqrt{-h}%
{\displaystyle\sum\limits_{i=0}^{n-1}}
{\displaystyle\sum\limits_{j=0}^{i}}
{\displaystyle\sum\limits_{m=0}^{i-j}}
\frac{\left(  -1\right)  ^{n-j-m}n!}{2^{j}j!m!\left(  n-i-1\right)  !\left(
i-j-m\right)  !\ell^{2\left(  i-j-m\right)  }}\times\nonumber\\
&  \delta_{\left[  2n-1\right]  }^{\left[  2n-1\right]  }\delta^{2\left(
i-j-m\right)  }W^{j}K^{2n-2i+2m-2}J~,
\end{align}
and using that%
\begin{equation}
\delta_{\left[  2n-1\right]  }^{\left[  2n-1\right]  }\delta^{2\left(
i-j-m\right)  }=\left(  2i-2j-2m\right)  !\delta_{\left[
2n-2i+2j+2m-1\right]  }^{\left[  2n-2i+2j+2m-1\right]  }~,
\end{equation}
we obtain%
\begin{align}
\delta B_{2n-1}  &  =\sqrt{-h}%
{\displaystyle\sum\limits_{i=0}^{n-1}}
{\displaystyle\sum\limits_{j=0}^{i}}
{\displaystyle\sum\limits_{m=0}^{i-j}}
\frac{\left(  -1\right)  ^{n-j-m}n!\left(  2i-2j-2m\right)  !}{2^{j}%
j!m!\left(  n-i-1\right)  !\left(  i-j-m\right)  !\ell^{2\left(  i-j-m\right)
}}\times\nonumber\\
&  \delta_{\left[  2n-2i+2j+2m-1\right]  }^{\left[  2n-2i+2j+2m-1\right]
}W^{j}K^{2n-2i+2m-2}J~. \label{triple_sum_even}%
\end{align}

For illustrative purposes, we now consider the $D=4$ case, and proceed to
simplify our expressions for $\delta B_{2n-1}$ and $\delta I_{EH}^{ren}$.

\subsubsection*{$D=4$ ($n=2$) case}

We consider the 4D bulk case and directly reproduce the known result for the
arbitrary variation. When $n=2$, the triple sum becomes%
\begin{align}
\delta B_{3}  &  =\sqrt{-h}%
{\displaystyle\sum\limits_{i=0}^{1}}
{\displaystyle\sum\limits_{j=0}^{i}}
{\displaystyle\sum\limits_{m=0}^{i-j}}
\frac{\left(  -1\right)  ^{2-j-m}2\left(  2i-2j-2m\right)  !}{2^{j}j!m!\left(
2-i-1\right)  !\left(  i-j-m\right)  !\ell^{2\left(  i-j-m\right)  }}%
\times\nonumber\\
&  \delta_{\left[  4-2i+2j+2m-1\right]  }^{\left[  4-2i+2j+2m-1\right]  }%
W^{j}K^{4-2i+2m-2}J~, \label{sum_B3}%
\end{align}
where the terms that contribute to the sum are $\left\{  \left(  i,j,m\right)
\right\}  =\left\{  \left(  0,0,0\right)  ,\left(  1,0,0\right)  ,\left(
1,0,1\right)  ,\left(  1,1,0\right)  \right\}  $.
Therefore, in that order, we have%
\begin{align}
\delta B_{3}  &  =\sqrt{-h}\left(  2\delta_{\left[  3\right]  }^{\left[
3\right]  }K^{2}J+\frac{4}{\ell^{2}}\delta_{\left[  1\right]  }^{\left[
1\right]  }J-2\delta_{\left[  3\right]  }^{\left[  3\right]  }K^{2}%
J-\delta_{\left[  3\right]  }^{\left[  3\right]  }WJ\right) \nonumber\\
&  =\sqrt{-h}\left(  \frac{4}{\ell^{2}}\delta_{\left[  1\right]  }^{\left[
1\right]  }J-\delta_{\left[  3\right]  }^{\left[  3\right]  }WJ\right)~.
\end{align}
Thus, we find that%
\begin{align}
\delta I_{EH}^{ren}  &  =%
{\displaystyle\int\limits_{\partial M}}
d^{3}x\frac{\sqrt{-h}}{16\pi G_{\mathrm{N}}}\left(  -\left[  \left(  h^{-1}\delta h\right)
_{j}^{i}K_{i}^{j}+2\delta K_{i}^{i}\right]  +\frac{\ell^{2}}{4}\left(
\frac{4}{\ell^{2}}\delta_{\left[  1\right]  }^{\left[  1\right]  }%
J-\delta_{\left[  3\right]  }^{\left[  3\right]  }WJ\right)  \right)
\nonumber\\
&  =\frac{\ell^{2}}{64\pi G_{\mathrm{N}}}%
{\displaystyle\int\limits_{\partial M}}
d^{3}x\sqrt{-h}\delta_{\left[  3\right]  }^{\left[  3\right]  }\left(
-W\right)  J~,
\end{align}
and therefore%
\begin{equation}
\delta I_{EH}^{ren}=\frac{\ell^{2}}{64\pi G_{\mathrm{N}}}%
{\displaystyle\int\limits_{\partial M}}
d^{3}x\sqrt{-h}\delta_{\left[  i_{1}i_{2}i_{3}\right]  }^{\left[  j_{1}%
j_{2}j_{3}\right]  }\left(  -W_{j_{1}j_{2}}^{i_{1}i_{2}}\right)  \left(
\left(  h^{-1}\delta h\right)  _{s}^{i_{3}}K_{j_{3}}^{s}+2\delta K_{j_{3}%
}^{i_{3}}\right)~.
\end{equation}
This last expression can be simplified further considering that, as explained
in eq.(\ref{tr_dM_W}), $W_{ij}^{ij}=0$, and that $W_{ij}^{ij}=-W_{ji}^{ij}$.
We therefore find%
\begin{equation}
\delta I_{EH}^{ren}=\frac{\ell^{2}}{64\pi G_{\mathrm{N}}}%
{\displaystyle\int\limits_{\partial M}}
d^{3}x\sqrt{-h}\left(  4W_{jl}^{il}\right)  \left(  \left(  h^{-1}\delta
h\right)  _{s}^{j}K_{i}^{s}+2\delta K_{i}^{j}\right)~.
\end{equation}

Now, we consider that $W_{jl}^{il}=-W_{j\rho}^{i\rho}=-\epsilon_{j}^{i}$,
where $\epsilon_{j}^{i}$ is the electric part of the bulk Weyl tensor as
defined in eq.(\ref{electric_W}). Also, by the ACF condition (Appendix \ref{C}), we have that
$\epsilon_{j}^{i}\sim O\left(  \rho^{\frac{3}{2}}\right)  $ already has the
order of the normalizable mode, so that any term of higher order in $\rho$
vanishes at the conformal boundary. Finally, we consider that $K_{j}^{i}%
=\frac{1}{\ell}\delta_{j}^{i}+\rho\ell\left(  S_{\left(  0\right)  }\right)
_{j}^{i}+\ldots$ and $\delta K_{j}^{i}\sim O\left(  \rho\right)  $, such that
to the lowest order of $O\left(  1\right)  $, $K_{j}^{i}\sim\frac{1}{\ell
}\delta_{j}^{i}$ and $\delta K_{j}^{i}\sim0$. Thus we have that, at the
conformal boundary,%
\begin{equation}
\delta I_{EH}^{ren}=\frac{\ell^{2}}{16\pi G_{\mathrm{N}}}%
{\displaystyle\int\limits_{\partial M}}
d^{3}x\sqrt{-h}\left(  -\epsilon_{j}^{i}\right)  \left(  h^{-1}\delta
h\right)  _{i}^{j}~.
\end{equation}

This result is in agreement with the expression for the arbitrary variation
presented in \cite{Miskovic:2008ck,Miskovic:2010ui}. It is also in agreement
with the definition of conformal mass given by Ashtekar and Das in \cite{Ashtekar:1999jx}, such that the Noether prepotential for computing the
asymptotic charges of the geometry is proportional to the electric part of the
Weyl tensor. Considering that, asymptotically, $\left(  h^{-1}\delta h\right)
_{i}^{j}=\left(  g_{\left(  0\right)  }^{-1}\delta g_{\left(  0\right)
}\right)  _{i}^{j}$, it is clear that in the $D=4$ case the arbitrary
variation of $I_{EH}^{ren}$ is both finite and compatible with a Dirichlet
condition at the conformal boundary. We also note that in the sum of
eq.(\ref{sum_B3}), the $\left(  i,j,m\right)  =\left(  0,0,0\right)  $ and
$\left(  1,0,1\right)  $ terms, which have non-zero powers of $K$ (in
particular $K^{2}$), cancel each other exactly. We will see that this feature
is general, such that the arbitrary variation for even dimensional bulks can
be written as a certain polynomial of contractions of the Weyl tensor with
boundary indices.

\subsubsection*{Simplification of the general case}

We now proceed to simplify the general even-dimensional case in order to
reproduce the known results for the arbitrary variation. It can be explicitly
verified (by brute force using an algebra software) that, in the triple
summation of (\ref{triple_sum_even}), all non-zero powers of $K$ vanish
identically. Considering this fact, we can impose the following condition on
the summation indices:%
\begin{equation}
2n-2i+2m-2=0~.
\end{equation}
This implies that $i-m=n-1$, which together with the fact that $i,m\geq0$ and
$i\leq n-1$ fixes the values of $i=n-1$ and $m=0$. Thus, the triple summation
becomes a single sum and we have
\begin{equation}
\delta B_{2n-1}=\sqrt{-h}%
{\displaystyle\sum\limits_{j=0}^{n-1}}
\frac{\left(  -1\right)  ^{n-j}n!\left(  2n-2j-2\right)  !}{2^{j}j!\left(
n-j-1\right)  !\ell^{2\left(  n-j-1\right)  }}\delta_{\left[  2j+1\right]
}^{\left[  2j+1\right]  }W^{j}J~.
\end{equation}
Now, isolating the $j=0$ term, we have that%
\begin{equation}
\delta B_{2n-1}=\sqrt{-h}\left(  \frac{1}{c_{2n-1}}\delta_{\left[  1\right]
}^{\left[  1\right]  }J+%
{\displaystyle\sum\limits_{j=1}^{n-1}}
\frac{\left(  -1\right)  ^{n-j}n!\left(  2n-2j-2\right)  !}{2^{j}j!\left(
n-j-1\right)  !\ell^{2\left(  n-j-1\right)  }}\delta_{\left[  2j+1\right]
}^{\left[  2j+1\right]  }W^{j}J\right)~,
\end{equation}
where $c_{2n-1}=\frac{\left(  -1\right)  ^{n}\ell^{2\left(  n-1\right)  }%
}{n\left(  2n-2\right)  !}$. Therefore, the $j=0$ term exactly cancels the
$\left(  -\Theta_{EH}\right)  $ term. We thus have that%
\begin{align}
\delta I_{EH}^{ren}  &  =%
{\displaystyle\int\limits_{\partial M}}
d^{2n-1}x\left(  -\Theta_{EH}\right)  +\frac{c_{2n-1}}{16\pi G}%
{\displaystyle\int\limits_{\partial M}}
d^{2n-1}x\delta B_{2n-1}\nonumber\\
&  =\frac{1}{16\pi G}%
{\displaystyle\int\limits_{\partial M}}
d^{2n-1}x%
{\displaystyle\sum\limits_{j=1}^{n-1}}
\frac{\left(  -1\right)  ^{j}\left(  n-1\right)  !\left(  2n-2j-2\right)
!\ell^{2j}}{2^{j}j!\left(  n-j-1\right)  !\left(  2n-2\right)  !}%
\delta_{\left[  2j+1\right]  }^{\left[  2j+1\right]  }W^{j}J~.
\label{DeltaIrenEven}%
\end{align}

Then, considering the ACF condition, which states that $W_{kl}^{ij}\sim
O\left(  \rho^{\frac{d}{2}}\right)  $ where $d=2n-1$, we have that only the
$j=1$ term has a non-vanishing and finite contribution at the conformal
boundary. Thus we obtain%
\begin{equation}
\delta I_{EH}^{ren}=\frac{1}{16\pi G_{\mathrm{N}}}%
{\displaystyle\int\limits_{\partial M}}
d^{2n-1}x\frac{\ell^{2}}{4\left(  2n-3\right)  }\delta_{\left[  3\right]
}^{\left[  3\right]  }\left(  -W\right)  J~.
\end{equation}

Finally, as we did in the previous section, we consider that%
\begin{align}
W_{ij}^{ij}  &  =0~,~W_{kl}^{ij}=-W_{lk}^{ij},\nonumber\\
W_{jl}^{il}  &  =-W_{j\rho}^{i\rho}=-\epsilon_{j}^{i},\nonumber\\
K_{j}^{i}  &  =\frac{1}{\ell}\delta_{j}^{i}+\rho\ell\left(  S_{\left(
0\right)  }\right)  _{j}^{i}+\ldots,\nonumber\\
\delta K_{j}^{i}  &  \sim O\left(  \rho\right)  ,
\end{align}
and we obtain
\begin{equation}
\delta I_{EH}^{ren}=\frac{\ell}{16\pi G\left(  2n-3\right)  }%
{\displaystyle\int\limits_{\partial M}}
d^{2n-1}x\left(  -\epsilon_{j}^{i}\right)  \left(  h^{-1}\delta h\right)
_{i}^{j}~,
\end{equation}
where $\epsilon_{j}^{i}=W_{j\rho}^{i\rho}$ is the electric part of the bulk
Weyl tensor. As in the $D=4$ case, this result is in agreement with the
expressions for the general variation given in
\cite{Miskovic:2008ck,Miskovic:2010ui} and with the definition of conformal
mass of \cite{Ashtekar:1999jx}. Because, asymptotically, $\left(  h^{-1}\delta
h\right)  _{i}^{j}=\left(  g_{\left(  0\right)  }^{-1}\delta g_{\left(
0\right)  }\right)  _{i}^{j}$, the finiteness and compatibility of
$I_{EH}^{ren}$ with the Dirichlet condition at the conformal boundary are
explicitly verified for the $D=2n$ case.

\subsection*{Variational principle for odd bulk dimension (even CFT
dimension)}

We now consider the case of odd bulk spacetime dimension, $D=2n+1$, where we have that
\begin{equation}
I_{EH}^{ren}=I_{EH}+\frac{c_{2n}}{16\pi G_{\mathrm{N}}}%
{\displaystyle\int\limits_{\partial M}}
d^{2n}xB_{2n}~,
\end{equation}
where%
\begin{equation}
c_{2n}=\frac{\left(  -1\right)  ^{n}\ell^{2n-2}}{2^{2n-2}n\left[  \left(
n-1\right)  !\right]  ^{2}}%
\end{equation}
and%
\begin{equation}
B_{2n}=-2n\sqrt{-h}
{\displaystyle\int\limits_{0}^{1}}
ds
{\displaystyle\int\limits_{0}^{s}}
dt\delta_{\left[  2n\right]  }^{\left[  2n\right]  }\delta K\left(  \frac
{1}{2}\mathcal{R}-s^{2}KK+\frac{t^{2}}{\ell^{2}}\delta\delta\right)  ^{n-1}~.
\end{equation}
Then, the arbitrary variation of $I_{EH}^{ren}$ is given by
\begin{align}
\delta I_{EH}^{ren}  &  =
{\displaystyle\int\limits_{\partial M}}
d^{2n}x\left(  -\Theta_{EH}\right)  +\frac{c_{2n}}{16\pi G_{\mathrm{N}}}%
{\displaystyle\int\limits_{\partial M}}
d^{2n}x\delta B_{2n}~,\\
\Theta_{EH}  &  =\frac{1}{16\pi G_{\mathrm{N}}}\sqrt{-h}\left[  \left(  h^{-1}\delta
h\right)  _{i}^{s}K_{s}^{i}+2\delta K_{i}^{i}\right]~.
\end{align}

In order to find $\delta B_{2n}$
we first expand $B_{2n}$. In particular, we have
\begin{align}
B_{2n}  &  =-2n\sqrt{-h}\delta_{\left[  2n\right]  }^{\left[  2n\right]
}\delta K
{\displaystyle\int\limits_{0}^{1}}
ds
{\displaystyle\int\limits_{0}^{s}}
dt
{\displaystyle\sum\limits_{i=0}^{n-1}}
\frac{\left(  n-1\right)  !}{i!\left(  n-i-1\right)  !2^{i}}\mathcal{R}%
^{i}\times
\nonumber\\
&
{\displaystyle\sum\limits_{k=0}^{n-i-1}}
\frac{\left(  n-i-1\right)  !\left(  -1\right)  ^{k}s^{2k}}{k!\left(
n-i-k-1\right)  !}K^{2k}\frac{t^{2\left(  n-i-k-1\right)  }}{\ell^{2\left(
n-i-k-1\right)  }}\delta^{2\left(  n-i-k-1\right)  }
\nonumber\\
&  =\sqrt{-h}%
{\displaystyle\sum\limits_{i=0}^{n-1}}
{\displaystyle\sum\limits_{k=0}^{n-i-1}}
\frac{n!\left(  -1\right)  ^{k+1}}{i!k!2^{i-1}\ell^{2\left(  n-i-k-1\right)
}\left(  n-i-k-1\right)  !}\times
\nonumber\\
&  \left[
{\displaystyle\int\limits_{0}^{1}}
ds\,s^{2k}%
{\displaystyle\int\limits_{0}^{s}}
dt\,t^{2\left(  n-i-k-1\right)  }\right]  \mathcal{R}^{i}K^{2k+1}\delta_{\left[
2n\right]  }^{\left[  2n\right]  }\delta^{2n-2i-2k-1}\nonumber\\
&  =%
{\displaystyle\sum\limits_{i=0}^{n-1}}
{\displaystyle\sum\limits_{k=0}^{n-i-1}}
\frac{n!\left(  -1\right)  ^{k+1}}{i!k!2^{i}\left(  n-i-k-1\right)  !\left(
n-i\right)  \left(  2n-2i-2k-1\right)  \ell^{2\left(  n-i-k-1\right)  }}%
\delta_{\left[  2n\right]  }^{\left[  2n\right]  }A_{ik}~,
\end{align}
where%
\begin{equation}
A_{ik}\equiv \sqrt{-h}\delta^{2n-2i-2k-1}\mathcal{R}^{i}K^{2k+1}~.
\end{equation}

Now, we proceed with the variation of the different terms in $A_{ik}$.
We
consider that
\begin{align}
\delta\sqrt{-h}  &  =\sqrt{-h}\frac{1}{2}\left(  h^{-1}\delta h\right)
_{s}^{s},\nonumber\\
\delta\mathcal{R}_{j_{1}j_{2}}^{i_{1}i_{2}}  &  =-\left(  h^{-1}\delta
h\right)  _{s}^{i_{2}}\mathcal{R}_{j_{1}j_{2}}^{i_{1}s}+2D^{i_{2}}D_{j_{1}%
}\left(  h^{-1}\delta h\right)  _{j_{2}}^{i_{1}}~,
\end{align}
where we have used some symmetry properties in the simplification of
$\delta\mathcal{R}ie$, just like in the case of Eq.~(\ref{eq.Rie}). Then, the
variation $\delta A_{ik}$ is given by%
\begin{align}
\delta A_{ik}  &  =\sqrt{-h}\delta^{2n-2i-2k-1}\left(  \frac{1}{2}\left(
h^{-1}\delta h\right)  _{s}^{s}\mathcal{R}^{i}K^{2k+1}+\left(  2k+1\right)
\mathcal{R}^{i}K^{2k}\delta K\right. \nonumber\\
&  \left.  +i\mathcal{R}^{i-1}K^{2k+1}\left(  -\left(  h^{-1}\delta h\right)
_{s}^{i_{2}}\mathcal{R}_{j_{1}j_{2}}^{i_{1}s}+2D^{i_{2}}D_{j_{1}}\left(
h^{-1}\delta h\right)  _{j_{2}}^{i_{1}}\right)  \right) \nonumber\\
&  =\sqrt{-h}\delta^{2n-2i-2k-1}\left(  \frac{1}{2}\left(  \left(
h^{-1}\delta h\right)  _{s}^{s}\mathcal{R}^{i}K^{2k+1}-2i\left(  h^{-1}\delta
h\right)  _{s}^{i_{2}}\mathcal{R}_{j_{1}j_{2}}^{i_{1}s}\mathcal{R}%
^{i-1}K^{2k+1}\right)  \right. \nonumber\\
&  \left.  +\left(  2k+1\right)  \mathcal{R}^{i}K^{2k}\delta K+2i\mathcal{R}%
^{i-1}K^{2k+1}D^{i_{2}}D_{j_{1}}\left(  h^{-1}\delta h\right)  _{j_{2}}%
^{i_{1}}\right)~.
\end{align}

In order to simplify $\delta A_{ik}$, we first consider the double derivative
term. We define%
\begin{equation}
\delta A_{DD,ik}=2i\sqrt{-h}\delta^{2n-2i-2k-1}\mathcal{R}^{i-1}%
K^{2k+1}D^{i_{2}}D_{j_{1}}\left(  h^{-1}\delta h\right)  _{j_{2}}^{i_{1}}%
\end{equation}
and%
\begin{align}
\delta I_{DD}  &  =\frac{c_{2n}}{16\pi G_{\mathrm{N}}}%
{\displaystyle\int\limits_{\partial M}}
d^{2n}x%
{\displaystyle\sum\limits_{i=0}^{n-1}}
{\displaystyle\sum\limits_{k=0}^{n-i-1}}
\frac{n!\left(  -1\right)  ^{k+1}}{i!k!2^{i}\left(  n-i-k-1\right)  !\left(
n-i\right)  \left(  2n-2i-2k-1\right)  \ell^{2\left(  n-i-k-1\right)  }}%
\times\nonumber\\
&  \delta_{\left[  2n\right]  }^{\left[  2n\right]  }\delta A_{DD,ik}~.
\end{align}
By analogy with (\ref{deltaIDD}), considering the Bianchi identity for
$\mathcal{R}ie$, integration by parts and the vanishing of total derivatives
(as the conformal boundary itself has no boundaries), it can be shown that
$\delta I_{DD}=0$. Thus, this term does not contribute to the variation of the
renormalized action and therefore $\delta A_{DD,ik}$ can be excluded from
$\delta A_{ik}$.

Now, we consider that%
\begin{align}
\delta_{\left[  2n+1\right]  }^{\left[  2n+1\right]  }\delta^{2n-2i-2k-1}%
\left(  h^{-1}\delta h\right)  K^{2k+1}\mathcal{R}^{i}  &  =\delta_{\left[
2n\right]  }^{\left[  2n\right]  }\left(  -\left(  2n-2i-2k-1\right)
\delta^{2n-2i-2k-2}\delta_{j_{1}}^{s}\left(  h^{-1}\delta h\right)
_{s}^{i_{1}}K^{2k+1}\mathcal{R}^{i}\right. \nonumber\\
&  -\left(  2k+1\right)  \delta^{2n-2i-2k-1}\left(  h^{-1}\delta h\right)
_{s}^{i_{1}}K_{j_{1}}^{s}K^{2k}\mathcal{R}^{i}\nonumber\\
&  +\delta^{2n-2i-2k-1}\left(  h^{-1}\delta h\right)  _{s}^{s}\mathcal{R}%
^{i}K^{2k+1}\nonumber\\
&  \left.  -\delta^{2n-2i-2k-1}2i\left(  h^{-1}\delta h\right)  _{s}^{i_{2}%
}\mathcal{R}_{j_{1}j_{2}}^{i_{1}s}\mathcal{R}^{i-1}K^{2k+1}\right)~,
\end{align}
and also%
\begin{equation}
\delta_{\left[  2n+1\right]  }^{\left[  2n+1\right]  }\delta^{2n-2i-2k-1}%
\left(  h^{-1}\delta h\right)  K^{2k+1}\mathcal{R}^{i}=0~.
\end{equation}
This last relation holds because the mannifold is $\left(  2n\right)
-$dimensional, and therefore the generalized Kronecker delta $\delta_{\left[
2n+1\right]  }^{\left[  2n+1\right]  }$ in the antisymmetric contraction
necessarily has repeated indices, which identically sets it to zero. Then, we have%
\begin{gather}
\delta^{2n-2i-2k-1}\left(  \left(  h^{-1}\delta h\right)  _{s}^{s}%
\mathcal{R}^{i}K^{2k+1}-2i\left(  h^{-1}\delta h\right)  _{s}^{i_{2}%
}\mathcal{R}_{j_{1}j_{2}}^{i_{1}s}\mathcal{R}^{i-1}K^{2k+1}\right)
=
\nonumber\\
\left(  2n-2i-2k-1\right)  \delta^{2n-2i-2k-2}\delta_{j_{1}}^{s}\left(
h^{-1}\delta h\right)  _{s}^{i_{1}}K^{2k+1}\mathcal{R}^{i}\nonumber\\
+\left(  2k+1\right)  \delta^{2n-2i-2k-1}\left(  h^{-1}\delta h\right)
_{s}^{i_{1}}K_{j_{1}}^{s}K^{2k}\mathcal{R}^{i}~.
\end{gather}

Then, $\delta A_{ik}$ simplifies to%
\begin{align}
\delta A_{ik}  &  =\frac{\sqrt{-h}}{2}\delta^{2n-2i-2k-2}\left(  \delta\left(
2k+1\right)  \mathcal{R}^{i}K^{2k}\left(  \left(  h^{-1}\delta h\right)
_{s}^{i_{1}}K_{j_{1}}^{s}+2\delta K_{j_{1}}^{i_{1}}\right)  \right.
\nonumber\\
&  \left.  +\left(  2n-2i-2k-1\right)  \delta_{j_{1}}^{s}\left(  h^{-1}\delta
h\right)  _{s}^{i_{1}}K^{2k+1}\mathcal{R}^{i}\right) \nonumber\\
&  =\frac{\sqrt{-h}\left(  2n-2i-2k-1\right)  }{2}\delta^{2n-2i-2k-2}%
\times\nonumber\\
&  \left(  \left(  \left(  \frac{\left(  2k+1\right)  }{\left(
2n-2i-2k-1\right)  }+1\right)  -1\right)  \delta\mathcal{R}^{i}K^{2k}\left(
\left(  h^{-1}\delta h\right)  _{s}^{i_{1}}K_{j_{1}}^{s}+2\delta K_{j_{1}%
}^{i_{1}}\right)  \right. \nonumber\\
&  \left.  +\delta_{j_{1}}^{s}\left(  h^{-1}\delta h\right)  _{s}^{i_{1}%
}K^{2k+1}\mathcal{R}^{i}\right) \nonumber\\
&  =\frac{\sqrt{-h}\left(  2n-2i-2k-1\right)  }{2}\delta^{2n-2i-2k-2}%
\times\nonumber\\
&  \left(  \left(  \frac{2\left(  n-i\right)  }{2n-2i-2k-1}\right)
\delta\mathcal{R}^{i}K^{2k}\left(  \left(  h^{-1}\delta h\right)  _{s}^{i_{1}%
}K_{j_{1}}^{s}+2\delta K_{j_{1}}^{i_{1}}\right)  \right. \nonumber\\
&  \left.  -K^{2k}\mathcal{R}^{i}\left(  \left(  h^{-1}\delta h\right)
_{s}^{i_{1}}\left(  K_{j_{1}}^{s}\delta_{j_{2}}^{i_{2}}-\delta_{j_{1}}%
^{s}K_{j_{2}}^{i_{2}}\right)  +2\delta_{j_{1}}^{i_{1}}\delta K_{j_{2}}^{i_{2}%
}\right)  \right)~.
\end{align}
Therefore, the variation $\delta B_{2n}$ can be written as the sum of two
terms, such that%
\begin{equation}
\delta B_{2n}=\delta B_{2n}^{\left(  0\right)  }+\delta B_{2n}^{\left(
W\right)  }~,%
\end{equation}
where%
\begin{equation}
\delta B_{2n}^{\left(  W\right)  }=\sqrt{-h}%
{\displaystyle\sum\limits_{i=0}^{n-1}}
{\displaystyle\sum\limits_{k=0}^{n-i-1}}
\frac{n!\left(  -1\right)  ^{k+1}\left(  2n-2i-2k-2\right)  !}{i!k!2^{i}%
\left(  n-i-k-1\right)  !\ell^{2\left(  n-i-k-1\right)  }}\delta_{\left[
2i+2k+1\right]  }^{\left[  2i+2k+1\right]  }\mathcal{R}^{i}K^{2k}J~,
\end{equation}
with%
\begin{equation}
J=\left(  \left(  h^{-1}\delta h\right)  _{s}^{i_{1}}K_{j_{1}}^{s}+2\delta
K_{j_{1}}^{i_{1}}\right)
\end{equation}
and%
\begin{equation}
\delta B_{2n}^{\left(  0\right)  }=\sqrt{-h}%
{\displaystyle\sum\limits_{i=0}^{n-1}}
{\displaystyle\sum\limits_{k=0}^{n-i-1}}
\frac{n!\left(  -1\right)  ^{k}\left(  2n-2i-2k-2\right)  !}{i!k!2^{i+1}%
\left(  n-i-k-1\right)  !\left(  n-1\right)  \ell^{2\left(  n-i-k-1\right)  }%
}\delta_{\left[  2i+2k+2\right]  }^{\left[  2i+2k+2\right]  }K^{2k}%
\mathcal{R}^{i}Q~,
\end{equation}
with%
\begin{equation}
Q=\left(  \left(  h^{-1}\delta h\right)  _{s}^{i_{1}}\left(  K_{j_{1}}%
^{s}\delta_{j_{2}}^{i_{2}}-\delta_{j_{1}}^{s}K_{j_{2}}^{i_{2}}\right)
+2\delta_{j_{1}}^{i_{1}}\delta K_{j_{2}}^{i_{2}}\right)~.
\label{Q_def}%
\end{equation}
Here we have used the fact that%
\begin{equation}
\delta_{\left[  2n\right]  }^{\left[  2n\right]  }\delta^{p}=p!\delta_{\left[
2n-p\right]  }^{\left[  2n-p\right]  }~.
\end{equation}

As we show in what follows, this decomposition of $\delta B_{2n}$ is relevant
because $\delta B_{2n}^{\left(  W\right)  }$ can be written entirely in terms
of a polynomial of contractions of the bulk Weyl tensor, while $\delta
B_{2n}^{\left(  0\right)  }$ can be directly related to the vacuum energy
contribution to the Noether prepotential (from which the asymptotic charges
for the geometry are computed).

We first proceed to simplify $\delta B_{2n}^{\left(  W\right)  }$ considering
the Gauss-Codazzi relation and the on-shell Einstein condition for the bulk
Weyl tensor. In particular, we have that%
\begin{align}
\mathcal{R}  &  =\left(  W+2KK-\frac{2}{\ell^{2}}\delta\delta\right)
,\nonumber\\
\mathcal{R}^{i}  &  =%
{\displaystyle\sum\limits_{j=0}^{i}}
{\displaystyle\sum\limits_{m=0}^{i-j}}
\,\frac{i!\left(  -1\right)  ^{i-j-m}2^{i-j}}{j!m!\left(  i-j-m\right)
!\ell^{2\left(  i-j-m\right)  }}W^{j}K^{2m}\delta^{2\left(  i-j-m\right)  }~.
\end{align}
Therefore,%
\begin{align}
\delta B_{2n}^{\left(  W\right)  }  &  =%
{\displaystyle\sum\limits_{i=0}^{n-1}}
{\displaystyle\sum\limits_{k=0}^{n-i-1}}
{\displaystyle\sum\limits_{j=0}^{i}}
{\displaystyle\sum\limits_{m=0}^{i-j}}
\frac{\sqrt{-h}n!\left(  -1\right)  ^{k+i-j-m+1}\left(  2n-2k-2j-2m-1\right)
!}{k!j!m!2^{j}\left(  n-i-k-1\right)  !\left(  i-j-m\right)  !\left(
2n-2i-2k-1\right)  \ell^{2\left(  n-k-j-m-1\right)  }}\times\nonumber\\
&  \delta_{\left[  2k+2j+2m+1\right]  }^{\left[  2k+2j+2m+1\right]  }%
W^{j}K^{2\left(  k+m\right)  }J~,
\label{QuadrupleSum}%
\end{align}
where we used that%
\begin{equation}
\delta_{\left[  p\right]  }^{\left[  p\right]  }\delta^{q}=\frac{\left(
2n-p+q\right)  !}{\left(  2n-p\right)  !}\delta_{\left[  p-q\right]
}^{\left[  p-q\right]  }~.
\end{equation}

Now, it can be checked (by brute force using an algebra software) that the
expression for $\delta B_{2n}^{\left(  W\right)  }$ given in
(\ref{QuadrupleSum}) simplifies to%
\begin{equation}
\delta B_{2n}^{\left(  W\right)  }=\frac{\sqrt{-h}}{c_{2n}}%
{\displaystyle\sum\limits_{i=0}^{n-1}}
\frac{\left(  -1\right)  ^{i}\left(  n-i-1\right)  !\ell^{2i}}{2^{3i}\left(
n-1\right)  !i!}\delta_{\left[  2i+1\right]  }^{\left[  2i+1\right]  }W^{i}J~,
\end{equation}
where%
\begin{equation}
c_{2n}=\frac{\left(  -1\right)  ^{n}\ell^{2n-2}}{2^{2n-2}n\left[  \left(
n-1\right)  !\right]  ^{2}}~.
\end{equation}

We now proceed to smplify $\delta B_{2n}^{\left(  0\right)  }$. We have
\begin{align}
\delta B_{2n}^{\left(  0\right)  }  &  =\sqrt{-h}%
{\displaystyle\sum\limits_{i=0}^{n-1}}
{\displaystyle\sum\limits_{k=0}^{n-i-1}}
\frac{n!\left(  -1\right)  ^{k}\left(  2n-2i-2k-2\right)  !}{i!k!2^{i+1}%
\left(  n-i-k-1\right)  !\left(  n-1\right)  \ell^{2\left(  n-i-k-1\right)  }%
}\delta_{\left[  2i+2k+2\right]  }^{\left[  2i+2k+2\right]  }K^{2k}%
\mathcal{R}^{i}Q\nonumber\\
&  =\sqrt{-h}%
{\displaystyle\sum\limits_{i=0}^{n-1}}
{\displaystyle\sum\limits_{k=0}^{n-i-1}}
\frac{n!\left(  -1\right)  ^{k}}{i!k!2^{i}\left(  n-i-k-1\right)
!\ell^{2\left(  n-i-k-1\right)  }}\frac{1}{2n-2i}\delta_{\left[  2n\right]
}^{\left[  2n\right]  }\delta^{2n-2i-2k-2}K^{2k}\mathcal{R}^{i}Q\nonumber\\
&  =n\sqrt{-h}\delta_{\left[  2n\right]  }^{\left[  2n\right]  }%
{\displaystyle\int\limits_{0}^{1}}
dtt%
{\displaystyle\sum\limits_{i=0}^{n-1}}
\frac{\left(  n-1\right)  !\left(  -1\right)  ^{n-i-1}t^{2n-2i-2}}{i!\left(
n-1-i\right)  !2^{i}}\mathcal{R}^{i}\times\nonumber\\
&
{\displaystyle\sum\limits_{k=0}^{n-i-1}}
\frac{\left(  n-1-i\right)  !\left(  -1\right)  ^{n-i-k-1}}{k!\left(
n-i-k-1\right)  !\ell^{2\left(  n-i-k-1\right)  }}\delta^{2n-2i-2k-2}%
K^{2k}Q\nonumber\\
&  =n\sqrt{-h}\delta_{\left[  2n\right]  }^{\left[  2n\right]  }%
{\displaystyle\int\limits_{0}^{1}}
ds\,s\left(  \frac{1}{2}\mathcal{R}-s^{2}\left(  KK-\frac{1}{\ell^{2}}%
\delta\delta\right)  \right)  ^{n-1}Q~.
\end{align}

Finally, we consider that%
\begin{equation}
\delta I_{EH}^{ren}=\frac{1}{16\pi G_{\mathrm{N}}}%
{\displaystyle\int\limits_{\partial M}}
d^{2n}x\left[  16\pi G\left(  -\Theta_{EH}\right)  +c_{2n}\left(  \delta
B_{2n}^{\left(  W\right)  }+\delta B_{2n}^{\left(  0\right)  }\right)
\right]~,
\end{equation}
where%
\begin{equation}
-16\pi G_{\mathrm{N}}\Theta_{EH}=\sqrt{-h}J
\end{equation}
and%
\begin{equation}
\delta B_{2n}^{\left(  W\right)  }=\frac{\sqrt{-h}}{c_{2n}}\left(  1+%
{\displaystyle\sum\limits_{i=1}^{n-1}}
\frac{\left(  -1\right)  ^{i}\left(  n-i-1\right)  !\ell^{2i}}{2^{3i}\left(
n-1\right)  !i!}\delta_{\left[  2i+1\right]  }^{\left[  2i+1\right]  }%
W^{i}\right)  J~.
\end{equation}
Therefore, the first term in $\delta B_{2n}^{\left(  W\right)  }$ exactly
cancels the EH boundary term, and we have that%
\begin{equation}
\delta I_{EH}^{ren}=\delta I^{\left(  W\right)  }+\delta I^{\left(  0\right)
}~, \label{Delta_ren_odd}%
\end{equation}
where%
\begin{equation}
\delta I^{\left(  W\right)  }=\frac{1}{16\pi G_{\mathrm{N}}}%
{\displaystyle\int\limits_{\partial M}}
d^{2n}x\sqrt{-h}%
{\displaystyle\sum\limits_{i=1}^{n-1}}
\frac{\left(  -1\right)  ^{i}\left(  n-i-1\right)  !\ell^{2i}}{2^{3i}\left(
n-1\right)  !i!}\delta_{\left[  2i+1\right]  }^{\left[  2i+1\right]  }%
W^{i}\left(  \left(  h^{-1}\delta h\right)  _{s}^{i_{1}}K_{j_{1}}^{s}+2\delta
K_{j_{1}}^{i_{1}}\right)
\end{equation}
and%
\begin{align}
\delta I^{\left(  0\right)  }  &  =\frac{nc_{2n}}{16\pi G_{\mathrm{N}}}%
{\displaystyle\int\limits_{\partial M}}
d^{2n}x\sqrt{-h}\delta_{\left[  2n\right]  }^{\left[  2n\right]  }%
{\displaystyle\int\limits_{0}^{1}}
dtt\left(  \frac{1}{2}\mathcal{R}-t^{2}\left(  KK-\frac{1}{\ell^{2}}%
\delta\delta\right)  \right)  ^{n-1}\times\nonumber\\
&  \left(  \left(  h^{-1}\delta h\right)  _{s}^{i_{1}}\left(  K_{j_{1}}%
^{s}\delta_{j_{2}}^{i_{2}}-\delta_{j_{1}}^{s}K_{j_{2}}^{i_{2}}\right)
+2\delta_{j_{1}}^{i_{1}}\delta K_{j_{2}}^{i_{2}}\right)~.
\label{DeltaIZero}%
\end{align}

The $\delta I^{\left(  W\right)  }$ term can be analyzed in the same way as
$\delta I_{EH}^{ren}$ in the even-dimensional bulk case (see
Eq.~(\ref{DeltaIrenEven})). In particular, considering the ACF condition, only
the $i=1$ term will contribute to $\delta I^{\left(  W\right)  }$ at the
conformal boundary, as all higher order terms vanish faster than the
normalizable mode. Thus, we have%
\begin{equation}
\delta I^{\left(  W\right)  }=\frac{1}{16\pi G_{\mathrm{N}}\left(  2n-2\right)  }%
{\displaystyle\int\limits_{\partial M}}
d^{2n}x\sqrt{-h}\frac{\ell^{2}}{4}\delta_{\left[  3\right]  }^{\left[
3\right]  }\left(  -W\right)  \left(  \left(  h^{-1}\delta h\right)
_{s}^{i_{1}}K_{j_{1}}^{s}+2\delta K_{j_{1}}^{i_{1}}\right)~.
\end{equation}
Then, considering the symmetry properties of $W_{kl}^{ij}$ and the fact that
its full trace $W_{ij}^{ij}$ is zero, we have%
\begin{equation}
\delta I^{\left(  W\right)  }=\frac{1}{16\pi G_{\mathrm{N}}\left(  2n-2\right)  }%
{\displaystyle\int\limits_{\partial M}}
d^{2n}x\sqrt{-h}\ell^{2}\left(  W_{jl}^{il}\right)  \left(  \left(
h^{-1}\delta h\right)  _{s}^{j}K_{i}^{s}+2\delta K_{i}^{j}\right)~.
\end{equation}

Finally, considering that $\delta K_{i}^{j}$ is of $O\left(  \rho\right)  $,
$K_{j}^{i}=\frac{1}{\ell}\delta_{j}^{i}+\ldots$ and the definition of the
electric part of the Weyl tensor as $\epsilon_{j}^{i}=W_{j\rho}^{i\rho
}=-W_{jl}^{il}$, we obtain that the non-vanishing contribution to $\delta
I^{\left(  W\right)  }$ at the conformal boundary is given by%
\begin{equation}
\delta I^{\left(  W\right)  }=\frac{\ell}{16\pi G_{\mathrm{N}}\left(  2n-2\right)  }%
{\displaystyle\int\limits_{\partial M}}
d^{2n}x\sqrt{-h}\left(  -\epsilon_{j}^{i}\right)  \left(  h^{-1}\delta
h\right)  _{i}^{j}~,
\end{equation}
which is consistent with the definition of conformal mass of \cite{Ashtekar:1999jx}, where the asymptotic charges (excluding the vacuum energy
contribution) are computed from the electric part of the Weyl tensor.

We now analyze the $\delta I^{\left(  0\right)  }$ term. By considering the
Gauss-Codazzi relation and the on-shell Einstein condition once more, we have
that the term inside the parenthesis in (\ref{DeltaIZero}) can be rewritten as%
\begin{equation}
\frac{1}{2}\mathcal{R}-s^{2}KK+\frac{s^{2}}{\ell^{2}}\delta\delta=\frac{1}%
{2}W+\left(  1-s^{2}\right)  KK-\frac{\left(  1-s^{2}\right)  }{\ell^{2}%
}\delta\delta~.
\end{equation}
Then, considering that $W\sim O\left(  \rho^{\frac{d}{2}}\right)  $ through
the ACF condition, and that $K=\frac{1}{\ell}\delta+\ell\rho S_{\left(
0\right)  }+\ldots$, we see that the $O\left(  1\right)  $ part of this term
cancels out. Therefore, to the lowest order,%
\begin{equation}
\left(  \frac{1}{2}\mathcal{R}-s^{2}KK+\frac{s^{2}}{\ell^{2}}\delta
\delta\right)  \sim O\left(  \rho\right)~.
\end{equation}
Now, we consider the $Q$ term, given by (\ref{Q_def}). Because $\left(
h^{-1}\delta h\right)  \sim O\left(  1\right)  $, $\delta K\sim O\left(
\rho\right)  $ and because the $O\left(  1\right)  $ part of $\left(
K_{j_{1}}^{k}\delta_{j_{2}}^{i_{2}}-\delta_{j_{1}}^{k}K_{j_{2}}^{i_{2}%
}\right)  $ cancels, we have that $Q$ is of order $\rho$. Thus,%
\begin{equation}
\left[  \left(  h^{-1}\delta h\right)  _{k}^{i_{1}}\left(  K_{j_{1}}^{k}%
\delta_{j_{2}}^{i_{2}}-\delta_{j_{1}}^{k}K_{j_{2}}^{i_{2}}\right)
+2\delta_{j_{1}}^{i_{1}}\delta K_{j_{2}}^{i_{2}}\right]  \left(  \frac{1}%
{2}\mathcal{R}ie-t^{2}KK+\frac{t^{2}}{\ell^{2}}\delta\delta\right)  ^{n-1}\sim
O\left(  \rho^{n}\right)~,
\end{equation}
where $n=\frac{d}{2}$, and therefore $\delta I^{\left(  0\right)  }\sim
O\left(  1\right)  $, giving another finite contribution to the variation. It
can be seen that this contribution is related to the casimir energy of the CFT
in the gauge/gravity correspondence.

In order to verify the Dirichlet condition at the conformal boundary, we note
that in $\delta I^{\left(  0\right)  }$ both the terms that depend on
$h^{-1}\delta h$ and $\delta K$ contribute at the same order, and therefore,
one cannot simply neglect the contribution due to $\delta K$, not even
asymptotically. However, one can show that the total variation can still be
written as a variation of $\delta g_{\left(  0\right)  }$, and therefore, the
action is consistent with a Dirichlet condition at the conformal boundary. To
see this, we consider that%
\begin{equation}
\delta K_{j}^{i}=\rho\ell\delta\left(  S_{\left(  0\right)  }\right)  _{j}%
^{i}+\ldots,
\end{equation}
where $\left(  S_{\left(  0\right)  }\right)  _{j}^{i}$ is the Schouten tensor
of $g_{\left(  0\right)  }$~. Also, we have that%
\begin{align}
\left(  S_{\left(  0\right)  }\right)  _{j}^{i}  &  =\frac{1}{\left(
2n-2\right)  }\left(  R_{j}^{i}-\frac{R}{2\left(  d-1\right)  }\delta_{j}%
^{i}\right) \\
&  =\frac{1}{\left(  2n-2\right)  }\left(  \delta_{l}^{k}g^{ip}R_{pkj}%
^{l}-\frac{\delta_{j}^{i}\delta_{l}^{n}g^{km}R_{mnk}^{l}}{2\left(  d-1\right)
}\right)~,
\end{align}
and therefore, $\delta S_{\left(  0\right)  }$ will contain a part that is
proportional to $\delta g_{\left(  0\right)  }$ and another part proportional
to $\delta Rie_{\left(  0\right)  }$. Then, considering that $\delta
Rie_{\left(  0\right)  }$ gives a term that depends on the covariant
derivative of the variation of the connection (with respect to $g_{\left(
0\right)  }$), and that, after a suitable integration by parts, the boundary
terms vanish (by the previous argument that the conformal boundary has no
boundary), it is apparent that the total variation $\delta K$ can be written
entirely in terms of a variation of $\delta g_{\left(  0\right)  }$.

Therefore, assuming the ACF condition, $\delta I_{EH}^{ren}$ for
odd-dimensional bulk manifolds given in (\ref{Delta_ren_odd}) is both
finite and consistent with a Dirichlet condition for $g_{\left(  0\right)  }$
at the conformal boundary. The resulting expression is also in agreement with
the form for the arbitrary variation given in
\cite{Miskovic:2008ck,Miskovic:2010ui}.

As a final remark we mention that, when comparing the variation of the
Kounterterm-renormalized Einstein-AdS action to that obtained through the
standard holographic renormalization procedure, they may differ by a finite
term, which should not spoil the holographic properties of the theory. This may
happen because, as shown by Graham, in the case of even-dimensional bulk
manifolds the finite part of the on-shell action is non-universal \cite{Graham:1999jg}. An example of this type of ambiguity may be seen in the
computation of the holographic stress tensor by Balasubramanian and Kraus
\cite{Balasubramanian:1999re}, where quadratic counterterms at the boundary of
AdS$_{5}$ may be introduced, which would modify the stress tensor, but whose
contribution to the anomaly is simply a total derivative (proportional to
$\square R$), which matches the usual ambiguity of the conformal anomaly with
respect to total derivative terms in the CFT side (i.e., the last term in (\ref{tii})), and can therefore be
neglected as being scheme-dependent.

\bibliographystyle{JHEP}
\bibliography{oddbib}

\end{document}